\documentclass[11pt,preprint,flushrt,eqsecnum]{aastex}

\def\eqq#1{Equation~(\ref{#1})}
\newcommand\etal{{\it et al.\/}}
\newcommand{\bfx}{\mbox{\bf x}}
\newcommand{\bfu}{\mbox{\bf u}}
\newcommand{\bfS}{\mbox{\boldmath $\Sigma$}}
\newcommand{\bfe}{\mbox{\boldmath $\eta$}}
\newcommand{\bfd}{\mbox{\boldmath $\delta$}}

\begin{document}

\slugcomment{CVS Revision: 2.2 2001/10/10, Accepted to AJ}

\title{Shapes and Shears, Stars and Smears:  Optimal Measurements for
Weak Lensing}

\author{G. M. Bernstein \& M. Jarvis}
\affil{Dept. of Astronomy, University of Michigan, Ann Arbor, MI 48109}
\email{garyb, jarvis@astro.lsa.umich.edu}

\begin{abstract}
We present the theoretical and analytical bases of optimal
techniques to measure weak gravitational shear from images of
galaxies.  We first characterize the geometric space of shears and
ellipticity, then use this geometric interpretation to analyse images.
The steps of this analysis include: measurement of object
shapes on images, combining measurements of a given galaxy on
different images, estimating the underlying shear from an ensemble
of galaxy shapes, and compensating for the systematic effects of image
distortion, bias from PSF asymmetries, and ``dilution'' of the signal
by the seeing. 
These methods minimize the ellipticity measurement noise,
provide calculable shear uncertainty estimates, and allow removal of
systematic contamination by PSF effects to arbitrary precision.
Galaxy images and PSFs are expressed as ``Laguerre expansions,'' a 2d
generalization of the Edgeworth expansion,
making the PSF
correction and shape measurement relatively straightforward and
computationally efficient.
We also discuss sources of noise-induced bias in weak lensing
measurements---selection biases, and ``centroid'' biases arising from
noise rectification---and provide a solution for these and previously
identified biases.
\end{abstract}

\keywords{gravitational lensing---methods: data analysis---techniques:
image processing}

\section{Introduction}
Gravitational lensing is a powerful tool for studying the
distribution of matter in the Universe, because photons are deflected
by all forms of matter, regardless of luminosity, composition or
dynamical state.  Dramatic manifestations of lensing---multiple
images, Einstein rings, and giant arcs, so-called {\em strong}
lensing---provide much information on the highest overdensities in the
Universe, namely rich galaxy clusters, cores of individual galaxies,
or collapsed objects.  To characterize the more typical mass
structures, or those without a fortuitously aligned bright background
source, we may use {\em weak} gravitational lensing, in which we
analyze the low-order distortions of the ubiquitous background galaxies
in order to infer the mass distribution.  Weak gravitational lensing
signals are extraordinarily subtle, even by astronomical standards:
one seeks a shear (or magnification) of the galaxy images
amounting to a few percent at most, more typically 0.2--1\% in current
studies.  Because the undistorted image is
not observable, the lensing distortions must be detected as a
perturbation to the 
intrinsic distribution in galaxy shapes (or sizes), which have
variation of 30\% or more, giving a signal-to-noise ratio ($S/N$) of
$\sim1/30$ from observation of a single galaxy.
Hence a very large number of galaxies must be
observed before the weak lensing becomes detectable over this
intrinsic {\em shape noise}.  Weak lensing analyses could not even be
attempted until automated means of measuring very large numbers of
galaxy shapes became available \citep{Va83, Ty84}.
Furthermore, optical and atmospheric
distortions in a typical sky image cause coherent shape (and size)
distortions that can masquerade as a lensing signal.  Such systematic
errors are 1--10\% in a typical image, up to 50 times larger than the
weak lensing signals.  A means to remove this contamination is
crucial; the necessary analyses can only be conducted with
well-calibrated, linear detectors.

Successful detection of a weak lensing signal did not occur until CCD
images of sufficient depth and field were available \citep{TVW}, and
early detections were of the $\approx10\%$ shears that are found
in the inner regions of rich clusters of galaxies \citep{Fa94, Bo94,
Sm95}. In regions of strong shear, the $S/N$ is 
sufficiently high that a map of the lensing mass can be created
\citep{KSB}.  \citet{Me99} includes a review of results from
cluster-lensing studies.

With the increase in collecting area of CCD imagers, sufficient
background galaxies can be measured to allow convincing detection of smaller
shear signals around weaker overdensities:  around individual
weak clusters \citep{Fi97} or collections of galaxy groups
\citep{Ho01}; around individual galaxies \citep{F00,
Sm01, Wi01}.  Most dramatically, lensing signals on random
lines of sight,
caused by the background matter fluctuation spectrum, have now been
detected and are one tool for ``precision cosmology'' \citep{Wi00,
VW00, Ba00, Wi01a}.  As technology has advanced, weaker and
weaker shears have become detectable under the shape noise, sometimes
as small as a few tenths of a percent ({\it e.g.} Fischer \etal\ 
2000, Jarvis \etal, in prep.).
As a consequence, the demands for rejection of systematic errors have
become more stringent. In many current weak lensing publications, it
is clear that the uncorrected systematic effects are only slightly
smaller than the signals under study.  It is therefore fair to say
that, at present, it is the analysis techniques, rather than the
ability to collect galaxy images, that bar the way to higher precision
in many weak-lensing studies. 

This paper describes the techniques for extraction of weak-lensing
signals from imaging data, which we have developed over the past few
years to meet these increasing demands.  As described below, our
efforts focus on the shear rather than the magnification of the galaxy
images by the lens, and hence we are measuring galaxy ellipticities.
The desiderata for a weak-lensing methodology include:
\begin{enumerate}
\item Shapes of individual galaxy images are determined with the highest
possible accuracy in the presence of measurement noise on the image.
\item Each shape measurement should have a known error distribution.
\item Individual galaxy shapes should be combined to
yield an estimate of the underlying lens shear with maximal $S/N$.
\item The shear estimator should have an error level and a calibration
that can be derived directly from the data, without recourse to
Monte-Carlo simulations. 
\item The galaxy shapes should be corrected for the systematic biases
due to the point-spread function (PSF), to arbitrary precision.
\item The scheme must allow for a PSF that varies continuously across
the image and is different in each exposure.
\end{enumerate}

Given the intrinsic floor on weak lensing accuracy because of shape
noise, one might ask why we should expend much effort on goal (1),
which is to minimize the effects of measurement noise---normally, we
consider that once the ellipticity measurement noise $\sigma_e$ is $\ll
\sigma_{SN}\equiv \langle e^2/2\rangle^{1/2}\approx0.3$,  further
gains do not increase the shear 
estimation accuracy---the error $\sigma_\delta$ on the lensing
distortion $\delta$ will just become $\sigma_{SN}/\sqrt{N}$, with $N$
the number of measured galaxies.  We
note first that the sky density of galaxies scales with apparent
magnitude $m$ as $10^{\alpha m}$ with $0.3\lesssim\alpha\lesssim0.4$.  
If we can cut the shape measurement error for a given image noise
level, then we can either use fainter galaxies in our lensing
measurement (increasing $N$), or cut the required exposure time.
Second, note that 
convolution with a PSF suppresses the measured lensing signal {\em and} the
intrinsic shape noise.  Hence the level to which we aim to reduce
$\sigma_e$ must, for poorly resolved galaxies, be well below the
canonical 0.3.  Thirdly, we will see in \S\ref{combineshapes} that it
may be possible to measure the shear to an accuracy much better than
$\sigma_{SN}/\sqrt{N}$, in cases where the distribution of intrinsic galaxy
shapes in the ellipticity plane has a cusp or pole at
$e=0$.  In simple cases such as a population of circular disk galaxies,
the accuracy to which we can measure the applied shear can increase
without limit as the measurement noise is decreased.

The need for traceable uncertainties is also critical as weak lensing
is used to measure the power spectrum of mass fluctuations in the
Universe.  In this application, the measurement uncertainties
(including shape noise) contribute to the power spectrum and must be
accurately estimated and subtracted to reveal the true cosmic power
spectrum.  Of course an accurate calibration is also necessary for
most applications to precision cosmology; if one must rely on
simulated data for the calibration, there is always the danger that
the simulations do not properly incorporate some aspect of the real
world.

Finally, the need for removal of the systematic PSF ellipticities to
arbitrary precision is extremely strong.  In the course of this paper
we will try to describe other approaches to the problem and compare to
our own.
The methods in most common
current use ({\it e.g.} \citet{KSB}, KSB) are formally valid only
certain special 
cases of PSF.  While heuristic adjustment and testing has demonstrated
that the method works to nominal accuracy in more general cases, 
the absence of a generally valid method is troubling.  A formally
exact PSF correction scheme has been put forward by \citet{K00}[K00],
which is based upon a Fourier-domain calculation of the effects of
shear and of PSF convolution.  Our approach will be to decompose the
image and the PSF into a vector over orthogonal polynomials, and treat
the deconvolution as a matrix operation carried to desired order.  A
very similar approach has been independently put forth by
\citet{Re01}. 

This is a longwinded paper, likely to be read in detail only by
practitioners of weak lensing.   A more casual reading will be
beneficial to those who wish to understand the methods and limitations
of past and future weak lensing analyses.  Some of the techniques we
develop may be useful beyond the weak lensing analysis, for example
our deconvolution method (\S\ref{laguerredeconv}) and the methods for
rapid convolution with spatially varying kernels (\S\ref{kernel}).
As discussed by \citet{Re01}, our orthogonal-function decomposition can be
a useful means for compression of galaxy images.

The paper outline is as follows:  the following section describes the
mathematical space occupied by ellipticities and shears.
Understanding the geometry of this space makes it easier to see how
our (and other) measurement schemes work.  In \S\ref{noseeing}, we
describe a scheme which uses our geometrical conceptualization of
ellipticity to produce measurements with maximal $S/N$
in images with infinitesimal PSF; a formula for the resultant
uncertainty in each ellipticity is also derived.
Next, \S\ref{combineexp} discusses several schemes for combining shape
measurements of a given galaxy from different exposures and/or filter
bands, to obtain the shape estimate that again offers the best
possible $S/N$ and a closed-form error estimate.  \S\ref{shearsection}
describes the means to combine shape estimates from different galaxies
to form an optimal estimate of the underlying lensing shear.  In the
absence of measurement noise, this takes a simple closed form; in the
presence of measurement noise, some approximations must be made to
obtain a closed form for the calibration and error of the shear, and
hence we do not fully satisfy goal (4) above.  \S\ref{psf} is a very
extensive discussion of the effects of the PSF on the image, other
approaches to the problem, and our method for
optimal extraction of the intrinsic shape in this case.  In this
section we introduce the Laguerre decomposition technique.
\S\ref{kernel} uses the Laguerre formalism to construct convolution
kernels that can add symmetries to the PSF of an image; this is one
means of removing the ellipticity biases due to the PSF.
\S\ref{centroidbias} discusses two very important effects that can give
rise to biased lensing measurements {\em even when a perfect
deconvolution for PSF effects is possible.}  It is likely that these
biases are present in all previously published data.

Finally, \S\ref{procedures} puts together all of the methods developed
in the paper in a flowchart form describing how raw image data are
converted into optimized, calibrated lensing shear data.  We reserve
for a succeeding paper \citep{Paper2} the detailed discussion of the
code that implements these methods, and a verification of its
performance on real and simulated data.  In Appendices to this paper,
we present the formulae for invoking various transformations on the
Laguerre-decomposition representation of an image, and derive some
approximate PSF-correction formulae that were used for the analyses of
\citet{Sm01} but which are superseded by the full Laguerre
methodology. 

\section{Geometry of Shape and Shear}
\subsection{Linear Approximation to Lensing}
The goal of weak gravitational lensing studies is to infer a distant
gravitational potential via the distortions that the potential's
deflection of light imparts upon the population of
galaxies in the background.  
The lensing is fully characterized by the map $\bfu(\bfx)$ from the
observed angular position \bfx\ to the source angular position \bfu.
The surface brightness observed at \bfx\ is equal to that which would
have been observed at \bfu\ in the absence of the lens.
For an individual background galaxy that is not near a lensing caustic,
the map can be accurately approximated by a Taylor expansion
\begin{equation}
\bfu = { d\bfu \over d\bfx} \bfx + \bfu_0.
\end{equation}
The displacement $\bfu_0$ carries no information 
(unless the source is multiply imaged) because the source
plane is unobservable.
The $2\times2$ amplification matrix has a unique decomposition
of the form
\begin{equation}
{ d\bfx \over d\bfu} = \mu {\bf S} {\bf R},
\end{equation}
where ${\bf R}$ is an orthogonal matrix (rotation); ${\bf S}$ is a
symmetric matrix with unit determinant (shear); and $\mu$ is a scalar
magnification. 

The rotation ${\bf R}$ is not useful for lensing studies because the unlensed
orientation is not known, and the ensemble of background galaxies should be
isotropic and hence any collective statistic should be
unchanged by rotation.  Furthermore the rotation
is absent in the limits of single-screen or weak lensing.

The magnification $\mu$ increases the angular size by $\mu$ and the
galaxy flux by $\mu^2$.  While the unlensed quantities are not observable,
the magnification is still detectable because the mean flux and size of the
population will shift.  The magnification also
reduces the sky-plane density of sources
by $\mu^2$.  The magnification thus modulates the number vs magnitude
relations for a given class of background galaxies, in a manner which
depends upon the size/magnitude/redshift distribution  of the original population.

The shear ${\bf S}$ has two degrees of freedom, corresponding to the
ellipticity and position angle imparted on a circular source galaxy.
For weak lensing this shear is undetectable on a single galaxy because
the unlensed shape is not necessarily circular and is not observed.
The collective distribution of galaxy shapes is assumed to be intrinsically
isotropic, and the applied shear breaks this symmetry, rendering it
detectable and measurable.

Both the shear and the magnification thus produce measurable effects
on the ensemble of galaxies and can in theory be used to quantify the
potential.  Shear measurements have been used for numerous
quantitative studies, but magnification methods still yield at best
marginal detections \citep{Dy01}.  There are several factors that favor
the shear method:  first, the two effects of magnification (increased flux and
reduced areal density) push the counts of background sources in
opposite directions, weakening the signal.  More importantly, the
shear is manifested as a variation in the mean orientation of galaxy
shapes, and this mean is zero in the absence of lensing; the
magnification signal is a modulation of $N(m)$ or some other non-zero
quantity.  It is always far easier to measure a small change from zero
than a small change in a non-zero quantity.  For example, exploitation
of the magnification effect in the weak-lensing regime would require
absolute photometry to much better than 0.01~mag accuracy.
Magnification measurements, on the other hand, give a direct measure
of the projected mass, whereas mass reconstructions from shear data
are degenerate under the addition of a constant-density mass sheet.
Hence magnification data are very useful when there is no {\it a
priori} means of determining the mean mass overdensity in the image.

Henceforth we will ignore the magnification effect and describe how to
optimally measure the shear ${\bf S}$.

\subsection{Parameterizations of Shear}
\subsubsection{Diagonal Shears}
The simplest shear matrix is a small perturbation aligned with the
coordinate axes:
\begin{equation}
{\bf S}_\eta =
\left(
\begin{array}{cc}
1 + \eta/2 & 0 \\ 0 & 1-\eta/2
\end{array}
\right), \qquad \eta\ll 1.
\label{smallshear}
\end{equation}
The effect of this transformation upon a circular source-plane object is
to induce an elongation along the $x$ axis, creating an elliptical
image with axis ratio $q\equiv b/a = 1-\eta$.  We can use this matrix
as a generator for the full family of diagonal shear matrices with
arbitrary $\eta$ to obtain
\begin{equation}
{\bf S}_\eta =
\left(
\begin{array}{cc}
e^{\eta/2} & 0 \\ 0 & e^{-\eta/2}
\end{array}
\right), \quad -\infty<\eta<\infty.
\end{equation}
The set of diagonal shear matrices forms a group under simple matrix
multiplication.  The operation is commutative, and clearly corresponds to simple
addition of the $\eta$ parameters:
\begin{equation}
{\bf S}_{\eta_3} = {\bf S}_{\eta_2} \times {\bf S}_{\eta_1}
\quad \iff \quad \eta_3 = \eta_2 + \eta_1.
\end{equation}
For this reason we will call $\eta$ the {\em conformal shear} and will
find it a useful parameterization of shear.  Other common
parameterizations of shear include the axis ratio $q$, the {\em
distortion} $\delta$ \citep{Mi91}, and the {\em reduced shear} 
$g=\gamma/(1-\kappa)$
\citep{Sc95}, which are related to $\eta$ via
\begin{eqnarray}
q & \equiv & b/a  = e^{-\eta}, \\
\delta & \equiv & {{a^2 - b^2} \over{a^2 + b^2} } = \tanh\eta, \\
g & \equiv & { {1-q} \over {1+q} } = \tanh\eta/2.
\end{eqnarray}
\citet{Bo95} define a further set of shear parameterizations, also
easily expressed in terms of $\eta$:
\begin{eqnarray*}
e \equiv 1-q & = & 2e^{-\eta/2}\sinh\eta/2 \\
\varepsilon \equiv {{1-q^2} \over {1+q^2}} & = & \tanh\eta \\
\delta_B \equiv { {1+q^2}\over{2q}} & = & \cosh\eta \\
\tau \equiv \varepsilon\delta_B & = & \sinh\eta.
\end{eqnarray*}
Note that for $\eta\ll1$, the parameters $\eta, e, \varepsilon, \tau$,
and the distortion $\delta$ are all equal.
Note also that most other author's formulations of shear do not define
the matrix ${\bf S}$ to have unit determinant, so do not form a group.

\subsubsection{General Shear}
A general (non-diagonal) shear matrix can be decomposed into a
diagonal shear and rotations as
\begin{equation}
\label{smatrix}
{\bf S}_{\eta,\beta} = {\bf R}_\beta {\bf S}_\eta {\bf R}_{-\beta}
 = \left( \begin{array}{cc}
\cosh\case{\eta}{2} + \sinh\case{\eta}{2} \cos \theta 
	& \sinh\case{\eta}{2} \sin \theta \\
\sinh\case{\eta}{2} \sin \theta 
	& \cosh\case{\eta}{2} - \sinh\case{\eta}{2} \cos \theta 
\end{array} \right).
\end{equation}
${\bf S}_{\eta,\beta}$ transforms a circular source to an ellipse with
axis ratio $q=e^{-\eta}$ at positional angle $\beta=\theta/2$.  The shear can
be represented as a 2-dimensional vector
\begin{equation}
\bfe \equiv (\eta_+, \eta_\times) \equiv (\eta\cos \theta,
\eta\sin\theta).
\end{equation}
Likewise a shear may be represented as a two-dimensional distortion
$(\delta_+, \delta_\times)$, etc.  A shear $(\eta_+,0)$ creates
ellipses oriented to the $x$ or $y$ axes, while $(0,\eta_\times)$
aligns circular sources to axes at 45\arcdeg\ to the coordinate axes.
The shear \bfe\ is not a vector in the image space, but rather is a
vector in a non-Euclidean shear space that we describe below.

The full set of shear matrices do not form a group under matrix
multiplication as ${\bf S}_{\bfe_2} {\bf S}_{\bfe_1}$ may be asymmetric (two-screen
lenses can effect a rotation for this reason).  But we can form a group
with an addition operation for 2-dimensional shears defined as
\begin{equation}
\label{defoplus}
\bfe_3 = \bfe_2 \oplus \bfe_1 \quad \iff \quad 
{\bf S}_{\bfe_3} {\bf R} = {\bf S}_{\bfe_2} {\bf S}_{\bfe_1},
\end{equation}
where {\bf R} is the unique rotation matrix that allows ${\bf
S}_{\bfe_3}$ to be symmetric. The geometric meaning of the
shears is preserved since ${\bf R}$ will leave a circular
source unchanged.  The simplest expression of the composition
operation in terms of components is
\begin{eqnarray}
\label{addition}
\cosh \eta_3 & = & \cosh \eta_2 \cosh \eta_1 +
	\sinh \eta_2 \sinh \eta_1 \cos (\theta_2-\theta_1) \\
\sinh\eta_3 \sin (\theta_3 - \theta_2) & = & 
	\sinh\eta_1 \sin (\theta_1 - \theta_2). \nonumber
\end{eqnarray}
Note that the second equation is {\em not} symmetric in the two
operands and hence the shear matrix group is non-Abelian. The identity
element is $\eta=0$ and the inverse of $\bfe=(\eta_+,\eta_\times)$ is
$-\bfe=(-\eta_+,-\eta_\times)$.
The addition formula in terms of distortion components is derivable
from (\ref{addition}), and is given by \citet{Mi91}:
\begin{eqnarray}
\label{distadd}
\delta_{3+} & = & { {\delta_{1+} + \delta_{2+} +
(\delta_{2\times}/\delta_2^2) [ 1 - \sqrt{1-\delta_2^2} ] 
(\delta_{1\times}\delta_{2+} - \delta_{1+}\delta_{2\times}) }
\over { 1 + \bfd_1 \cdot \bfd_2} }, \\
\delta_{3\times} & = & { {\delta_{1\times} + \delta_{2\times} +
(\delta_{2+}/\delta_2^2) [ 1 - \sqrt{1-\delta_2^2} ] 
(\delta_{1+}\delta_{2\times} - \delta_{1\times}\delta_{2+}) }
\over { 1 + \bfd_1 \cdot \bfd_2} }. \nonumber.
\end{eqnarray}
We omit the derivations of these equations, which are
straightforwardly but tediously executed by composing the
transformation matrices.  A more elegant derivation follows from
noting that the transformation \eqq{smatrix} transforms the complex
plane as\footnote{
We thank the anonymous referee for this derivation.}
\begin{equation}
z \rightarrow \cosh {\eta \over 2} z + e^{i\theta}\sinh {\eta\over 2}
\bar z.
\end{equation}
It will be useful to consider the limit where $\delta_2\ll1$:
\begin{equation}
\label{smalldist1}
\left.
\begin{array}{ccc}
(d\bfd \oplus \bfd)_+ & \approx &
\delta + (1-\delta^2)d\delta_+ \\
(d\bfd \oplus \bfd)_\times & \approx &
d\delta_\times \\
\end{array}
\right\} \quad d\delta\ll1, \quad \delta_\times=0.
\end{equation}
If we instead make $\delta_1\ll1$, the asymmetry of shear addition is
manifested as a change to the azimuthal component formula:
\begin{equation}
\label{smalldist2}
\left.
\begin{array}{ccc}
(\bfd \oplus d\bfd)_+ & \approx &
\delta + (1-\delta^2)d\delta_+ \\
(\bfd \oplus d\bfd)_\times & \approx &
\sqrt{1-\delta^2}\,d\delta_\times \\
\end{array}
\right\} \quad d\delta\ll1, \quad \delta_\times=0.
\end{equation}
\subsection{The Shear Manifold}
\begin{figure}
\epsscale{0.8}
\plotone{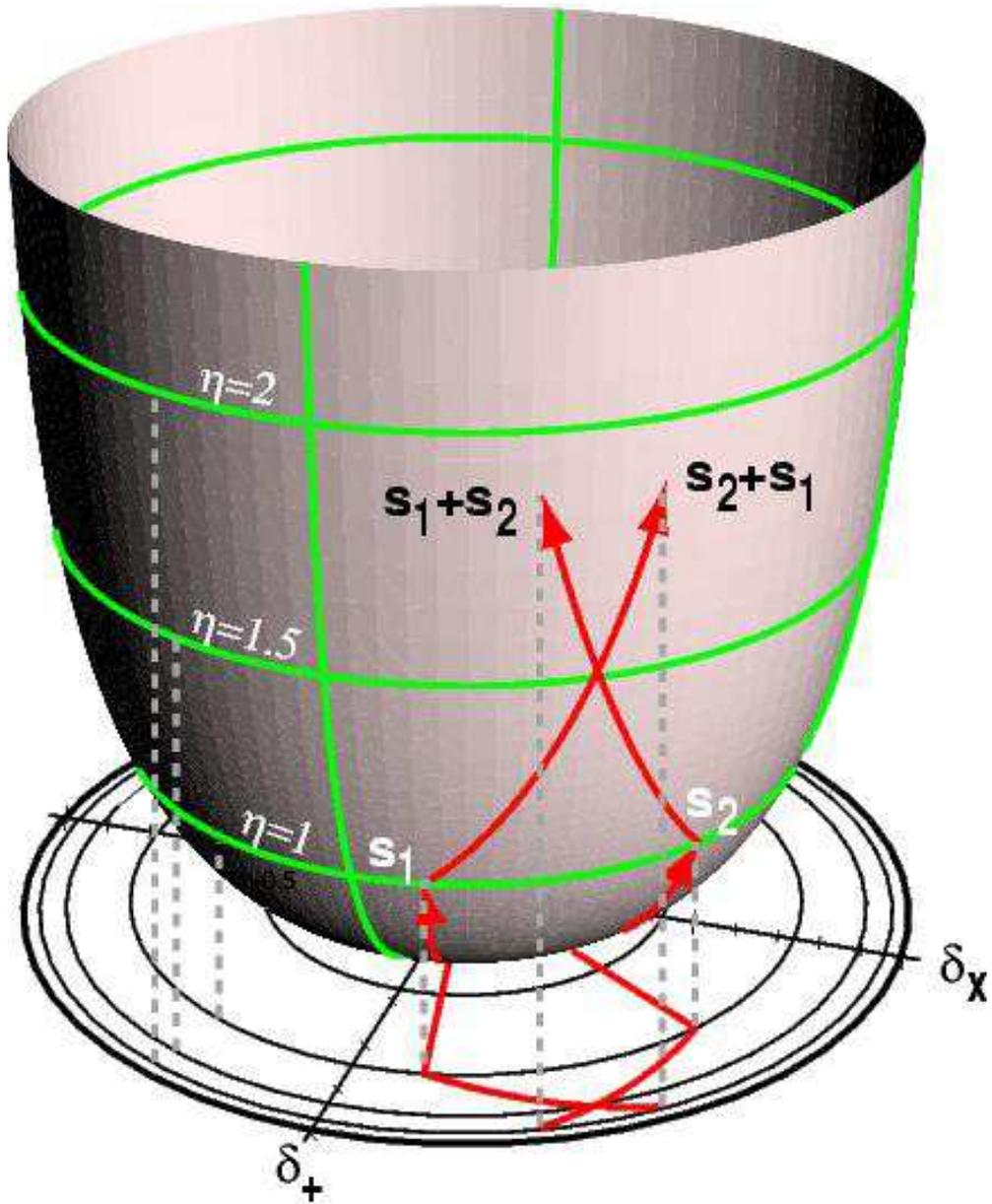}
\epsscale{1.0}
\caption[]{
\small The shaded surface is an embedding of the 
shear manifold into Euclidean space.  The radius vector along this
surface is the conformal shear $\eta$; the radius upon projection onto
the $xy$ plane is the distortion $\delta$ (or ellipticity $e$).  At
small $\eta$ the manifold is tangent to the \bfd\  plane, and at large
$\eta$ approaches the unit cylinder.  Two shear vectors ${\bf s}_1$
and ${\bf s}_2$ of length $\eta=1$ are plotted from the origin, both
on the true shear
manifold and in the \bfd\ plane.  The result of adding
the two vectors is also plotted; displacements do not commute in this
non-Euclidean space.
}
\label{surface}
\end{figure}

We define a metric distance between two points $\bfe_3$ and $\bfe_1$
in shear space as 
\begin{equation}
s \equiv |\bfe_3-\bfe_1| = |\eta_2|, \quad \bfe_2 \equiv \bfe_3 \oplus
(-\bfe_1).
\end{equation}
The differential form of the metric can be derived by specializing
Equations~(\ref{addition}) to the case $ds=\eta_2\ll1$, $\theta_1=0$,
$\theta_2=\theta$, 
yielding
\begin{eqnarray}
\eta_3 & = & \eta_1 + ds\cos\theta \\
 \theta_3 \, \tanh\eta_1 & = & ds\sin\theta
\end{eqnarray}
which means that the metric is
\begin{eqnarray}
ds^2 & = &  (\eta_3 - \eta_1)^2 + \tanh^2\eta_1\,
(\theta_3 - \theta_1)^2 \\
\label{metric}
 & = & d\eta^2 + \tanh^2\eta\,
d\theta^2 \\
 & = & (1-\delta^2)^2 d\delta^2 + \delta^2 d\theta^2.
\end{eqnarray}
Note that the $\eta$ version of the metric has the normal Euclidean
form for the radial component, and the $\delta$ parameterization has
the Euclidean form for the tangential component of the metric, but
neither representation gives a fully Euclidean metric---the shear
space is curved.
The 2-dimensional shear manifold defined by this metric can be
embedded in Euclidean 3-space as illustrated in Figure~\ref{surface}.
This geometric depiction of shear is helpful in understanding the
transformations of shears.
Near $\eta=0$ the surface is tangent to the Euclidean plane, so small
shears add with Euclidean component-wise addition.  The
shear-space surface then curves upwards and as the conformal radius
$\eta$ grows large, the surface approaches a cylinder of unit radius
about the $z$ axis.  If we project the shear surface onto the $z=0$
plane, the radius vector in this plane is equal to the distortion
$\delta$.  The \bfd\ vector is confined within the unit circle.

\subsection{Definition of Shape}
A {\em shear} is a transformation of the image plane; we next need a
quantity to describe the {\em shape} of an arbitrary galaxy image.
Let $I$ represent some object whose isophotes are a family of similar
ellipses.  We can simply parameterize the shape of $I$ by the shear
\bfe\ which produces this object from some object $I_0$ having
circular isophotes, {\it i.e.}
\begin{equation}
I_{\bfe}  = {\bf S}_{\bfe}  I_0 \quad \Rightarrow \quad {\bf S}_{-\bfe}
I_{\bfe}  = I_0.
\label{etacirc}
\end{equation}
We could thus call \bfe\ the {\em conformal shape} of the object, and
can think of a given ellipse as a location on the shear manifold.  More
commonly the distortion is used to define the shape; an object is said
to have {\em ellipticity} ${\bf e}$ if a shear with distortion
$\bfd=-{\bf e}$ makes it circular.  
We will use the symbol ${\bf e}$ since this quantity agrees with the
traditional second-moment definition of ellipticities for truly
elliptical objects.  
Equation~(\ref{etacirc}) makes it
obvious how an ellipse with shape $\bfe_1$ will be
transformed under the action of a shear $\bfe_2$:  we
simply add the shear to the shape using the addition rules of shear
space [Equation~(\ref{addition})]:
\begin{equation}
\label{etamap}
{\bf S}_{\bfe_2} I_{\bfe_1} = I_{\bfe_3}, \quad{\rm with}\;
\bfe_3=\bfe_2\oplus\bfe_1.
\end{equation}
Likewise we can also say that a distortion \bfd\ maps the ellipticity
${\bf e}\rightarrow \bfd\oplus{\bf e}.$  In general, an applied shear may
be viewed as a shift of all shapes along the shear manifold.
We will use ${\bf e}$ to represent the {\em
shape} of an image, whereas \bfd\ represents a {\em shear},
which is a transformation of the image plane.  The shape and shear
spaces, however, transform identically under an applied shear.

A real galaxy has some image intensity distribution $I({\bf x})$ which
may not have elliptical isophotes; we would like to define a shape for
an arbitrary image.  By analogy to Equation~(\ref{etacirc}), all that
is needed is some definition of a ``round'' image.  Let ${\bf M}(I)$
be any measurement applied to the image which has the simple property
that if ${\bf M}(I)=0$, then $M({\bf S}_{\bfe} I)\ne0$ for any non-zero
shear \bfe.  Then the condition ${\bf M}(I)=0$ is our definition of
roundness, and we can assign a unique shape \bfe\ to an arbitrary
object $I({\bf x})$ by the condition
\begin{equation}
\label{etaarb}
{\bf M}({\bf S}_{-\bfe} I) = 0.
\end{equation}
Any shape defined by such a rule clearly transforms under an applied
shear just as an ellipse does, namely via Equation~(\ref{etamap}).
With any roundness measure ${\bf M}$ we therefore have a definition of
shapes and their mapping under shears that follow the rules of
addition in shear space.  We do not attempt to prove that the solution
to \eqq{etaarb} exists or is unique.

\subsection{Shear in Fourier Space}
\label{ftshearsec}
For consideration of the effects of convolution upon sheared images it
will be useful to ponder the action of shears in Fourier space.
We first note that shearing an image $I({\bf x})$ by
\bfe\ is equivalent to shearing its Fourier transform $\tilde I({\bf
k})$ by $-\bfe$.  For a diagonal shear:
\begin{eqnarray}
\label{ftshear}
\widetilde{{\bf S}_{\bfe} I}(k_x,k_y) & = &
	(2\pi)^{-1}\int d^2x\, I(e^{-\eta/2}x, e^{\eta/2}y)
	\exp[2\pi i(k_x x +  k_y y)] \\
	& = & (2\pi)^{-1}\int d^2x\, I(x, y)
	\exp[2\pi i(e^{\eta/2}k_x x + e^{-\eta/2} k_y y)] \\
	& = & I(e^{\eta/2}k_x, e^{-\eta/2} k_y).
\end{eqnarray}
A nondiagonal shear must also satisfy this relation since rotation of
the real-space function corresponds to rotation in $k$-space.
We can therefore just as easily define a galaxy's
shape by a roundness criterion in $k$-space as in real space.  This is
useful when considering finite resolution (\S\ref{psf}) or when
analyzing interferometric images with limited Fourier coverage.

\section{Optimal Measurements (without Seeing)}
\label{noseeing}
In this section we derive an optimum method of measuring galaxy shapes
in the case where the angular resolution and sampling of the
instrument are assumed to be perfect.  In \S\ref{psf} we will treat
the more complex case of seeing-convolved images.

\subsection{The Ideal Test for Roundness}
\label{idealround}
We have defined the shape of an image in Equation~(\ref{etaarb}) by
asking what coordinate shear is needed to make the object appear
round.  We need to choose a measurement ${\bf M}$ which detects with
the highest possible signal-to-noise ratio ($S/N$) any small departure
of the image from its round state ${\bf M}=0$.  It can be shown that,
under some sensible simplifying assumptions about $M$, the solution of
\eqq{etaarb} becomes equivalent to finding the best least-squares fit
of an elliptical-isophote model to the galaxy image.

We assume first that
the measurement will be a linear function of $I({\bf x})$.  Any
non-linear method will prove extremely difficult to apply to the case
where the image has been convolved with a PSF---it will be hard to use
the measurements of bright stars to correct the shapes of faint
galaxies for convolution.  The most general form of ${\bf M}$ is then
\begin{equation}
\label{wt1}
{\bf M}(I) = \int\! d^2\!x\, {\bf W(x)}I({\bf x}),
\end{equation}
where ${\bf W}$ is some weight function.  The weight will be two
dimensional, as the measurement must test for departure from roundness
in both the $\eta_+$ and $\eta_\times$ directions in shear space.

We consider first the weight component to detect a small change in
$\eta_+$.  We can decompose the image $I(r,\theta)$ into multipole
elements $I_m(r)$ via 
\begin{eqnarray}
\label{multipole}
I({\bf x})=I(r,\theta) & = & 
	\sum_{m=-\infty}^\infty I_m(r) e^{im\theta} \\
I_m(r) & = & \frac{1}{2\pi} \int_0^{2\pi} d\theta\,I(r,\theta) e^{-im\theta}
\end{eqnarray}
We are interested in the change in our measurement upon mapping of the
image $I({\bf x}) \rightarrow \tilde I({\bf x}) =
I({\bf Sx})$ where ${\bf S}$ is a shear of amplitude $\eta\ll 1$
oriented on the $x$-axis, as in Equation~(\ref{smallshear}). 
The quantity for which we wish to optimize the $S/N$ can thus be
written as 
\begin{equation}
\label{deltam}
\delta M = M(\tilde I) - M(I) = \sum_{m=-\infty}^\infty\int_0^\infty
r\,dr\,w_m(r) [\tilde I_m(r) - I_m(r)],
\end{equation}
with $w_m$ an arbitrary radial function for each multipole.
A little
bit of algebra yields the transformation of multipoles
\begin{equation}
\label{mpoleshear}
\tilde I_m(r) - I_m(r) =  {\eta \over 4} \left[
	(m-2)I_{m-2} - rI^\prime_{m-2} - (m+2)I_{m+2} -
	rI^\prime_{m+2} \right]
	+ O(\eta^2),
\end{equation}
where the primes denote derivatives with respect to $r$.
For an object with truly circular isophotes, we have $I_m(r)=0$ for
$m\ne0$, and the only effect of the shear is to induce a quadrupole term
$\tilde I_2=-\eta r I^\prime_0/4$.  For objects without perfect
circular symmetry, there are terms beyond the monopole.  But for
an object to be thought of as ``round,'' the monopole term $I_0$
should dominate the higher multipoles.  The monopole is also the only
term guaranteed to be positive for all galaxies.
Hence the largest effect of
the shear will be to alter the $m=\pm2$ quadrupole intensities (which
are conjugates of each other as $I$ is real).  The optimal sensitivity
to small shear should therefore weight only the quadrupole term:
\begin{equation}
\label{w2}
M(I) = \int_0^\infty r\,dr\, I_2(r) r^2 w(r) 
 = {1 \over 2\pi} \int\int r\,dr\,d\theta I(r,\theta)
 w(r) r^2 e^{-2i\theta}.
\end{equation}
It is clear that this quadrupole test is the optimal linear
measurement for objects with circular symmetry; for more general
shapes, the shear has effects on other multipoles that can be measured
and used to enhance signal-to-noise (related to the suggestions of
\citet{Re01}).  But this would require a knowledge 
of $I_4(r)$ and the other multipoles to construct the ideal formula;
we settle on the simple quadrupole as the best general solution, as we
are always guaranteed that $I_0$ is present and positive for any real
source.  The measurement of some weighted quadrupole is also the
normal definition of ellipticity for weak lensing measurements
\citep{Mi91}.

Combining Equations~(\ref{deltam}), (\ref{mpoleshear}), and
(\ref{w2}) we obtain 
\begin{equation}
\label{msignal}
\delta M = {-\eta \over 4} \int r\,dr\, r^2 w(r) [rI^\prime_0 + 4 I_4 +
rI^\prime_4]. 
\end{equation}
The noise in the measure of $M$ can be derived in two limits:  the
most common case will be sky-dominated observations, for which the
variance of the flux in area $A$ is $nA$, where $n$ is
the number of sky photons per unit area (it is assumed that $I$ is in
units of photons).  In this case we have, from Equation~(\ref{w2}),
the variance of each component of $\delta M$
\begin{equation}
\label{varm}
{\rm Var}(M) = {n \over 4\pi^2} \int\int r\, dr\, d\theta r^4 w^2(r) 
\cos^2 2\theta.
\end{equation}
If we ignore the $I_4$ terms in Equation~(\ref{msignal}) as being dominated
by $I_0$ terms, then the choice of weight function which optimizes the
detectability of the shear $\eta$ is
\begin{equation}
\label{wopt}
w_{\rm opt}(r) = {-I^\prime_0 \over r} = -{1 \over r} {dI_0 \over dr}.
\end{equation}
With this optimal weight, the variance of the measurement $M$
would lead to an error in each component of \bfe\ equal to
\begin{equation}
\label{snopt}
\sigma^2_\eta = {4 n \over \pi} \left[ \int r^3 dr\,
	[I^\prime_0(r)]^2\right]^{-1}.
\end{equation}

If the object is much brighter than the night sky, then the noise is
no longer uniform and the optimization becomes
\begin{eqnarray}
\label{shotnoise}
{\rm Var}(M) & = & {1 \over 4\pi^2} \int\int r\, dr\, d\theta r^4 w^2(r) I(r)
\cos^2 2\theta \\
w_{\rm opt}(r) & = & {-I^\prime_0 \over r I_0} = -{1 \over r} {d\ln
I_0 \over dr}, 
\nonumber \\
\sigma^2_\eta & = & {4 \over \pi} \left[ \int r^3 dr\,
	(I^\prime_0)^2/I_0 \right]^{-1}. 
\nonumber
\end{eqnarray}

\subsubsection{Gaussian Objects}
An elliptical Gaussian object, when sheared to be circular, will obey
\begin{equation}
\label{gaussi}
I(r,\theta) = I_0(r) = { f \over {2\pi\sigma^2}} e^{-r^2/2\sigma^2}.
\end{equation}
In the sky-limited case the optimal weight is the same Gaussian:
\begin{equation}
\label{gaussopt}
w_{\rm opt}(r) = {-I^\prime_0 \over r} \propto e^{-r^2/2\sigma^2}.
\end{equation}
Note that the optimal weight for shape measurement is in this case
equal to the optimal filter for detection, {\it i.e.} a matched filter.
If we define the detection significance $\nu$ as the signal-to-noise
ratio for detection of the object with the matched filter, we find
\begin{eqnarray}
\nu^2 & = & { { [\int\! d\!A\,w(r)I(r)]^2} \over {\int\! d\!A\,n w^2(r)}} \\
	& = & {{f^2} \over {4\pi n \sigma^2}};\\
\sigma^2_\eta & = & {16\pi n \sigma^2 \over f^2} \\
	& = & \left( {2 \over \nu} \right)^2.
\end{eqnarray}
We therefore end up with the simple result that the error in each
component of the shear is 2 over the detection significance.

The above derivations assumed that the center and the size $\sigma$ of
the Gaussian were known in advance.  If there were a sky filled with
Gaussian galaxies, we likely would not know in advance the size and
location of each.  We can determine the centroid in the usual manner
by forcing the weighted first moments to vanish:
\begin{equation}
\label{centroid}
\int\! dA\, w I\, re^{i\theta} = 0.
\end{equation}
The weight for centroiding does not necessarily have to match that
used for the shape 
measurement, but it is convenient to do so.  The proper size
$\sigma_w$ for the weight can be forced to match the size $\sigma$ of
the object by requiring the significance to be maximized:
\begin{equation}
\label{sizecriterion}
0={\partial \nu \over \partial \sigma_w} \propto \int\! dA\, w I
(1-r^2/\sigma_2^2). 
\end{equation}

In the limit of a Gaussian with low background noise, the
Equations~(\ref{shotnoise}) apply and the optimal weight is {\em
uniform}.  The detection significance in this case is just $\nu=\sqrt
f$ (with the flux $f$ in photons), and we find again that the standard
error in $\eta$ is equal to $2/\nu$.  In practice this situation can
never be realized because the weight extends to infinity, and at large
radii the sky or read noise will dominate the shot noise from the
galaxy, and neighboring objects will impinge upon the integrations.
We will henceforth confine our discussion to background-limited
observations.  

\subsubsection{Exponential or Other Profiles}
To obtain an optimized measurement of a real galaxy we would have to
measure its radial profile and construct a custom optimized weight
using Equation~(\ref{wopt}).  The majority of galaxies are spirals or
dwarfs which are typically described by exponential profiles:
\begin{equation}
\label{expwt}
I(r) \propto e^{-\alpha r} \quad \Rightarrow \quad 
w_{\rm opt}(r) = {e^{-\alpha r} \over r}.
\end{equation}
This weight diverges at the origin, though all the necessary integrals
of the weight are convergent.  If the galaxy is truly cusped in the
center, then the intensity near the center is very sensitive to small
shears and is weighted heavily.

In practice it is simpler to adopt
a weight function that is universal (up to a scale factor), especially
at low $S/N$ where attempts to measure each individual profile would be
pointless.  There are a number of reasons to prefer a Gaussian weight:
\begin{itemize}
\item The Gaussian drops very quickly at large radii, minimizing
interference from neighboring objects.  Integrals of all moments are
convergent. 
\item Weights with central divergences or cusps are difficult to use
in data with finite sampling, and also amplify the effects of seeing
on the galaxy shapes.  The Gaussian is flat at $r=0$.
\item Gaussian weights are analytically convenient, allowing many
useful formulae to be rendered in closed form.
\item Gaussian weights allow construction of the family of
orthonormal basis functions that we will use in later chapters to
compensate measured shapes for finite resolution.
\item The Gaussian is not far from optimal for most galaxy shapes.
For a well-resolved galaxy with an exponential profile, the Gaussian
weight measures 
\bfe\ with only 7\% higher noise than the optimal weight in
Equation~(\ref{expwt}).  In the presence of seeing, the difference
between the Gaussian and optimal weight is even smaller.  Recall that
{\em any} weight we choose yields a valid definition of roundness and hence
of shape; the Gaussian just incurs a small penalty in noise level.
\end{itemize}

The procedure for measuring galaxy shapes is therefore as follows:
\begin{enumerate}
\item Estimate a shape \bfe\ for the image $I$ and apply the shear ${\bf
S}_{-\bfe}$ to obtain $\tilde I$.
\item Iterate the center and size of the Gaussian weight function
until the centroid condition (\ref{centroid}) and the
maximum-significance condition (\ref{sizecriterion}) are satisfied.
\item Compute the second moments with the Gaussian weight function
\begin{equation}
M(I) = \int\! dA\, w(r) I(r,\theta) r^2 e^{-2i\theta}, \quad
w(r)=e^{-r^2/\sigma^2}. 
\label{gausstest}
\end{equation}
\item If the real and imaginary parts of $M$ are zero, then \bfe\ is
the shape of the object.  If not, then we use the measured $M$ to
generate another guess for \bfe\ and return to step 1.
\end{enumerate}
The process is mathematically equivalent to measuring the second
moments of $I({\bf x})$ with an elliptical Gaussian weight, and
iterating the weight ellipticity, center, and size until they match
the measured object 
shape.  It is therefore an {\em adaptive} second-moment measurement.
The method is also mathematically equivalent to finding the
elliptical Gaussian that provides the best least-squares fit to the
image.  

\subsection{Uncertainties in Shape Estimates}
\label{etaerrors}
Once we have settled upon a weight of the form $w=e^{-r^2/2\sigma^2}$, 
we can integrate $\delta M$ from Equation~(\ref{msignal}) by parts and use
(\ref{varm}) to calculate the variance in \bfe.  We will again ignore
the $I_4$ terms; this means we may have a small tendency to over- or
under-estimate our shape errors if galaxies tend to be boxy or disky.
We first define the weighted flux $f_w$, significance $\nu$, and
weighted radial moments $\langle Ir^m\rangle_w$ as
\begin{eqnarray}
f_w & = & \int\! dA\, w I \\
\nu^2 = f^2_w / {\rm Var}(f_w) & = & f_w^2 / \pi n \sigma^2 \\
\langle Ir^m\rangle_w & = & {1 \over f_w} \int\! dA\,w I r^m.
\end{eqnarray}
The condition (\ref{sizecriterion}) for optimal significance is
\begin{equation}
\langle Ir^2\rangle_w = \sigma^2
\end{equation}
and under this condition the variance in each component of the shape is
\begin{eqnarray}
\label{sigeta}
\sigma^2_\eta & = & {{4\pi n \sigma^2} \over {f_w^2 (1-a_4)^2}} + O(\nu^{-4})\\
	& = & {4 \over {\nu^2 (1-a_4)^2}}  + O(\nu^{-4}) \\
a_4 & \equiv & {{ \langle Ir^4\rangle_w } \over {2\sigma^4} } -1.
\end{eqnarray}
The quantity $a_4$ is a form of kurtosis which is zero for a Gaussian
image. The terms of order $\nu^{-4}$ arise from errors in the centroid
determination, and are discussed further in \S\ref{centroidbias}.

\begin{figure}
\epsscale{0.3}
\plotone{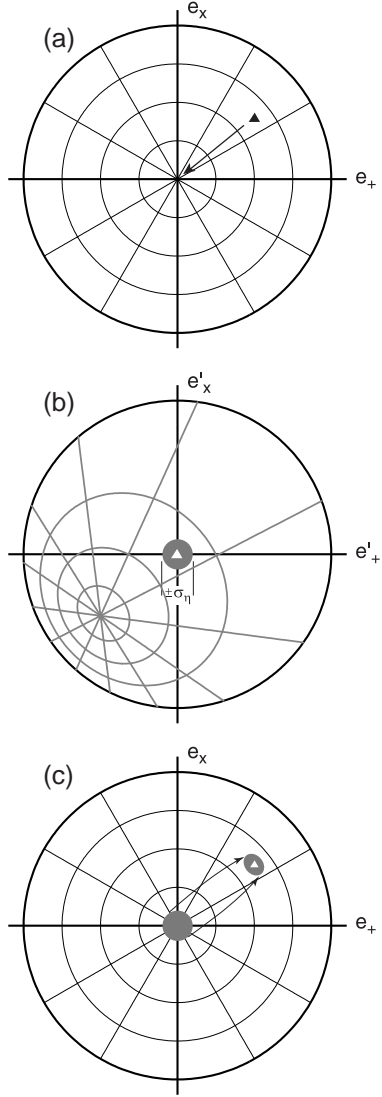}
\epsscale{1.0}
\caption[]{
\small The scheme for ellipticity measurement and its errors are
illustrated schematically:  in (a), the triangle marks the true shape
of a target galaxy, located in the {\bf e} plane.  The shape is
determined by shearing the image until the galaxy appears round.  In
the ellipticity plane, we are moving the object along the vector to
the origin.  Panel (b) shows the location of the target (and the
original coordinate grid) in the {\bf e} plane {\em after} application
of the shear that makes it round.  The shaded region represents the
uncertainty region for the shape in the sheared view---the error
region must be circular because the image is round.  Finally in (c) we
undo the applied shear, shifting the target and the error region back
to the original shape.  This mapping, however, shrinks the error
region by a factor $1-e^2$ ($\sqrt{1-e^2}$) in the
radial (azimuthal) axes.
}
\label{errormap}
\end{figure}

The procedure for measuring an object of shape \bfe\ requires applying
a shear $-\bfe$ to the image coordinates ${\bf x}$ to produce a
coordinate system ${\bf x^\prime}$ in which the object appears round.
The uncertainties in Equation~(\ref{sigeta}) apply in this sheared
coordinate system.  Because the object is round in this frame, there
is no preferred direction in the shear space and the uncertainty
region is circular, with an uncertainty of $\pm\sigma_\eta$ on each
component $(\eta_+^\prime,\eta_\times^\prime)$ measured in the sheared
frame.  We must reapply a shear $+\bfe$ to restore our measurement
to the original coordinate system.  This process is illustrated in
Figure~\ref{errormap}. The $\eta$ coordinate transformations defined in 
Equations~(\ref{addition}), in the limit of
$\eta_1=\sigma_\eta\ll 1$, indicate that the uncertainty region on
\bfe\ will be elliptical, with a shrunken principal axis in the
circumferential direction of the shear manifold:
\begin{equation}
\sigma_\theta  =  {{\sigma_\eta} \over {\sinh\eta}}
 \quad\Rightarrow\quad ds = \tanh\eta \, \sigma_\theta =
 \sigma_\eta\,{\rm sech}\,\eta.
\end{equation}
If we instead use Equations~(\ref{smalldist2}), we can find the
uncertainty ellipse in the ellipticity plane to be
\begin{equation}
\label{errore}
\begin{array}{rcl}
\sigma_e & = & (1-e^2) \sigma_\eta, \\
e\, \sigma_\theta & = & \sqrt{1-e^2} \sigma_\eta.
\end{array}
\end{equation}
So on the ${\bf e}$ unit circle, the uncertainty ellipse shrinks
radially by $1-e^2$ and tangentially by $\sqrt{1-e^2}$ as we transport
the error region from the origin back to the original ellipticity
${\bf e}$.

Note that our derivation assumes that the noise characteristics of the
image are unchanged when we apply a shear.  This is true in the
background-limited case because the noise spectrum is flat, and our
shear matrices ${\bf S}$ have unit determinant.  For an image that has
been smoothed or deconvolved, the power spectrum will have structure
and the shear will alter the noise statistics.  We will discuss this
in \S\ref{psf} in the context of finite image resolution.

\subsection{Comparison with Other Methods}
\label{compare}
Galaxy ellipticities for weak lensing were first determined by
computing unweighted second moments of the intensity \citep{TVW}.  If
the moment integrals are taken to infinity, then the measured
ellipticities transform under shear using the addition rules in
Equation~(\ref{distadd}), and furthermore the correction for PSF
effects is extremely simple.  It is clearly impractical, however, to
carry the integrals to infinity, since neighboring objects will
interfere, the noise is divergent, and, as noted by K00, many common
PSFs have 
divergent second moments.  So the initial methods generally used some
sort of isophotal cutoff to the moments.  This has the disadvantage of
creating moments which are non-linear in the object flux.  An
alternative would be to use unweighted moments within a fixed circular
aperture, but as noted by KSB, the noise properties of unweighted
moments are far from optimal.  

In the KSB method, the measured
ellipticity ${\bf \hat e}_{\rm KSB}$ is computed from the second moments
measured with a circular Gaussian weight with size selected to
maximize the detection significance $\nu$.  [A different weight
function was suggested by \citet{Bo95}.]
The distinction between
the KSB method and ours is that KSB always apply a weight which is
circular in the original image plane; in our adaptive method, the
weight is circular in the sheared image plane which makes the object
round.  Or, as viewed in the original image coordinates, 
the weight is an ellipse with shape iterated to match that of the
object.  This distinction has two consequences:  first, our adaptive
method yields lower uncertainties for non-circular objects because the
weight is a better match to the image.  
This effect is minor, though,
for objects with $e\lesssim0.5$, and a minority of images are more
elliptical then this.

The second, more important advantage over the KSB method is that our
definition of shape via Equation~({\ref{etaarb}) guarantees that our
measured ellipticities ${\bf \hat e}$ are transformed by an applied
shear via Equations~(\ref{distadd}).  The circular-weight ${\bf \hat
e}_{\rm KSB}$ does not have this property---indeed for an object with
elliptical isophotes, ${\bf \hat e}_{\rm KSB}$ does not equal the true
ellipticity.  In the KSB method, it is therefore necessary to
calculate a ``shear polarizability'' for each object, describing the
response of ${\bf \hat e}_{\rm KSB}$ to a small shear \bfd; this
polarizability depends upon the radial profile and $m=4$ moments of
the object.  The ``polarizability'' of our measured ellipticities is
just the $\delta\ll 1$ limit of Equations~(\ref{distadd}):
\begin{equation}
\label{smalldist}
{\bf e}  \rightarrow  { {{\bf e} + \bfd } \over {1+\bfd\cdot{\bf e}}}
  \approx  {\bf e} + \bfd - {\bf e}(\bfd\cdot{\bf e}).
\end{equation}
This transformation rule, and the shapes of our uncertainty
regions in ellipticity space, arise solely from the geometry of the
shear manifold and are independent of the details of the
galaxy images.  This will simplify the following discussions of
methods to derive a shear from an ensemble of measured galaxy shapes. 

We implement the adaptive-weighted-moments scheme in the program {\tt
ellipto}, described further in \citet{Paper2}.

\section{Combining Exposures}
\label{combineexp}
In a typical observing program, a given background galaxy is imaged
in a number of different exposures, in one or more bandpasses.  This
is done to increase exposure time, permit rejection of cosmic rays,
and/or gather color information.  Multiple exposures can also reduce
systematic effects by placing data for a given galaxy on different
parts of the detector and in different seeing conditions.

We hence encounter the question of how to combine data on a
given galaxy in different images to an optimal single measure of the
shape.  There are two possible approaches:
\begin{enumerate}
\item Measure the shape on each exposure, then create a weighted
average of the {\em measurements} as the final shape.
\item Register and average the {\em images}, then measure the object
on the combined image.
\end{enumerate}
We first consider which offers the lowest noise on the final shape.
Consider the task of combining $N$ exposures, with the object having
significance $\nu_0$ on each exposure.  Following
Equation~(\ref{sigeta}), the uncertainty in the shape of a
nearly-round object measured from a single image will be
\begin{equation}
\label{singexp}
\sigma^2_{\eta,0} = {4 \over \nu_0^2} + C^\prime\, [{\rm Var}(x_0)]^2
	= 4\nu_0^{-2} + C\nu_0^{-4}.
\end{equation}
The second term is the uncertainty due to centroiding error, and $C$
and $C^\prime$ are constants of order unity.  If we average
measurements (Method 1), then we decrease $\sigma_\eta$ by $\sqrt N$.
If we average images (Method 2), then we increase $\nu$ by $\sqrt N$.  
The net error on the shape in the two cases is then
\begin{equation}
\sigma_\eta = {2 \over {\sqrt N \nu_0}} \cases{
	\sqrt{1 + C/2\nu_0^2} & \hbox{Method 1} \cr
	\sqrt{1 + C/2N\nu_0^2} & \hbox{Method 2} \cr
}
\end{equation}
The two methods are equivalent, except for the centroiding noise.  If
$\nu_0$ is not $\gtrsim 5$ than averaging images will produce better
accuracy on $\eta$.  Keep in mind that (a) the galaxy will not even be
{\em detected} on the individual exposures unless $\nu_0\gtrsim3$, and
(b) the galaxy is useless for weak lensing unless
$\sigma_\eta\lesssim0.5$, which requires $N\nu_0^2\gtrsim 16$.  When
$N\lesssim 3$, the centroiding penalty is small for any object that will be
useful, so a combined image is extraneous.
When
$N\gtrsim5$, there are many galaxies detectable on the summed image
that are not detectable on the individual images, and a summed image
has detectability and centroid-noise advantages.

There is a compromise,
``Method 1.5,'' which has the practical advantages (delineated below)
of Method 1, while retaining the small $S/N$ edge of Method 2:  that
is to create a summed image and use it for object detection and
centroid determination, so that ${\rm Var}(x_0)\approx 1/N\nu_0^2$.
Then this centroid is used to measure shapes on individual exposures,
and the shape errors are equivalent to Method 2.  In practice we will
combine deconvolved Laguerre coefficients (\S\ref{laguerre}) rather
than measured shapes.

There are several reasons why it may be preferable to average
catalogs instead of images:
\begin{itemize}
\item Correction of shapes for PSF effects is paramount, and only
possible if the PSF is constant or slowly varying across the image. 
If the different exposures in a summed image overlap only partially,
then the PSF (and noise level) will jump discontinuously as one
crosses the boundaries of component exposures.  It is therefore
preferable to correct for PSF effects on an exposure-by-exposure
basis.  If the PSF is very stable ({\it e.g.} a space telescope) or if
the exposures all have nearly the same pointing (a single deep field),
then a summed image will have well-behaved PSF variations.
\item For the smallest objects, the exposures with the best seeing
will contain nearly all the useful shape information and should be
weighted heavily \citep{KF94}.  Large objects are, on the other hand,
measured equally well in every exposure.
Averaging catalogs allows one to adjust the weights of different
exposures on an object-by-object basis, whereas this is not possible
when combining images.
\item If there are exposures in different bands, then the optimal
weighting of the exposures is dependent upon the color of the object.
This is easily done when averaging cataloged shapes but not easily
done by summing images.
\item Creation of a summed image requires registration and
interpolation of pixels.  The latter process smoothes the noise field
and causes subtle variations in the PSF, both of which complicate
later analyses.
\item An especially pernicious hazard to creating a summed image is
that slight mis-registration of the component images will cause
coherent elongations of the images, which if not corrected will mimic
a lensing signal.  This is discussed by KSB; in theory such effects
are handled by a proper PSF correction scheme.  This is a danger for
Method 1.5 as well.
\end{itemize}
Some practical advantages to Method 2, combining images, are that:
\begin{itemize}
\item The data storage and processing requirements can be lower for a
single combined image if $N$ is large.
\item In Method 1, outliers (from cosmic rays) are rejected on an
object-by-object basis, whereas in Method 2 rejection is
pixel-by-pixel.  If the galaxies are very oversampled and the
cosmic-ray rate is high, Method 2 could salvage the un-contaminated
parts of galaxy images that Method 1 discards.
\end{itemize}
For the simplest circumstances (a single-filter stack of images with
common pointing), image averaging is easier and has few drawbacks.  For
multi-filter or mosaicked data, catalog averaging is needed.  The
hybrid Method 1.5 is best for such cases, though more work.  In the
rest of this section we detail procedures for each Method.

\subsection{Combining Images}
\label{combineimage}
There are standard tools for combining exposures into a single image.
We remark here upon a few special considerations when doing this for
weak lensing observations.

First, accurate registration is paramount.  Our scheme for image
registration is described in \citet{Paper2}.

Second, the use of median algorithms is commonplace but dangerous.
Proper correction for PSF effects will require that the images of
bright stars have precisely the same PSF as do the faint galaxies.
But with a median algorithm, the bright, high-$S/N$ stars will be
constructed with a PSF which is a {\em median} of all the exposures.  The
images of faint objects, however, will tend toward a PSF that is the
{\em mean} of all the exposures, because the noise fluctuations will
dominate PSF variations.  The final PSF will therefore vary with
magnitude. A sigma-clipping average is much preferred over the median
for the necessary task of cosmic-ray rejection when combining.

Similarly one must be careful about rejecting saturated pixels.  There
will be many stars which saturate only on the best-seeing exposures;
if the saturated pixels are rejected, these stars will have final PSFs which
are broader than the PSF for faint objects.  One must take care to
ignore stars which are saturated in {\em any one} of the exposures.

\subsection{Combining Shape Measurements}
\label{combineshapes}
Suppose that a given galaxy has been measured to have ellipticity
${\bf e}_i$ in images $i\in\{1,2,\ldots,N\}$.  
We desire the ${\bf e}$ which best estimates the true
ellipticity of the object.
Using the results of
\S\ref{etaerrors},  we see that in the absence of PSF distortions, the 
minimum-variance estimate of ${\bf e}$ will be
that which minimizes the $\chi^2$ given by
\begin{equation}
\chi^2 = \sum_i ({\bf e} \ominus {\bf e}_i) {\bf S}_i^{-1} 
({\bf e} \ominus {\bf e}_i)
\end{equation}
Here, ${\bf e} \ominus {\bf e}_i$ is equivalent to 
${\bf e} \oplus (-{\bf e}_i)$, where $\oplus$ corresponds to the addition 
operator introduced in Equation~(\ref{defoplus}), and ${\bf S}_i$ is a
covariance matrix which is $\sigma_\eta^2{\bf I}$ in simple cases.

Note that if the ${\bf e}_i$ are measured in
different filters, than the galaxy may have no single well-defined
ellipticity.  By ``best estimate,'' then, we must mean that which
offers the best sensitivity to a weak lensing distortion, and the
minimum-variance combination of the ${\bf e}_i$ is still the desired
quantity. 

The $\ominus$ is a non-linear operator, 
so we could use a non-linear minimization algorithm to find the 
value of ${\bf e}$ at which $\chi^2$ is minimized.  However, this is both
impractical for time considerations and unnecessary since the values 
of $\sigma_\eta$ 
are usually small.  Thus, we can linearize the subtraction operator
\begin{equation}
{\bf e} \ominus {\bf e}_i = {\bf T}_i ({\bf e} - {\bf e}_i)
+ O(({\bf e} - {\bf e}_i)^2)
\end{equation}

T can be derived from Equations~(\ref{distadd})
\begin{equation}
{\bf T} = {1 \over {1 - e^2}} \left[ I - {{1 - \sqrt{1-e^2}} \over {e^2}}
\left( 
\begin{array}{cc}
e_\times^2 & -e_\times e_+ \\ -e_\times e_+ & e_+^2
\end{array}
\right) \right]
\end{equation}

The linearized $\chi^2$ becomes
\begin{eqnarray}
\chi^2 & = & \sum_i ({\bf e} - {\bf e}_i) \bfS_i 
({\bf e} - {\bf e}_i), \\
\bfS_i & \equiv & {\bf T}_i^T {\bf S}_i^{-1} {\bf T}_i
\end{eqnarray}
which has a minimum at
\begin{eqnarray}
\label{minimizee}
{\bf e} & = & \left( \sum_i \bfS_i 
\right)^{-1}
\left( \sum_i \bfS_i {\bf e}_i \right) \\
{\rm Cov}({\bf e}) & = & \left( \sum_i \bfS_i 
\right)^{-1}
\nonumber
\end{eqnarray}

This is a standard least-squares solution for the mean of the ${\bf
e}_i$ given covariances $\bfS_i$ in a Euclidean ${\bf e}$ space.
In the simple case of ${\bf S} = \sigma_\eta^2 {\bf I}$, the expression 
for $\bfS_i$ simplifies considerably to
\begin{equation}
\bfS = {1 \over {\sigma_\eta^2 (1-e^2)^2}}
\left(
\begin{array}{cc}
1-e_\times^2 & e_\times e_+ \\ e_\times e_+ & 1-e_+^2
\end{array}
\right), 
\end{equation}
which is equivalent to \eqq{errore}. 
However, when we apply corrections for PSF dilution, we will find that
the covariance 
matrix is more generally an ellipse with axes aligned in the radial
and tangential directions. That is, 
\begin{equation}
\bfS = {\bf R}_\theta 
\left(
\begin{array}{cc}
\sigma_e^2 & 0 \\ 0 & e^2 \sigma_\theta^2
\end{array}
\right) 
{\bf R}_{-\theta}
\end{equation}
where ${\bf R}_{\theta}$ is a rotation matrix with $\theta = \arctan \left( 
{e_\times \over e_+} \right)$.  In either case the minimization
(\ref{minimizee}) is numerically straightforward, and we are left with
an uncertainty ellipse for the mean ${\bar e}$.  It is wise to
implement some outlier-rejection algorithm in this process as well.
%

\section{Estimating Shear from a Population of Shapes}
\label{shearsection}
Now we presume to have measured ellipticities ${\bf e}_i$ for a set of
$N$ distinct galaxies, with known measurement uncertainties for each.
Our final task is to create a statistic $\hat{\bfd}$ from the ${\bf e}_i$
which best estimates the lensing distortion \bfd\ that has been
applied to this ensemble.  There are three main effects which must be
considered in constructing the estimator:  first, the ${\bf e}_i$
respond differently to an applied distortion \bfd, as embodied by
Equation~(\ref{smalldist}) for true ellipticities, or by shear
polarizabilities for the KSB estimators, so we need to know the {\em
responsivity} ${\cal R}\equiv\partial\hat{\bfd} / \partial\bfd$ of our
statistic. Second, the variety of
ellipticities in the parent (unlensed) galaxy population causes {\em
shape noise} in the shear estimate.  In most weak lensing projects
this is the dominant random error, and we wish to minimize its effects.
Third, there is {\em measurement error} in each ellipticity, which we
also wish to minimize in our shear estimator.

Most practitioners have adopted a simple arithmetic mean of $e_+$ and
$e_\times$ as estimators for the applied distortion
({\it e.g.} Fischer \& Tyson 1997). Using the weak-distortion
Equations~(\ref{smalldist}), it is easy to see that, in the absence of
measurement error, this estimator has
a responsivity ${\cal R}=1-\sigma^2_{\rm SN}$ and a variance ${\rm
Var}(\hat\delta_+)=\sigma^2_{\rm SN}/N$, where we have defined the
shape noise $\sigma^2_{\rm SN}\equiv\langle e^2_+ \rangle$ (the
$\times$ component has the same properties and shape noise).

Others have realized, however, that rare, highly elliptical galaxies
have too much influence on the arithmetic mean and should be
deweighted.  Cutoffs on $|{\bf e}|$ \citep{Bo95} or other weighting functions
$w(e)$ \citep{VW00} have been applied to the ellipticities and
tested with simulations, but without any sort of analytic optimization
or justification.  \citet{LB98} consider the optimization of a general
weighted sum of second moments (rather than ellipticities); this
unfortunately couples the ellipticity measurement to the distribution
of {\em sizes} of the galaxies and leads them to consider only weights
which are power-law functions of the moments.

\citet{Ho00} present a weighting scheme which incorporates both
measurement error and the shape noise, and
K00 gives a detailed discussion of optimal weighting for distortion
measurements.  Both are similar to our method in many respects, which
we comment upon at the end of this section.

\subsection{Without Measurement Error}
\label{combineperfect}
We start with an unlensed background galaxy population with ellipticities
distributed within the unit circle according to
\begin{equation}
\label{dnde}
{{dN} \over N} = P({\bf e^\prime}) d^2\!e^\prime = 
P(e^\prime) e^\prime\,de^\prime\,d\theta^\prime.
\end{equation}
A fundamental assumption of weak lensing is that the background is
isotropic so that the unlensed population can have $P$ depend only
upon the amplitude of ${\bf e^\prime}$, not its orientation.
The effect of a distortion \bfd\ is to map the background population
to a new, anisotropic distribution $P_{\bfd}({\bf e})$, as illustrated
in Figure~\ref{pprime}.  We are given a sample of $N$ galaxies from
the new distribution, and our task is to estimate the \bfd\ which gave
rise to the distribution from the original $P$.

\begin{figure}[t]
\plotone{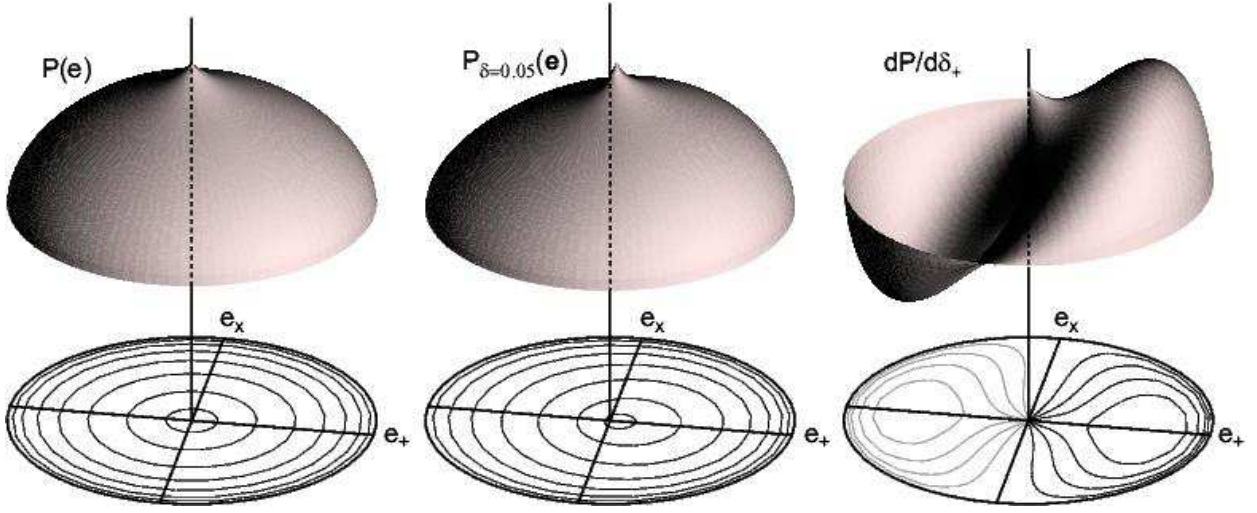}
\epsscale{1.}
\caption[]{
\small The left panel shows a model for intrinsic distribution
$P(e)$ of galaxy shapes over the ${\bf e}$ plane---it must have
circular symmetry.  When each galaxy is
sheared by $\delta_+=0.05$, the galaxy distribution shifts to the
right as in the middle panel.  The right-hand panel shows the
change in population under the applied distortion; this is the signal
which we wish to detect.  Shape noise arises from the Poisson
fluctuations in the population, which is proportional to the left-hand
panel's $P(e)$.  The optimal weight for $\delta_+$ determination is
the ratio of the right-hand to the left-hand panel.
}
\label{pprime}
\end{figure}

One approach is to find the value of \bfd\ which maximizes the 
likelihood of the observed ${\bf e}_i$.  This is true when
\begin{equation}
\label{maxl}
0 = {\partial \over {\partial \delta_+}} \sum \log P_{\bfd}({\bf e}_i)
 = {\partial \over {\partial \delta_+}} \sum \log P(|{-\bfd}\oplus {\bf e}_i|).
\end{equation}
A similar condition holds for $\delta_\times$. These equations define the
maximum-likelihood \bfd\ even for strong distortions---though there is
not in general a closed-form solution for strong \bfd.

For weak lensing ($\delta\ll 1$), Equation~(\ref{smalldist}) and 
the conservation of number can be used to derive $P_{\bfd}({\bf e})$ to 
first order in $\delta$.
\begin{eqnarray}
P_{\bfd}({\bf e})\,d^2\!e 
& = & P(e^\prime)\,d^2\!e^\prime \\
& = & P(|{-\bfd}\oplus {\bf e}|)
	\left| {\partial {\bf e^\prime}}
	\over {\partial {\bf e}} \right| d^2\!e \\ 
& \approx & P\left(e - \bfd\cdot{\bf e} \left( {1-e^2 \over e}\right) \right)
	\left( 1 + 3\bfd\cdot{\bf e} \right) d^2\!e \\
P_{\bfd}({\bf e}) & = & P(e)\left(1 + \bfd\cdot{\bf e} \left( 3 - 
	{1-e^2 \over e} {d\log P \over de} \right) \right) \\
\label{pxfrm}
\log P_{\bfd}({\bf e}) & = & \log P(e) + \bfd\cdot{\bf e}
	\left( 3 - { {1-e^2} \over e} {{d\log P} \over de} \right)
\end{eqnarray}
It can further be shown that the maximum-likelihood
estimator for \bfd\ takes the form
\begin{equation}
\label{maxlsoln}
\bfd = {1 \over N} \sum {\bf e}_i \left( 3 - { {1-e^2} \over e}
{{d\log P} \over de} \right).
\end{equation}
The parenthesized expression is thus a weight
function for combining ellipticities into a distortion.  We can also
show that this weight function is optimal in terms of $S/N$ for weak
distortions, as follows.  Let us create an estimator $\hat{\bfd}$
which is a general weighted sum of the ellipticities,
\begin{equation}
\label{ewt1}
\hat{\bfd} = {{ \sum w(e_i) {\bf e}_i} \over {\sum w(e_i)}} 
= {{\int\! d^2\!e\, w(e){\bf e} P_{\bfd}({\bf e})} \over 
  {\int\! d^2\!e\, w(e) P(e)}}.
\end{equation}
The response of this statistic to a small applied shear is
\begin{eqnarray}
{\cal R} \equiv {{\partial \hat\delta_+} \over {\partial\delta_+}}
& = & {{ \sum \left(w^\prime {{\partial e} \over {\partial \delta_+}} e_+
	+ w { {\partial e_+} \over {\partial \delta_+}} \right)
	\, \sum w - 
	\sum w e_+ \, \sum w^\prime {\partial e \over \partial \delta_+}}
	\over {\left(\sum w\right)^2}} \\
 & = & { \sum \left[ w(1-e^2_+)
	+ w^\prime {e_+^2 \over e} (1-e^2) \right]
	\over {\sum w}}
\label{resp1}
\end{eqnarray}
where in the last line, we have dropped terms linear in $e_+$ or
$e_\times$ which average to zero over an isotropic population.  With
an isotropic population, the derivative 
$\partial\hat\delta_\times/\partial\delta_\times={\cal R}$ as well,
and the off-diagonal elements of $\partial\hat{\bfd}/\partial\bfd$ are
zero.

We may use Equation~(\ref{resp1}) to calculate the response of any
weighted estimator by summing over the {\em observed} $e_i$, because
the small difference between observed and intrinsic distributions does
not alter ${\cal R}$ to first order.  In the case where we have some
analytic form for $P(e)$, we may replace the sums with integrals over
the distribution to obtain
\begin{eqnarray}
\langle {\cal R} \rangle & = & 
	{ {\int\! d^2\!e\, w(e) e_+ (\partial P_{\bfd} / \partial \delta_+)}
	\over {\int\! d^2\!e\, w(e) P(e)} } \\
& = & { {\int\! d^2\!e\, w(e) e_+^2 P(e) \left(3 - 
	{ {1-e^2} \over e} { {d\log P} \over {de} } \right)}
	\over {\int\! d^2\!e\, w(e) P(e)} } 
\label{resp2}
\end{eqnarray}
In the absence of measurement noise, the variance in $\hat{\bfd}$ is
due to shot noise.  Assuming that the background galaxies obey Poisson
statistics and their shapes are randomly assigned, we can propagate
the Poisson errors through Equation~(\ref{ewt1}) to get the expected error
\begin{eqnarray}
\label{rvar1}
{\rm Var}(\hat\delta_+) & = & { { \int\! d^2\!e\, w^2(e) e_+^2 P(e) }
 \over {N [\int\! d^2\!e\, w(e) P(e)]^2} } \\
 & = & { { \sum w^2(e) e_+^2 } \over { \left[\sum w(e)\right]^2} }.
\label{rvar2}
\end{eqnarray}
In (\ref{rvar2}) it is assumed that the sum is over a sufficiently
large ensemble of background galaxies to sample the distribution
$P(e)$.  Any weak lensing measurement has thousands of background
galaxies, so this gives a direct estimate of the error in the shear.

The optimal weight is that which minimizes ${\rm Var}(\delta_+)/{\cal
R}^2$, which is 
\begin{eqnarray}
\label{wopt1}
w_{\rm opt}(e) & \propto & 3 - 
	{ {1-e^2} \over e} { {d\log P} \over {de} }  \\
\label{wopt2}
 & = & 3 + 2 {{ d\log P} \over {d\log(1-e^2)}} \\
\Rightarrow\quad \sigma_\delta & = & \left[ N \int\! d^2\!e \, P(e) e_+^2
	\left( 3 + 2 {{ d\log P} \over {d\log(1-e^2)}}\right)^2
\right]^{-1/2}.
\label{sigdelta}
\end{eqnarray}
where the last line gives the optimized error in $\hat\delta_+ / {\cal
R}$, which is our calibrated estimate of the distortion.
Equation~(\ref{wopt1}) reproduces the maximum-likelihood solution in
Equation~(\ref{maxlsoln}).  This may be compared to the distortion
uncertainty for equal weighting $w=1$,
\begin{equation}
\label{eqwts}
\sigma_\delta = 
	{ 1 \over \sqrt N } 
	\left( {{ \langle e_+^2 \rangle^{1/2} } \over {1-\langle e_+^2
	\rangle}} \right)
	= \left( {{ \sigma_{\rm SN}} \over {1-\sigma^2_{\rm SN}} }\right).
\end{equation}

We first see that a simple arithmetic mean of the ellipticities is the
optimum estimator only if $P\propto (1-e^2)^\alpha$ for some exponent
$\alpha$. For the real galaxy population, there can be a
significant gain in accuracy through use of $w_{\rm opt}$ over equal
weighting.  An extreme case is a population of randomly oriented
circular disks, for which
\begin{eqnarray}
\label{disk1}
P(e) & = & { 1 \over {2\pi e}} (1-e)^{-1/2} (1+e)^{-3/2} \\
\label{disk2}
\Rightarrow \quad w_{\rm opt}(e) & = & { {1+e} \over {e^2}}.
\end{eqnarray}
With $w=1$, we would have $\sigma_\delta=0.590/\sqrt N$.  The optimal weight
diverges at $e\rightarrow0$ to take advantage of the extreme
sensitivity of $P(e)$ to distortion near $e=0$.  The integral in
Equation~(\ref{sigdelta}) in fact diverges at $e=0$, driving
$\sigma_\delta$ to zero---which would be a significant improvement
over the equal-weighting case!  Unfortunately any small measurement
error or departures from circularity for the disks will smooth out the
central spike in $P(e)$, creating a finite value for
$\sigma_\delta$.  

Figure~\ref{pe} shows the $P(e)$ measured for well-measured galaxies in
the CTIO lensing survey (Jarvis \etal, in preparation).  These
shape histograms are derived from 230,000 galaxies which are well
resolved ($R>0.4$, under the definitions in Appendix~\ref{rfactorapp})
and have errors on the intrinsic ellipticity of
$\sigma_\eta\lesssim0.03$---primarily galaxies of magnitude
$17<m_R<22$.  The shape of $P(e)$ is observed to be highly dependent
upon the surface brightness (SB) of the galaxies.
The low-SB galaxies show the rise at $e>0.8$ expected of a disk
population, but there the distribution drops at $e>0.95$ instead of
diverging---this reflects the finite thickness of the disks.
There is also no pole at $e=0$ for the low-SB galaxies, showing that
the disks are not perfectly circular.  The high-SB galaxies are
presumably early types since there are very few with $e>0.5$.  While
the value of $P(e=0)$ increases with surface brightness, it always
remains finite, but with $dP/de<0$.  The ideal weight \eqq{wopt1}
therefore grows as $1/e$ as $e\rightarrow 0$, but the contribution to
the $S/N$ does not diverge at zero as for perfect disks.  
None of the $P(e)$ curves is well fit by a single Gaussian or power
law. 

\begin{figure}
\plotone{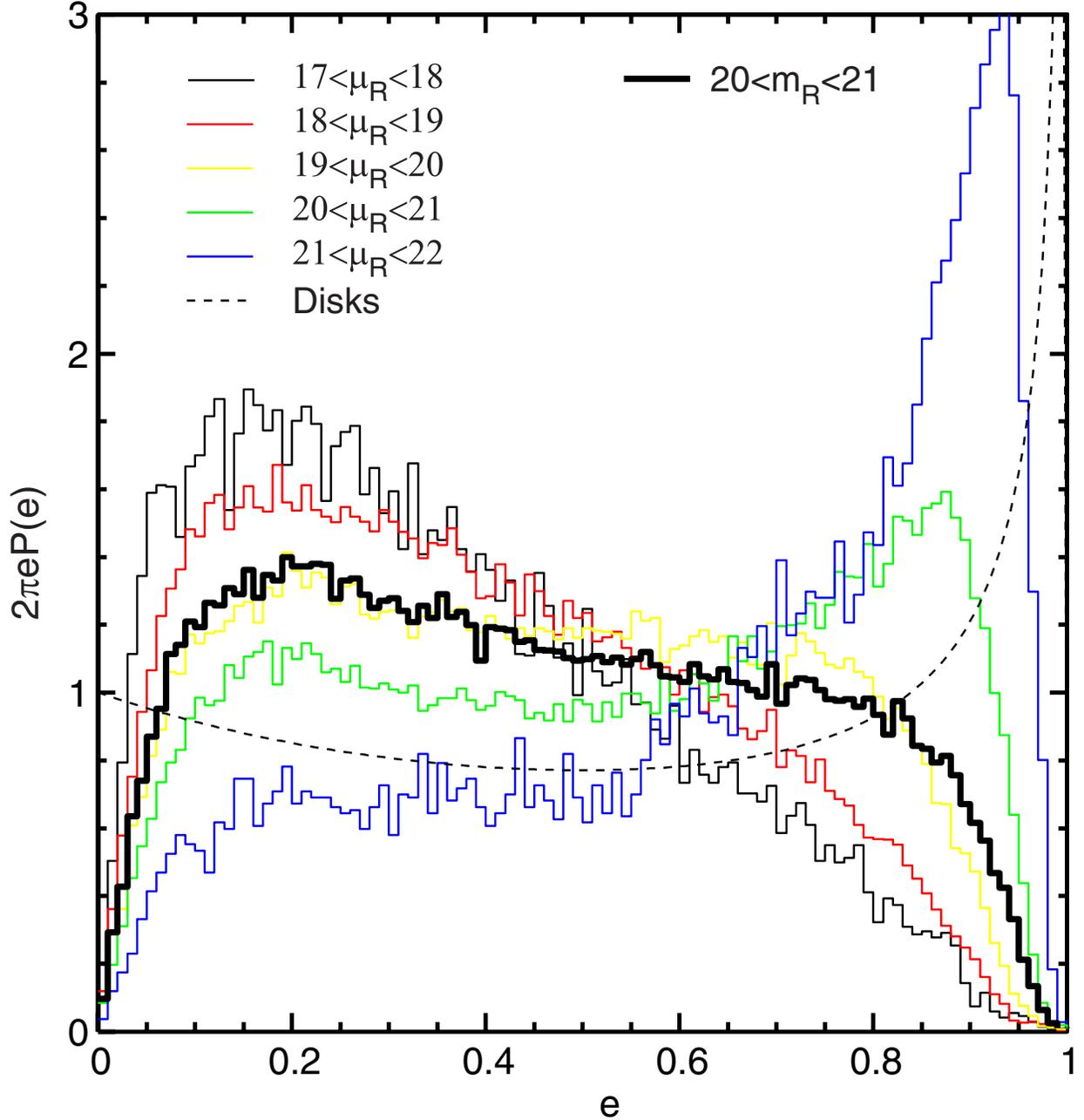}
\caption[]{
\small The distribution of intrinsic ellipticities for modestly bright
galaxies ($m_R\approx 20$) is plotted; we plot $2\pi e P(e)$ rather
than $P(e)$ as the nature of the population is more apparent.  The
distribution is highly dependent upon surface brightness $\mu_R$,
presumably reflecting the difference between spheroid- and
disk-dominated galaxies.  The dashed line is the distribution for an
isotropic population of circular disks.
The high-$\mu$ galaxies are more useful for
distortion measurements. The heavy histogram combines all surface
brightnesses in the magnitude range $20<m_R<21$.  Though it is
difficult to tell from this plot, $P(e)$ is finite and increasing as
$e\rightarrow 0$.  Optimal weighting takes advantage of this structure
to reduce the noise in the distortion measurement.
}
\label{pe}
\end{figure}

The intrinsic ellipticity variance $\sigma_{\rm SN}$ varies from
$0.30$ to $0.49$ between the highest and lowest SB bins.  The optimally
weighted distortion $S/N$ per galaxy for high-SB, early types is 2--3
times higher than for the lowest-SB galaxies, indicating the
desirability of incorporating some galaxy-type discriminant---surface
brightness, color, or concentration---into the weighting scheme.  The
only requirement is of course that the discriminant be independent of
ellipticity.  It seems likely that $P(e)$ will vary substantially with
magnitude.

The heavy histogram in Figure~\ref{pe} combines all well-measured
galaxies with $20<m_R<21$, which we henceforth use as a representative
measure of the real galaxy population.  The distribution has
$\sigma_{\rm SN}=0.38$, which would lead to
$\sigma_\delta=0.44/\sqrt{N}$ for an unweighted average.  The optimal
weighting gives $\sigma_\delta=0.33/\sqrt{N}$; the weighting therefore
gives a gain equivalent to a 1.8-fold increase in $N$.  The gain in
telescope time is at least as large.
This gain is reduced, however, in the presence of measurement noise,
which will tend to wash out the sharp feature in $P(e)$, as discussed
next.

We reiterate two favorable results of this section: first,
the responsivity ${\cal R}$ and the variance of $\hat{\bfd}$
can be expressed exactly as direct sums over the observed population, for
arbitrary choice of $w$; there is no need for calculation of
polarizabilities nor recourse to simulated images.  Second, we note
that the variance in $\hat\delta$ can be significantly below the
canonical $\sigma^2_{SN}/N$ if the ellipticity distribution $P(e)$ has
structure that is not washed out by measurement noise.

\subsection{With Measurement Error}
A galaxy image with {\rm true} ellipticity ${\bf e}$ will be {\em
measured} at some ellipticity ${\bf \tilde e}$, with a probability
distribution of $p({\bf \tilde e | e})$. We consider a population of
galaxies all having the same significance $\nu$ and resolution
parameter $R$ (see \S\ref{psf} and Appendix~\ref{rfactorapp}), so
that they all have a common 
$p({\bf \tilde e | e})$.  The measured distribution of ellipticities
observed under distortion \bfd\ will then be
\begin{equation}
\label{ptilde}
\tilde P_{\bfd}({\bf\tilde e}) = \int\! d^2\!e\, P_{\bfd}({\bf e}) 
p({\bf \tilde e | e}),
\end{equation}
where $P_{\bfd}$ is the distribution of true ellipticities as in the
previous section.  The symmetry of $P$ and $p$ in the ellipticity
plane guarantees that in the absence of distortion, 
the measured distribution $\tilde
P(\tilde e)$ must again depend only upon the magnitude, not the
direction, of the measured ellipticity.

Equation~(\ref{ptilde}) is not strictly a convolution because the
measurement function $p({\bf \tilde e | e})$ may depend upon ${\bf e}$
and not simply upon ${(\bf \tilde e - e)}$---for example
Equation~(\ref{errore}) describes how the error ellipses contract as
${\bf e}$ departs from zero, even if the significance of the detection
is held fixed.  In \S\ref{psf} we will show that the behavior of the
measurement error is different when the effects of PSF smearing upon
the image are important, so at this point we will consider $p({\bf
\tilde e | e})$ to be, most generally, some kind of Gaussian whose
$1\sigma$ ellipse depends only upon the magnitude $e$.  An important
point is that the functional form of $p({\bf \tilde e | e})$ is
unchanged by an applied distortion, since $p$ is determined by $\nu$
and $R$, which are unchanged by a pure shear.

Another fact to keep in mind is that, with finite resolution and noise, it is
possible to measure $\tilde e >1$, if the image noise makes the object
appear smaller than the PSF in some dimension.  Our formulae should
therefore be tractable even for $\tilde e >1$, and we cannot simply
discard such measurements without contemplating the consequences.

We proceed as in the previous section, by assuming a distortion
estimator of the form
\begin{eqnarray}
\label{ewt2}
\hat{\bfd} & = & {{ \sum w(\tilde e_i) {\bf \tilde e}_i} \over 
	{\sum w(\tilde e_i)}} 
	= { {\int\! d^2\!\tilde e \, w(\tilde e) \tilde P_{\bfd}(\tilde e) 
	{\bf \tilde e}} \over
	{\int\! d^2\!\tilde e \, w(\tilde e) \tilde P(\tilde e)}} \\
\Rightarrow \langle {\cal R} \rangle  =  
{ {\partial \hat\delta_+} \over { \partial \delta_+}} & = & 
	{ {\int\! d^2\!\tilde e \left[ w(\tilde e) \tilde e_+ 
	\int\! d^2\!e\, p({\bf \tilde e | e})
	(\partial P_{\bfd}({\bf e}) / \partial \delta_+) \right] }
	\over {\int\! d^2\!\tilde e\, w(\tilde e) \tilde P(\tilde e)} } \\
& = & { {\int\! d^2\!\tilde e \left[ w(\tilde e) \tilde e_+ 
	\int\! d^2\!e\, P(e) p({\bf \tilde e | e}) e_+ \left(3 - 
	{ {1-e^2} \over e} { {d\log P} \over {de} } \right) \right]}
	\over {\int\! d^2\!\tilde e\, w(\tilde e) \tilde P(\tilde e)} } 
\label{resp3} \\
\label{rvar3}
{\rm Var}(\hat\delta_+) & = & { { \int\! d^2\!\tilde e\, w^2(\tilde e)
 \tilde e_+^2 \tilde P(\tilde e) }
 \over {N [\int\! d^2\!\tilde e\, w(\tilde e) \tilde P(\tilde e)]^2} } \\
 & = & { { \sum w^2(\tilde e) \tilde e_+^2 } \over { \left[\sum
 w(\tilde e)\right]^2} }.
\label{rvar4} \\
\Rightarrow w_{\rm opt}(\tilde e) & = & 
	{ 1 \over \tilde P(\tilde e) }
	\int\! d^2\!e\, P(e) p({\bf \tilde e | e}) { {e_+}\over{\tilde
	e_+}}
	 \left(3 - { {1-e^2} \over e} { {d\log P} \over {de} } \right)
\label{wopt3}
\end{eqnarray}
Given functional forms for the {\em intrinsic} distribution $P(e)$ and
the uncertainty function $p$, we could use (\ref{wopt3}) to derive an
optimal weight, and use (\ref{resp3}) and (\ref{rvar3}) to get the
responsivity and noise for the estimator using this or any weight
function.  In most cases these integrals will not have analytic solutions.

The parenthesized quantity under the integral in \eqq{resp2} is a galaxy's
responsivity to shear, which depends upon the {\em intrinsic} shape.  
\eqq{resp3} is the average of this responsivity for the galaxies with
some {\em measured} shape.  The measurement noise can cause these two
quantities to differ; in other words, a naive determination of the
responsivity is biased by the measurement noise.  KSB-based methods
will also suffer a calibration error due to this effect; binning the
polarizabilities in parameter space can reduce the {\em noise} in the
polarizability, but will not remove biases.  Precision cosmology will
require that such calibration issues be addressed---there are
unfortunately no cosmic calibration standards for shear.

The need for $P(e)$ in the above formulae is an unfortunate
complication since $\tilde P$ is 
the directly observed quantity.  Note that the variance of the
estimator can be expressed as a closed sum over the observed shapes
(\ref{rvar4}), but the responsivity cannot.  A precise calibration of
the resultant shear/mass maps requires, therefore, that $P$ be
estimated either by deconvolving the observed $\tilde P$ with the
error distribution $p$, or by recourse to higher-quality images that
give $P$ directly.

Derivation of the the optimal weight also requires knowledge of the
intrinsic $P$, but we can explore some generic cases, and make some
approximations that give workable methods.

\subsubsection{Approximate Form For Responsivity with Errors}
We wish to have a form for ${\cal R}$ as a sum over the observed
objects and applied weights, as in \eqq{resp1}, for the case of finite
measurement errors.  Toward that end we can take the derivative of
\eqq{ewt2}, which is greatly simplified if we assume that the
measurement {\em error}, {\it i.e.} ${\bf \tilde e}_i - {\bf e}_i$,
does not have any first-order dependence on $\delta$.  While not
strictly valid it is a good approximation.  In this case
\begin{equation}
{\cal R}  = { \sum \left[ w (1 - \langle e^2_+ \rangle_{\tilde
e})
+ { \tilde e_+^2 \over \tilde e} {dw \over d\tilde e} \left(
	1 - \langle e^2_+ \rangle_{\tilde e}
	- \langle e_+ e_\times \rangle_{\tilde e} \tilde e_\times
	/ \tilde e_+ \right) \right] 
	\over
	\sum w },
\end{equation}
where the brackets indicate an average of the true quantity at a given
measured value, {\it e.g.} 
\begin{eqnarray}
\langle e^2_+ \rangle_{\tilde e} & = &\int d^2\!e\, p({\bf e} | {\bf \tilde
e}) e^2_+ \\
 & = & { \int d^2\!e\, p({\bf \tilde e} | {\bf e}) P(e) e^2_+
	\over 
	\int d^2\!e\, p({\bf \tilde e} | {\bf e}) P(e) }
\end{eqnarray}
Note that the weight function $w$ may depend upon $\tilde e$ directly,
but also indirectly through some dependence in its covariance matrix
\bfS.  If the measurement error function $p({\bf \tilde e} | {\bf e})$
and the intrinsic distribution $P(e)$ have circular symmetry, then we
must be able to write
\begin{equation}
\label{k0k1}
\langle e^2_+ \rangle_{\tilde e} = k_0(\tilde e) + k_1(\tilde e)
\tilde e^2_+
\end{equation}
where $k_0$ and $k_1$ are functions only of the magnitude, not
direction, of ${\bf \tilde e}$.  We must also have 
$\langle e_+e_\times \rangle_{\tilde e} = k_1(\tilde e)
\tilde e_+\tilde e_\times$.
Further manipulation, taking
advantage of the isotropy of the parent population, yields
\begin{equation}
\label{nresp1}
{\cal R}  = { \sum \left[ w \left(1 - k_0 - {k_1 \tilde e^2 \over 2}\right)
+ { \tilde e \over 2} {dw \over d\tilde e} \left(
	1 - k_0 - k_1 \tilde e^2 \right) \right]
	\over
	\sum w }.
\end{equation}
This form for ${\cal R}$ depends only upon the observed quantities and
the chosen weight scheme, except through the two functions $k_0$ and
$k_1$, which we will approximate below.  The resemblance to
\eqq{resp1} is clear.  With this equation and some integration by
parts, we may also derive a form for the optimal weighting function:
\begin{equation}
\label{nwt1}
w_{\rm opt}[\tilde e, \bfS(\tilde e)] =
	3 k_1 + {1 \over \tilde e} {dk_0 \over d\tilde e}
	+ \tilde e {dk_1 \over d\tilde e} - 
	{1 - k_0 - k_1 \tilde e^2 \over \tilde e} { d\log\tilde P
	\over d\tilde e }.
\end{equation}

\subsubsection{Special Case: Gaussians}
In general the functions $k_0(\tilde e)$ and $k_1(\tilde e)$ must be
calculated numerically using a presumed underlying $P(e)$ for the
background population, but analytic solutions are possible
in the case of a Gaussian $P(e)$ with variance $\sigma^2_{\rm
SN}$ in each component (the shape noise) and a constant measurement
error $\sigma^2_e$ on each component.  
We find that both $k_0$ and $k_1$ are independent of $\tilde e$:
\begin{equation}
\label{k0k1gauss}
\begin{array}{rcl}
k_0  & = & (1-f)\sigma^2_{\rm SN} \\
k_1 & = & f^2 \\
f & \equiv & { \sigma^2_{\rm SN} \over  \sigma^2_{\rm SN} +
\sigma^2_e}.
\end{array}
\end{equation}
The quantity $f$ is the fraction of the total ellipticity variance that is
attributable to shape noise.  When the measurement noise is small,
$f\approx1$, the ideal weight is close to $(1-\tilde
e^2)/(\sigma^2_{\rm SN} + \sigma^2_e)$.  This is quite similar to the
weight adopted by \citet{Ho00}.

For the Gaussian case, $d\log\tilde P/d\tilde e$ is also quite simple,
so \eqq{nresp1} can be used.
In our surveys to date (Smith \etal\ 2001; Jarvis \etal\ in prep.) we
have adopted a weight that results from optimizing the Gaussian case.

\subsubsection{Practical, Nearly Ideal Approximation}
We obtain an approximation to the correct responsivity ${\cal R}$ and
resultant ideal weight if we adopt the constant $k_0$ and $k_1$
functions in Equations~(\ref{k0k1gauss}) even for non-Gaussian $P(e)$
distributions.  The shape noise $\sigma^2_{\rm SN}$ may be defined as
the assumed variance of the underlying $e_+$ and $e_\times$, and may
be found by subtracting the measurement noise from the observed
$\langle \tilde e^2 \rangle$.  The measurement noise $\sigma^2_e$ is
known for each galaxy using the methods of this paper; since the
covariance matrix for ${\bf \tilde e}$ is generally anisotropic, some
representative scalar must be selected.

With these guesses for $k_0$ and $k_1$ in hand, ${\cal R}$ may be
estimated with a sum over the {\em observed} galaxies using
\eqq{nresp1} for any chosen weight function. 

For the real-Universe shape distributions measured in the CTIO survey,
we find that the following ``easy'' weight function offers very close to
optimal distortion measurements:
\begin{equation}
\label{goodwt}
w = \left[ e^2 + (1.5\sigma_\eta)^2\right]^{-1/2},
\end{equation}
where $\sigma_\eta$ is the shape uncertainty that the object would
have were it circular [{\it cf.} \eqq{errore}].

We can check the accuracy of our approximations numerically for chosen
$P(e)$ and $p({\bf \tilde e} | {\bf e})$ functions.  We examine the
case when $P(e)$ is that shown in Figure~\ref{pe} for galaxies with
$20<m_R<21$, and the measurement error follows \eqq{errore}.  We find
that the weight function given by \eqq{nwt1} is in fact very close to
optimal for all noise levels, even when the simple approximations
(\ref{k0k1gauss}) are used for the $k$ functions.  The ``easy'' weight
\eqq{goodwt} also performs nearly optimally, so most applications
could use this weight and need not attempt to determine $P(e)$.

A more critical question is whether the approximations
(\ref{k0k1gauss}) yield a proper estimate of the calibration factor
${\cal R}$ when used with \eqq{nresp1}.  Figure~\ref{resplot} shows
how the simple ${\cal R}$ estimator compares to the correct value in
\eqq{resp3} for our choice of underlying distribution and the ``easy''
weights (\ref{goodwt}).  The approximate form yields a responsivity
correct to better than 5\% for $\sigma_\eta\lesssim0.4$. 
It is clear from the Figure that some detailed knowledge of the
underlying $P(e)$ distribution will be needed in order to calibrate
lensing measurements to the one percent level.

\begin{figure}
\plotone{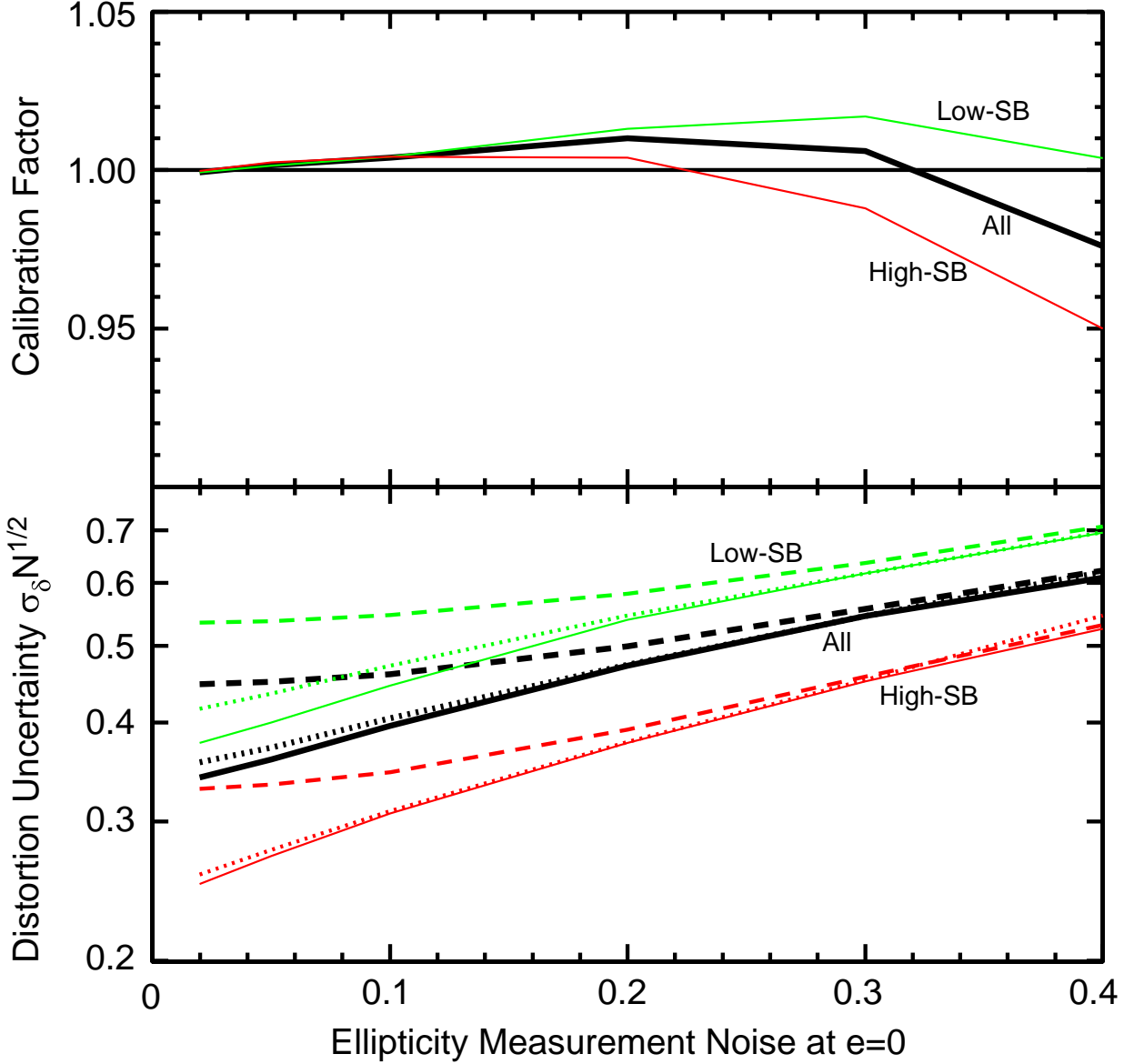}
\caption[]{
\small The upper panel plots the approximate, simplified
closed-sum estimate of the calibration factor ${\cal R}$ (Equation
[\ref{nresp1}]) relative to the exact form (Equation~[\ref{resp3}]),
as a function of the ellipticity measurement noise at $e=0$.
The weight function is the ``easy'' \eqq{goodwt} and the factors $k_0$ and
$k_1$ adopt the simple
heuristic approximation (\ref{k0k1gauss}).  The heavy (black) curve is
for the population of $20<m_R<21$ galaxies, the upper (green) curve is
for a low-SB sample ($20<\mu_R<21$), and the lower (red) curve for 
a high-SB sample ($17<\mu_R<19$), for which the intrinsic $P(e)$
distributions are plotted in Figure~\ref{pe}.  The simple formulation
yields a calibration accurate to 5\% or better in all cases, but 1\%
accuracy is difficult to achieve.  The lower panel shows the
uncertainty in the distortion determination when the galaxy shapes are
combined with optimum weights (solid lines), the ``easy'' weights
(\ref{goodwt}, dotted), and equal weighting (dashed).  
The line weights/colors code
the galaxy sample, as above.  Note that with optimal or easy
weighting, the distortion errors continue to shrink even when
measurement error is well below the canonical shape noise of $0.3$.
}
\label{resplot}
\end{figure}

The lower panel of Figure~\ref{resplot} shows the potential advantage
of optimal weighting.  When the measurement error is $\gtrsim0.2$,
there is little difference between various weighting schemes.  For
an unweighted distortion estimator, the accuracy levels out as the
measurement noise drops below $\sigma_e\approx0.2$.  When optimal
weights are used, however, the distortion errors continue to drop as
the measurement error is pushed toward zero---the optimal weights take
advantage of the $e=0$ cusp in the shape distribution.  Our
``easy'' weight scheme recovers nearly all of this potential gain.

To summarize, a practical method of weighting and calibrating the
response in the presence of measurement noise is:
\begin{enumerate}
\item Determine the underlying $\sigma^2_{\rm SN}=\langle
e^2_+\rangle$ of the the intrinsic distribution.  The measurement
error $\sigma_\eta$ (or the full covariance matrix) of each galaxy is
known from the formulae in previous sections.
\item Approximate the quantities $k_0$ and $k_1$ with \eqq{k0k1gauss}.
\item Choose a weight function, for example the ``easy'' form
\eqq{goodwt}, or preferably the optimal form \eqq{nwt1} which can be
derived from the observed distribution $\tilde P(\tilde e)$.
\item The distortion estimator is the sum (\ref{ewt2}).  The variance
in the estimator is the sum (\ref{rvar4}).
\item The estimator (and variance) must be scaled by the responsivity
which is calculated with \eqq{nresp1}.
\end{enumerate}
If more is known about the intrinsic shape distribution, then more
accurate functions for $k_0$ and $k_1$ may be derived and used in the
sums.

\subsection{Additional Weighting Considerations}
The optimal weight may depend upon parameters other than the
observed ellipticity $\tilde e$.  It must, for example, depend upon
the measurement errors as described above.  If intrinsic shape
distribution $P(e)$ depends upon galaxy type, for example, then it may
be advantageous to have different weight functions for each type---as
long as galaxy type can be determined independent of ${\bf e}$.
Ellipticals may be distinguished from spirals, for example, by a
concentration ratio such as the $b_{22}$ coefficient described below.

The expected shear will depend upon the redshift $z$ of the background
galaxy, and hence there may be a $z$-dependent weight determined from
photometric redshift estimates of the background population.  More
crudely the apparent magnitude may be taken as an indicator of $z$ and
used in the weight formulation.  The use of such additional weights
depends upon the problem being addressed:  see \citet{Sm01} as an
example. 

\subsection{Relation to Previous Methods}
The optimal weighting scheme of K00 differs from ours in several
respects. It assumes a different measure of ${\bf e}$ which does not
follow our geometric transformation relations, so the mean
polarizability of galaxies must be calculated in some parameter space,
{\it e.g.} of flux, size, and $e$.  Binning or smoothing in some such
space is common to most of the KSB-based weighting schemes as well
\citep{Er01,Ho00}.  The variance of the estimators $\hat\delta_i$ are
calculated within each bin, then weights are chosen inversely to the
variance of each bin to create a minimum-variance weighted mean
estimator.  There are three contributors to the variance from each
bin:  
\begin{enumerate}
\item The intrinsic ellipticities of galaxies within the bin are drawn
at random from the parent distribution. If
$|{\bf e}|$ is one parameter of the space, then only the direction
$\theta$ is allowed to vary within the bin.  This is of course the
shape noise.
\item The measured shape of a given galaxy is drawn from the
measurement-error distribution.
\item The number of galaxies within the bin is drawn from a Poisson
distribution.  If $|{\bf e}|$ is a parameter, then this Poisson noise
includes some elements of the shape noise (1) and measurement noise (2).
\end{enumerate}
All three of these effects are important to optimization; all are
included in our formulation and (implicitly) in that of K00, so we
expect them to be essentially equivalent in the long run---this is not
the case, though, for some of the heuristic or
parameter-space weight formulations in the literature.  The virtue of
our scheme is that the nature of the weight function is apparent given
the intrinsic shape distribution $P(e)$ and the measurement errors,
and there is no need to choose a parameter space for weight selection
and polarizability smoothing.  Our formulation tells us when further
parameters might be desirable, namely when $P(e)$ changes significantly.

\section{Measurements with Finite Resolution}
\label{psf}
The preceding sections outline a method for optimal recovery of weak
distortion from galaxy images, and rigorous estimation of the
uncertainties on these shears, for the case when the detector views
the galaxies with perfect resolution.  Unfortunately, the finite
resolution of real observations has a strong effect upon shape
measurements in every weak lensing observation to date, even those
using the Hubble Space Telescope.  Finite resolution produces two
deleterious effects:
\begin{enumerate}
\item A PSF which is not circularly symmetric can induce ellipticities on
the images, thus breaking the intrinsic isotropy of the background
galaxy population and mimicking a lensing distortion.  This is a {\em
bias} induced by asymmetric PSFs. Since present-day weak lensing
surveys are seeking distortion signals well below 1\%, measured shapes
must be compensated for even the slightest asymmetry in the PSF with
some {\em smear} correction.
\item Convolution by a circularly symmetric PSF will make most
galaxies appear rounder, driving ${\bf e}\rightarrow 0$.  This is
therefore a {\em dilution} of the true lensing signal.  While this
mechanism cannot create a lensing signal where there is none, it can
misleadingly modulate the lensing signal or cause a calibration error
in the inferred mass distribution.
\end{enumerate}
Most (but not all) approaches to PSF corrections treat the bias and
dilution effects in separate steps.  To our knowledge, {\em all
published weak-lensing observations have incomplete PSF
correction, leaving systematic
distortion errors of a fraction of a percent or higher.}
While in most cases these residual systematic effects have not altered
the validity of the authors' conclusions, the proper correction of PSF effects
is presently the largest barrier to the use of weak lensing for precision
cosmology. 

In \S\ref{oldpsf} we review some
existing approaches to these problems, in \S\ref{ponderpsf} we contemplate
how one would ideally wish to approach the problem, and in further
sections we develop two means of implementing a nearly-ideal
approach:  one which treats bias and dilution separately, and another
which corrects both problems simultaneously with a limited form of
deconvolution. 

\subsection{Existing Approaches to PSF Corrections}
\label{oldpsf}
\subsubsection{The Unweighted Ideal}
Unweighted second moments of galaxies are ideal measures of
ellipticity---not only do the derived ellipticities transform
according to the rules of shear space, but correction for PSF effects
is in principle quite simple because the unweighted second moment of the
image is just the sum of the original moment and the PSF moment.  Thus
by subtracting the PSF moments from the measured moments we
simultaneously correct for bias and dilution, and obtain the image
shape.  In the case where the PSF is round, the dilution of a true
(pre-seeing) image-plane
ellipticity ${\bf e}^i$ to an observed (post-seeing) ellipticity ${\bf
e}^o$ is described by the exact equation
\begin{eqnarray}
\label{unweightedR}
{\bf e}^i & = & {\bf e}^o / R \\
R & \equiv & { {\langle r^2\rangle_i} \over 
	{ \langle r^2\rangle_i + \langle r^2\rangle_\star}} \nonumber \\
 & = & 1- { {\langle r^2\rangle_\star} \over 
	{ \langle r^2\rangle_o } }.
\end{eqnarray}
The {\em resolution parameter} $R$ is determined by the unweighted
second radial moment of the measured image $\langle r^2\rangle_o$
relative to that of the PSF $\langle r^2\rangle_\star$.  Two things to
note:  first, the error ellipse on the dilution-corrected, pre-seeing
ellipticity ${\bf e}^i$ is magnified by $1/R$ from the original
measurement error \eqq{errore}, and is further stretched in the radial
direction by the uncertainty in $R$ itself.  Thus the error ellipse is
no longer simply described by a single $\sigma_\eta$.  Second,
\eqq{unweightedR} can give rise to $|{\bf e}^i|>1$, if the noise makes
the galaxy look smaller than the PSF about some axis.  We cannot
arbitrarily discard such measurements without creating a bias in our
mean shear.  These two phenomena are common to all modes of
PSF-dilution correction.

This blissfully simple dilution correction is spoiled by two major
problems: first, as 
discussed above, unweighted second moments have divergent noise
properties and for this and other reasons are not practical shape
estimators.  An equally serious problem noted by K00 is that
the second moments themselves are divergent for many realistic PSFs.
Further, many galaxies follow deVaucouleurs profiles, for which the
second moment converges very slowly.

The simple formulae (\ref{unweightedR}) are still valid under the
special circumstances that the object and PSF are both Gaussians.
The post-PSF object is again a Gaussian, and
deconvolution of Gaussians is a simple subtraction of second
moments.  Hence any shape-measuring algorithm which extracts the
proper ellipticity for a Gaussian ellipsoid would allow PSF dilution
correction via Equations~(\ref{unweightedR}) in this limited (and
unrealistic) case.

Some early weak lensing measurements \citep{Va83}
adopt second-moment subtraction as a means of PSF correction, despite
the fact that this method is not exact when isophotally-bounded or
weighted moments are used, and the images are not Gaussian.  This
would not suffice, however, for the more sensitive measurements being
done today.

\subsubsection{Heuristic Methods}
In the case of unweighted
moments or Gaussians, Equation~(\ref{unweightedR}) would indicate that
a regression of the lensing signal against $\langle r^2\rangle_o^{-1}$
would yield a distortion free of PSF effects as
$\langle r^2\rangle_o^{-1}\rightarrow 0$. 
\citet{Mo94} have attempted to measure very weak
shears in the presence of PSF effects using such a regression (though
against $\langle r\rangle^{-1}$, in which case a linear relation is not
expected).
Even with weighted moments, we expect the PSF dilution and bias to
decrease as the object becomes well-resolved, so there is some basis
to this method, even if it is not exact.  Other problems, however, are
that the distortion is not likely to be the same for all sizes of
galaxy as they likely lie at different distance.  Also the regression
will lead to substantially higher noise than a more direct dilution
correction. 

Another approach to the dilution correction is exemplified by
FT97, who attempt no analytic correction, instead calibrating
the dilution effect by measuring simulated background galaxies which
have been subjected to the same distortion, seeing, sampling and noise as 
the real images.  Such a simulation is an essential test of any
weak-lensing methodology.  The difficulty with sole reliance upon
simulated data is that the result is extremely sensitive to one's
ability to exactly match the size-magnitude distribution of the true
galaxy population, because the dilution correction is a strong
function of size (as in Equation~[\ref{unweightedR}]) in the typical
regime of slightly resolved galaxies.  Further, as we show below, the
dilution correction depends upon detailed higher-order moments of the
galaxy images, which would be very difficult to simulate faithfully.
One alternative is to use high-resolution, high-$S/N$ images from HST
instead of simulated galaxies---but the total sky area imaged to
sufficient $S/N$ by HST is a tiny fraction of a square degree, too
small for rigorous calibration tests.   It would be preferable to have
an analytic correction for dilution, and use the simulated data to
spot-check the accuracy of the analytic method.

\subsubsection{Perturbative Methods}
A step beyond the unweighted-mean approximation to the bias correction
is taken by KSB and by FT97.  Both make the assumption that
the anisotropy of the PSF can be described as a small anisotropic
convolution applied to a larger, circularly symmetric PSF.  In this
case, the effect of the tiny asymmetric deconvolution upon the
weighted second moments of a given image can be expressed as a
fourth-order weighted moment of the image, which KSB christen the {\em
smear polarizability.}  Given the smear polarizability of an image
and a measure of the anisotropy of the PSF, the measured second
moments are corrected analytically for the PSF bias.

The FT97 method differs in that the correction for PSF anisotropy is
applied to the image rather than to the measured moments:  a minimal
$3\times3$ convolution kernel is created which will ``circularize''
the PSF.  The galaxy shapes are measured after this kernel is applied
to the image.

The primary drawback to these methods (K00) is that the approximation upon
which they are based often fails:  a typical diffraction-limited PSF
in no way resembles a small convolution to a round PSF, and even a
simple aberration such as coma creates PSFs that violate this
condition.  

These methods have features, however, that we wish to retain in any
improved formulation.  They are easily adapted to a PSF that varies
across the image---the requisite moments of the PSF are measured
wherever a star falls upon the image, and interpolated to the location
of each galaxy.  The FT97 method, by fixing the image, frees us from
having to measure the higher-order moments that comprise the smear
polarizability, though at the expense of a slight reduction in image
resolution and/or an increase in the image noise.

The perturbative methods correct only the PSF bias, not the dilution,
because the dilution cannot be considered a small perturbation in any
extant data set.  FT97 calibrate the dilution with simulations, as
mentioned above.  The original KSB work made use of empirical
calibration as well, as the shear polarizability ({\it cf.}
\S\ref{compare}) measures only the susceptibility of the image to a
distortion which might be applied {\em after} the PSF convolution.
\citet{Wi96} suggest an empirical calibration by deconvolving the
real images, applying a shear, reconvolving, and remeasuring to
determine the response.
\citet{LK97} introduce the {\em pre-seeing
shear polarizability,} which approximates the
susceptibility of the KSB weighted moments to a shear applied {\em
before} the PSF.  The pre-seeing shear polarizability is, to its
lowest order, similar to the resolution factor $R$ introduced above for
unweighted moments.  There are, however, fourth-order moments of
galaxy and stellar shapes involved, and all are measured with Gaussian
weights so the noise does not diverge.

The KSB method updated to use the pre-seeing shear polarizability 
is exact for
Gaussians and in the limit of a compact anisotropy kernel, but is not
exact in the general case (K00).  There are additional ambiguities
regarding the appropriate size of the Gaussian weight to be applied
when measuring the PSF moments \citep{Ho98}.  The updated KSB method
is in wide use, and several papers have investigated how accurately
it performs for simulated galaxies and PSFs that are not Gaussian
\citep{Er01,Ba00}.  While it is clear that in many circumstances KSB is good
enough, we would prefer to understand and overcome its limitations.

We will demonstrate below that the KSB and FT97 bias corrections are the
first terms in a series expansion of the deconvolution of the galaxy
image. 

\citet{Rh00} investigate the KSB method in some detail, carrying forth
the transformations equations to higher order than do KSB.  The PSF
corrections, however, still require substantial approximations.  Below
we construct a hierarchy of {\em all} the weighted moments of the
image and give general formulae for their transformations under shear,
convolution, and other operations.

\subsubsection{Deconvolution Methods}
\citet{Ku99} suggests that one determine shear by summing the
images of all the galaxies in some cell, then comparing this
high-$S/N$ summed galaxy image to the PSF.  The summed galaxy image
should approach circular symmetry in the source plane; the shape of
the summed image can thus be modeled relatively simply as a
circular source with arbitrary radial profile, sheared by the lens to
have elliptical isophotes, then convolved with the PSF.  He adopts a
flexible parametric representation of this mean radial profile.  A
candidate profile with candidate lens distortion can be convolved with
the known PSF and compared to the measured mean image.  The radial
profile and distortion are then adjusted to give a best fit.  This
method can be viewed as a limited form of deconvolution:  the only
characteristic of the deconvolved image one cares about is the
ellipticity; the multipoles beyond quadrupole are irrelevant to the
measurement and are discarded.  

In Kuijken's method, the higher-order
multipoles are discarded by averaging over many galaxies.  This may be
difficult in practical situations, where the PSF and/or the distortion
signal are varying across the field too rapidly to gather sufficient background
galaxies to sum.  Another drawback is that summing galaxies' images
may not be the optimal way to combine their information on the shear.
Kuijken suggests applying the method to individual galaxies given
these potential problems, which amounts to an assumption that the
galaxy ellipticity is constant with radius.  The accuracy of this
method is not discussed in detail.
But the method does have the strong advantage of being able to cope
with arbitrary PSF behavior, and simultaneously removing both the bias
and dilution effects from the measurement.  

\subsubsection{Commutator Method}
K00 introduces a new approach to PSF correction, deriving from the PSF
an operator which can be applied to the {\em observed} image
to effect the transformation of applying a given {\em pre-seeing}
shear.  This operator is derived by considering the commutation of the
shear and convolution operators in Fourier space.
With the operator in hand, the response to a pre-seeing shear can be
determined directly from the post-seeing image, given sufficiently
detailed knowledge of the PSF.

The K00 formulation is the first, to our knowledge, to offer an exact
correction for PSF bias and dilution.  We take a different approach
below, and then offer some comparison of the two approaches.

\subsection{Optimal Methods in the Presence of a Convolution}
\label{ponderpsf}
In \S\ref{ftshearsec} we noted that a real-space shear is equivalent
to an opposing shear of the Fourier-space image, so that we can
conduct the roundness test in Fourier space.
One virtue of using a Gaussian weight function for our roundness
test, as in
Equation~(\ref{gausstest}), is that this test takes the exact same
form in $k$-space, namely
\begin{equation}
M(I) = 0 \iff 0 = \tilde M(\tilde I) = 
\int\! d^2\!k\, e^{-k^2\sigma^2/2} I({\bf k}) k^2 e^{-2i\phi}.
\label{kgausstest}
\end{equation}
The only difference is that the Gaussian weight has width $1/\sigma$
in $k$-space, {\it i.e.} a broader weight in real space is narrower in
Fourier space.  Furthermore, our assumed uniform white noise in real
space transforms to uniform white noise in Fourier space.
Therefore the entire derivation of the optimal weight in
\S\ref{idealround} could have been executed in $k$-space without any
change in the result.

The effect of the PSF convolution is to suppress the image by some
transfer function $\tilde T({\bf k})$.  With perfect knowledge of the
PSF, we can deconvolve the observed image $\tilde I^o$ to retrieve the
image-plane transform $\tilde I^i=\tilde I^o / \tilde T$ (here we mean
the image plane of the gravitational lens, before the application of
seeing).  A deconvolution would remove the PSF bias entirely, but the
noise is no longer homogeneous in $k$-space, having been amplified
(perhaps infinitely) by $1/\tilde T$.  If the PSF is close to
circularly symmetric, then $\tilde T$ is nearly independent of
direction.  If we now imagine making our roundness test on the
deconvolved $k$-space image, we adapt
Equations~(\ref{msignal})--(\ref{wopt}) to give
\begin{eqnarray}
\label{msignalk}
\delta \tilde M & = & {\eta \over 4} \int k\,dk\, k^2 w(k) k
{{d \tilde I^i_0} \over {dk}} \\
\label{varmk}
{\rm Var}(M) & = & {n \over 4\pi^2} \int\ k\,dk\,d\phi\, k^4 w^2(k) 
\cos^2 2\phi / \tilde T^2(k). \\
\label{woptk}
\Rightarrow w_{\rm opt}(k) & \propto & 
{ {-\tilde T^2(k)} \over { k }} {{d \tilde I^i_0} \over {dk}}.
\end{eqnarray}
The optimal filter is therefore narrower in $k$-space than both $d\tilde I
/ d(k^2)$ and $\tilde T(k)$. 
Hence in real space, on the deconvolved image, the
optimal filter is broader than both the object and the PSF.  This
means that we should, sensibly, restrict our roundness test to the
region of $k$-space that (a) has signal in the true galaxy image, and
(b) is not suppressed below the noise by the convolution.  

Our strategy might then be to create some kind of deconvolved image,
and apply a Gaussian weight to test for roundness in the deconvolved
$k$-space (which is equivalent to using a Gaussian weight in real
space). For a
Gaussian object with pre-seeing size $\sigma_i$ and Gaussian seeing
with size $\sigma_*$, the optimal weight in {\em deconvolved
$k$-space} is a Gaussian with size $(2\sigma_*^2+\sigma_i^2)^{-1/2}$.  
Such a roundness test is equivalent to a roundness test
on the {\em observed, real-space} image with a Gaussian weight of size
$\sqrt{\sigma_i^2+\sigma_*^2}=\sigma_o^2$.  So the optimal
size of the weight is again matched to the size of the {\em observed}
image. 

Recall that our algorithm for measuring \bfd\ requires that we shear
the coordinates until the object appears round ($\tilde M \rightarrow
0$).  We want to apply this shear to the deconvolved image.
If the object is not round to begin with, then in this
sheared coordinate system the transfer function $\tilde T({\bf k})$
will no longer have azimuthal symmetry, which will invalidate the
above derivation of the optimal weight.  We will still have a valid
measurement of the shape of the deconvolved object, but possibly with
sub-optimal noise level.  The increase in noise is a second-order
effect, however, so we will not bother to re-optimize the roundness
test for this asymmetric noise spectrum.

When the transfer function is anisotropic, then the noise spectrum in
the deconvolved image is also anisotropic.  There are subtle
second-order effects, described in \S\ref{centroidbias}, which bias the
orientation of the measured shear in the presence of an anisotropy in
the noise.  Such a noise anisotropy is present after any deconvolution
of an anisotropic PSF.  A noise anisotropy is also present in the {\em
observed} image if the PSF is anisotropic and we are not strictly
sky-limited.  Furthermore, K00 points out a selection bias that can
creep into the shear measurements even with a perfectly unbiased shape
measurement algorithm.  We will discuss means to defeat these biases
in \S\ref{centroidbias}.

Applying a Gaussian weight in deconvolved $k$-space leads, formally,
to divergent noise if $\tilde T(k)=0$ for some $k$.  This is a real
problem, as any finite-sized telescope must produce a transfer
function that is identically zero beyond some critical $k_c$.  The
deconvolved image hence has infinite noise for $k>k_c$, while the
Gaussian weight remains finite.  As we shear the $k$-space to make our
source appear round, the ``wall'' of infinite noise moves inwards to
$e^{-\eta/2} k_c$.  In order for our method to remain feasible, our
deconvolution algorithm must not attempt to fully deconvolve those
portions of $k$-space at or near $k_c$, so that the noise remains
bounded.  There is hence a balance to be struck in executing the
deconvolution:  we want the carry out the deconvolution to
sufficiently high order that the effects of the PSF upon the source
ellipticity are removed, but we do not want to deconvolve
high-order details that will increase the noise.  It would seem,
intuitively, that this is possible, since the ellipticity we seek, the
Gaussian-weighted quadrupole moment, is a low-order characteristic of
the deconvolved image.

The method we describe below is based upon an expansion of the image
and PSF into hierarchies of Gaussian-weighted moments---essentially
an eigenfunction
decomposition.  Convolution corresponds to a matrix operation on the
vector of moments.  The moment vector, and hence the convolution
matrix, are formally infinite, but we can choose to truncate the
description at some order which we believe to contain all the useful
information on the image ellipticity.  The convolution matrix is then
finite and can be inverted, and the deconvolution executed as a matrix
operation.  The high-order moments are not deconvolved, so the noise
in the deconvolved moments remains finite and in fact close to
optimally small.  This moment-based method reduces the deconvolution
(as well as other transformations) to a matrix multiplication, which
can be executed on an exposure-by-exposure basis even for objects with
very low $S/N$ on a single exposure, and is therefore very practical
for the purpose of weak lensing measurements.

\subsection{The Laguerre Expansion}
\label{laguerre}
\subsubsection{Definitions}
The simplicity of the formulae for deconvolution in the special case
of Gaussian objects, plus the utility of the Gaussian weight for shape
measurements, lead us to seek a description of the image in some
Gaussian-based expansion.  To maintain the simplest form for
convolutions, we look for a decomposition of our images into
components which are eigenfunction of the Fourier operator.  Such
functions will also be eigenfunctions of $(-\nabla^2 + r^2)$, hence we
are led to the eigenfunctions of the 2d quantum harmonic oscillator
(QHO).\footnote{
Thanks again to the anonymous referee for providing this logic.}
In one dimension, the QHO eigenfunctions are each a Gaussian
times a Hermite polynomial.  
The Edgeworth expansion, familiar to many astronomers, is a
decomposition into 1d QHO eigenfunctions.
The 2d version we call the ``Laguerre expansion,'' using
the QHO eigenfunctions
\begin{eqnarray}
I(r,\theta) & = & \sum_{p,q\ge0} b_{pq}\psi_{pq}^\sigma(r,\theta) \\
\label{rlaguerre}
\psi_{pq}^\sigma(r,\theta) & = & 
	{ {(-1)^q} \over {\sqrt\pi \sigma^2}}
	\sqrt{ {q!} \over {p!} }
	\left( {r \over \sigma} \right)^m e^{im\theta}
	e^{-r^2/2\sigma^2}
	L_q^{(m)}(r^2/\sigma^2) \qquad (p\ge q) \\
\psi_{qp}^\sigma & = & \bar\psi_{pq}^\sigma \\
m & \equiv & p-q.
\end{eqnarray}
$L_q^{(m)}(x)$ are the Laguerre polynomials, which are defined by
the generating function \citep{Abram}
\begin{equation}
(1-z)^{-q-1}\exp\left( {{xz} \over {z-1}} \right)
= \sum_{m=0}^\infty L_m^{(q)}(x) z^m.
\end{equation}
The Laguerre polynomials satisfy many recurrence relations; the
following provide a way to calculate them rapidly:
\begin{eqnarray}
L_0^{(m)}(x) & = & 1 \\
L_1^{(m)}(x) & = & (m+1)-x \\
(q+1) L_{q+1}^{(m)}(x) & = & [(2q+m+1)-x]L_{q}^{(m)}(x)
	- (q+m)L_{q-1}^{(m)}(x).
\end{eqnarray}
A QHO with wavefunction $\psi_{pq}$ has angular momentum $m=p-q$.
We will also make use of the quantum
number $N=p+q$, which is the excitation energy of the
state.  Any two of 
$\{N,m,p,q\}$ suffice to specify the state.  The intensity multipole
functions $I_m(r)$ are composed from the $\psi_{Nm}$ for
$N=|m|,|m|+2,|m|+4,\ldots$. The polynomial in
$\psi_{pq}$ has terms up to order $N$ in $r$, and $\psi_{pq}$ can also be
expressed as the Gaussian times a (complex) polynomial of order $N$ in
$x$ and $y$.  A few low-order $\psi_{pq}$ are plotted in
Figure~\ref{laguerreplot}. 

\begin{figure}
\epsscale{0.6}
\plotone{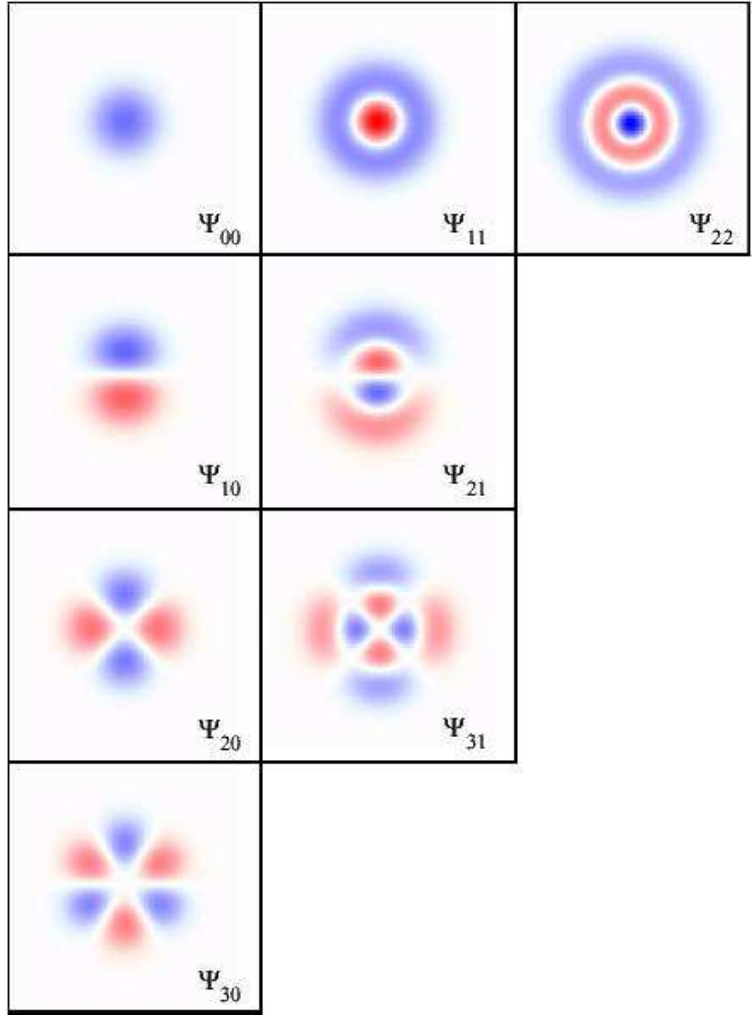}
\epsscale{1.}
\caption[]{
\small The first few of the orthogonal functions $\psi_{pq}$ are
plotted. Only the real parts are plotted, with blue (red) for positive
(negative) regions.
The characteristics are familiar from their use as
eigenfunctions of the 2d quantum harmonic oscillator:  each has
$2m=2|p-q|$ azimuthal nodes and $N-2m$ radial modes, where $N=p+q$ is
the number of quanta in the state.  The $\psi_{20}$ component is most
important as it responds in first order to shear, and hence its
absence is our test for ``roundness.''
}
\label{laguerreplot}
\end{figure}

\citet{Re01} has independently introduced the application of QHO
eigenfunctions to galaxy shape analysis.  Some of the results and
ideas presented in this section are presented therein, and applied in
\citet{RB01}, with different notation.  \citet{Re01} also presents
useful relations for a Cartesian-based family of 2d QHO eigenfunctions.
We make use only of the polar family, which are eigenstates of the
angular momentum and hence have strong rotational symmetry.

The Laguerre functions have many properties we will find useful.  As
eigenfunctions of the harmonic oscillator, they are orthonormal, up to
the factor $\sigma$ which we introduce in order to give
the $b_{pq}$ units of flux:
\begin{eqnarray}
\label{orthonorm}
\int\! d^2\!x\, \psi_{pq}^\sigma({\bf x}) 
	\bar\psi_{p^\prime q^\prime}^\sigma({\bf x}) 
& = & {1 \over {\sigma^2}} \delta_{pp^\prime} \delta_{qq^\prime} \\
\label{inversion}
\Rightarrow b_{pq} & = & \sigma^2
	\int\! d^2\!x\, I({\bf x}) \bar \psi_{pq}^\sigma({\bf x}) .
\end{eqnarray}
Thus $b_{pq}$ is a Gaussian-weighted moment of the intensity.
Since $I$ is real, we must have $b_{pq}=\bar b_{qp}$.  When measuring
the $b_{pq}$ from an image with uniform white noise,
Equation~({\ref{inversion}) yields a covariance matrix for the $b_{pq}$
that is diagonal:
\begin{equation}
\label{bpqcov}
{\rm Cov}(b_{pq} \bar b_{p^\prime q^\prime}) = n\sigma^2 
	\delta_{pp^\prime} \delta_{qq^\prime},
\end{equation}
where $n$ is the number of counts per unit area.  The variance in
$b_{pq}$ is shared between the real and imaginary parts, except for
$b_{pp}$, which must be real.  Non-uniform noise, {\it e.g.} shot
noise from the galaxy itself, produces a more complicated covariance
matrix, as described in \citet{Re01}. The significance $\nu$ of detection
with the Gaussian filter is given by
\begin{equation}
\label{signif}
\nu^2 = {{b_{00}^2} \over {n\sigma^2} }.
\end{equation}

Our algorithm for measuring object shapes requires finding the
coordinate system in which the centroid is zero
(Equation~[\ref{centroid}]) and the roundness test 
(Equation~[\ref{w2}]) yields zero.  When disregarding seeing, we also
set the weight size $\sigma$ by maximizing the significance
(Equation~[\ref{sizecriterion}]).  These conditions are succinctly
stated by the Laguerre coefficients:
\begin{eqnarray}
\label{bcentroid}
\hbox{Centroid:}\quad & & b_{10} = 0 \\
\label{bround}
\hbox{Roundness:}\quad & & b_{20}  =  0 \\
\label{bsig}
\hbox{Significance:}\quad & & b_{11}  =  0.
\end{eqnarray}
The first two of these equations involve both real and complex
components, the third is real.  We must satisfy these equations by
translation, shear, and dilation of the object (or of the underlying
coordinate system).  These operations can be expressed as
transformation matrices acting upon the vector ${\bf b}=\{b_{pq}\}$.  The
determination of shape is thus efficiently executed by measuring 
${\bf b}$ in the original coordinate frame, converting
Equation~(\ref{inversion}) to a sum over pixels.  Henceforth we can
iterate to a solution of our three conditions by manipulating
${\bf b}$, and there is no need to return to the pixel data.

In the case of finite resolution, we wish to satisfy
Equations~(\ref{bcentroid}) and (\ref{bround}) for the deconvolved
image.  We can express the deconvolution as a matrix operation on 
${\bf b}$ as well.  So we need to find the matrix equivalents of the
translation, dilation, shear, and convolution transformations.

\subsubsection{Raising and Lowering Operators, Point Transformations}
\label{xformsec}
The raising and lowering operators for the 2d harmonic oscillator
wavefunctions have the properties:
\begin{equation}
\label{raiseops}
\begin{array}{rrcr}
a_p^\sigma = { 1 \over 2}\left[ {{x-iy}\over\sigma} + 
	\sigma\left( {\partial \over {\partial x}} - i{\partial \over
	{\partial y}}	\right)\right] \qquad
	& a_p^\sigma \psi_{pq}^\sigma & = & \sqrt p \,\psi_{(p-1)q}^\sigma, \\[6pt]
a_p^{\sigma\dagger} = { 1 \over 2}\left[ {{x+iy}\over\sigma} - 
	\sigma\left( {\partial \over {\partial x}} + i{\partial \over
	{\partial y}}	\right)\right] \qquad
	& a_p^{\sigma\dagger} \psi_{pq}^\sigma & = & \sqrt{p+1}
	\,\psi_{(p+1)q}^\sigma, \\[6pt]
a_q^\sigma = { 1 \over 2}\left[ {{x+iy}\over\sigma} + 
	\sigma\left( {\partial \over {\partial x}} + i{\partial \over
	{\partial y}}	\right)\right] \qquad
	& a_q^\sigma \psi_{pq}^\sigma & = & \sqrt q \,\psi_{p(q-1)}^\sigma, \\[6pt]
a_q^{\sigma\dagger} = { 1 \over 2}\left[ {{x-iy}\over\sigma} - 
	\sigma\left( {\partial \over {\partial x}} - i{\partial \over
	{\partial y}}	\right)\right] \qquad
	& a_q^{\sigma\dagger} \psi_{pq}^\sigma & = & \sqrt{q+1}
	\,\psi_{p(q+1)}^\sigma. 
\end{array}
\end{equation}
Note that the commutators of the operators are all zero except for
$[a_p,a_p^\dagger]=[a_q,a_q^\dagger]=1$.
These operators can be used to formulate the transformation matrices
we need.  Consider first a dilation of the image $I$ by factor
$1 + \mu$ with $\mu\ll1$ to a new image $I^\prime$:
\begin{eqnarray}
I^\prime(x,y) & = & I[(1-\mu)x,(1-\mu)y] \\
 & \approx & \left\{1-\mu \left( x { \partial \over {\partial x}}
	+ y { \partial \over {\partial y}}\right)\right\} I \\
 & = & \left\{1 - \mu\left( a_p^\sigma a_q^\sigma 
	- a_p^{\sigma\dagger}a_q^{\sigma\dagger}-1\right)\right\} I \\
 & = & \sum b_{pq} \left\{(1+\mu)\psi_{pq}
 -\mu\sqrt{pq}\,\psi_{(p-1)(q-1)}
+\mu\sqrt{(p+1)(q+1}\,\psi_{(p+1)(q+1)} \right\} \\
\label{lindil}
\Rightarrow\quad b^\prime_{pq} & = &
	(1+\mu)b_{pq} - \mu\sqrt{(p+1)(q+1)}\,b_{(p+1)(q+1)}
	+ \mu\sqrt{pq}\,b_{(p-1)(q-1)}
\end{eqnarray}
An infinitesimal dilation $\mu$ thus mixes coefficients $(N\pm2,m)$
into the $(N,m)$ coefficient---as expected, multipole order $m$ is
conserved by the transformation.  The dilation transformations are
thus generated by the matrix
\begin{equation}
\label{dgen}
{\bf d} = 1 - a_p^\sigma a_q^\sigma 
	+ a_p^{\sigma\dagger}a_q^{\sigma\dagger},
\end{equation}
and the transformation matrix ${\bf D}$ for a finite dilation $e^\mu$
can be expressed as 
\begin{equation}
\label{dfinite}
{\bf D}_\mu = \exp[\mu{\bf d}].
\end{equation}
The finite dilation will preserve $m$ as the generator does, but will
mix terms of $N+2j$ together, with $j$ any integer.  The matrix
elements for the finite dilation are derivable in closed form and in a rapid
recursion form; the latter are derived in Appendix~\ref{finitetransforms}.

The generator {\bf s} for the shear transformation is also easily expressed in
terms of the raising and lowering operators.  For a shear $\eta\ll1$
along the $x$ axis,
\begin{eqnarray}
I^\prime(x,y) & = & I[(1-\eta/2)x,(1+\eta/2)y] \\
 & \approx & \left\{1-{\eta\over2} \left( x { \partial \over {\partial x}}
	- y { \partial \over {\partial y}}\right)\right\} I \\
 & = & (1 + \eta {\bf s})I, \\
\label{sheargen}
{\bf s} & = & { 1 \over 4} \left[ (a_p^\dagger)^2 + (a_q^\dagger)^2
	- a_p^2 - a_q^2\right] \\
\label{linshear}
\Rightarrow\quad b^\prime_{pq} & = &
	b_{pq}  + {\eta\over 4}\left(
	\sqrt{p(p-1)}\,e^{-2i\beta}b_{(p-2)q} + 
	\sqrt{q(q-1)}\,e^{2i\beta}b_{p(q-2)}  \right. \nonumber \\*
	& & \left. - \sqrt{(p+1)(p+2)}\,e^{2i\beta}b_{(p+2)q} -
	\sqrt{(q+1)(q+2)}\,e^{-2i\beta}b_{p(q+2)} \right), 
\label{shearbpq} \\
{\bf S}_\eta & = & \exp[\eta{\bf s}].
\end{eqnarray}
In (\ref{linshear}) we have inserted the phase factors which result
from a shear at arbitrary position angle $\beta$.
To leading order, the shear mixes in states with $p$ or $q$
changed by $\pm2$; in the finite case,
the shear transformation mixes together $b_{pq}$ for which $p$ and
$q$ each change by any even number. 
Note that this equation generalizes the weighted-moment
transformations in \citet{Rh00} to all orders.  Note also that
the KSB method identifies the quantity $b_{20}/b_{00}$ as the
ellipticity, so the ``shear polarizability'' is compactly expressed using
\eqq{shearbpq} with $p=2$, $q=0$.

An infinitesimal translation of the image by $(x_0,y_0)$ gives
\begin{eqnarray}
I^\prime(x,y) & = & I(x-x_0,y-y_0) \\
 & \approx & \left(1- x_0 { \partial \over {\partial x}}
	- y_0 { \partial \over {\partial y}}\right) I \\
\label{transgenerator}
 & = & (1 + z {\bf t} - \bar z {\bf t}^\dagger)I, \\
z & \equiv & { {x_0+iy_0} \over \sigma}, \\
{\bf t} & = & {1\over2}\left(a_q^\dagger-a_p\right), \\
\label{lintran}
\Rightarrow\quad b^\prime_{pq} & = &
	b_{pq}  + {z\over 2}
	\left(\sqrt q \, b_{p(q-1)} - \sqrt{p+1}\,b_{(p+1)q}\right)
	+ {\bar z\over 2}
	\left(\sqrt p \, b_{(p-1)q} - \sqrt{q+1}\,b_{p(q+1)}\right) \\
{\bf T}_z & = & \exp[z{\bf t} - \bar z {\bf t}^\dagger]
\end{eqnarray}
To leading order the translation changes $p$ or $q$ by $\pm1$, and the
finite translation ${\bf T}_{\bf z}$ mixes all the states together.
Appendix~\ref{finitetransforms} gives the matrix elements for all the
finite transformations.

The ``smear polarizability'' coefficients of KSB could be quickly
derived at this point by noting that an infinitesimal convolution
along the $x$ direction can be expressed as an average of two images
translated by $\pm x_0$, which will lead to a smear generator that is second
order in the raising and lowering operators. Appendix~\ref{rfactorapp}
gives a closely related derivation.

\subsubsection{Shape Measurement from Laguerre Decomposition}
Our algorithm for shape measurements (with perfect resolution) is to
find the translation z, dilation $\mu$, and shear \bfe\ which must
be applied to the image to satisfy the conditions
(\ref{bcentroid})--(\ref{bsig}).  If ${\bf b}$ is the vector of
Laguerre coefficients for the image $I$, we can write
\begin{equation}
{\bf b}^\prime = {\bf M}\cdot{\bf b} = ({\bf S}_{\bfe} {\bf D}_\mu {\bf
T}_z)\cdot {\bf b},
\end{equation}
and the elements of ${\bf b}^\prime$ with (pq) = (10), (11), and (20) must
vanish.  This is in general a complex non-linear equation, but for 
$\mu,\eta,z\ll 1$ the linearized Equations~(\ref{lindil}),
(\ref{linshear}), and (\ref{lintran}) yield
\begin{eqnarray}
b^\prime_{10} & = & b_{10} 
+ (b_{10} - \sqrt{2} b_{21}) \mu 
- \case{\sqrt 2}{4} b_{12} \bar\eta 
- \case{\sqrt{6}}{4}  b_{30} \eta 
\nonumber \\* & & \qquad
- \case{\sqrt{2}}{2} b_{20} z  + \case{1}{2}(b_{00} - b_{11}) \bar{z}  \\
b^\prime_{11} & = & b_{11} 
+ (b_{00} + b_{11} - 2b_{22}) \mu 
- \case{\sqrt{6}}{4} b_{13} \bar\eta 
- \case{\sqrt{6}}{4} b_{31} \eta 
\nonumber \\* & & \qquad
+ \case{1}{2}(b_{10} - \sqrt 2 b_{21}) z  
+ \case{1}{2}(b_{01} - \sqrt 2 b_{12}) \bar z \\
b^\prime_{20} & = & b_{20}
+ (b_{20} - \sqrt 3 b_{31}) \mu 
+ \case{\sqrt{2}}{4} (b_{00} - b_{22}) \bar\eta
- \case{\sqrt{3}}{2} b_{40} \eta
\nonumber \\* & & \qquad
- \case{\sqrt{3}}{2} b_{30} z
+ \case{1}{2} (\sqrt 2 b_{10} - b_{21}) \bar z,
\end{eqnarray}
writing $\eta = \eta_+ + i \eta_\times$.
These linear equations may be solved explicitly for the desired
transformation coefficients $\eta$, $\mu$, and $z$.  The solution
appears much simpler if the object is nearly round, such that
$b_{pq}\ll b_{00}$ for $p\ne q$.  In this case the transformation
parameters are, to leading order,
\begin{eqnarray}
\label{linearsoln}
\eta & = & {{ -2\sqrt{2} b_{02} } \over {b_{00}-b_{22}}} \\
z & = & {{ -2 b_{01} } \over {b_{00}-b_{11}}} \\
\mu & = & {{ - b_{11} } \over {b_{00}-2b_{22}}} .
\end{eqnarray}
From these equations it is clear that $b_{20}$, $b_{10}$, and $b_{11}$
are the primary carriers of information on shape, centroid, and size,
respectively.  Since the shape component $\eta_+$ is $(\eta + \bar
\eta)/2$, its uncertainty is to leading order
\begin{eqnarray}
\label{sigeta_laguerre}
\sigma^2_\eta \equiv {\rm Var}(\eta_+) & = & 
{ {2\left[ {\rm Var}(b_{20}) + {\rm Cov}(b_{20}b_{02}) 
	+ {\rm Var}(b_{02}) \right]} \over
	(b_{00}-b_{22})^2 } \\
 & = & {{4 n \sigma^2} \over {b_{00}^2}} \left[ 1 - b_{22}/b_{00}
\right]^{-2} \\
\Rightarrow\quad \sigma_\eta & = & {2 \over \nu} 
	\left[ 1 - b_{22}/b_{00}\right]^{-1}.
\end{eqnarray}
Here we have made use of Equation~(\ref{bpqcov}) for the covariance
matrix in the case of white noise, and Equation~(\ref{signif}) for the
definition of the significance $\nu$.  We see that the Laguerre
expansion easily reproduces the earlier result in
Equation~(\ref{sigeta}), with the $a_4$ parameter in that equation
being simply the strength of the $b_{22}$ coefficient.

With a more tedious procedure we may derive the solution for $\eta$ to
second order in $b_{pq}/b_{00}$.  We take $b_{pp}\ll b_{00}$ for $p>0$
this time in order to simplify the expression:
\begin{equation}
\label{etanonlinear}
\begin{array}{rl}
\displaystyle
\eta \approx & -2\sqrt{2} {b_{02} \over {b_{00}-b_{22}}}
 - 2\sqrt{3} { b_{20} b_{04} \over b_{00}^2 } \\*[4pt]
 & \quad + { 2 b_{01}^2 - 2\sqrt{6} b_{10}b_{03} - 2\sqrt{2}
b_{01}b_{12} \over b_{00}^2 } \\*[4pt]
  & \quad -  2\sqrt{6} {b_{11}b_{31} \over b_{00}^2} 
\end{array}
\end{equation}
The terms in the first line arise from solving for $\eta=0$; the terms
in the second and third lines arise when simultaneously solving the
centroid and size constraints, respectively.  From these terms one can
derive the $O(\nu^{-4})$ terms in the shape uncertainty if desired.
But more importantly, this expression uncovers a very significant
source of bias in shape measurements:  the presence of the
$b_{pq}$ in second order means that it is possible for noise to be
rectified in the determination of ${\bf e}$.  If the noise is
anisotropic---as is the case after correction for an anisotropic
PSF---then ${\bf e}$ will be biased.  We will explore this in more detail
in \S\ref{centroidbias}.

\subsubsection{Fourier Transforms and Convolution}
\label{foursec}
The observed galaxy intensity $I^o({\bf x})$ is the convolution of the
image-plane intensity $I^i({\bf x})$ and the stellar PSF 
$I^\star({\bf x})$:
\begin{equation}
I^o({\bf x}) = I^i({\bf x}) \circ I^\star({\bf x})
= \int\! d^2\!{x^\prime}\, I^i({\bf x^\prime})
I^\star({\bf x-x^\prime}).
\end{equation}
The post-seeing, pre-seeing, and stellar images can be expressed as
vectors ${\bf b}^o$, ${\bf b}^i$, and ${\bf b}^\star$, respectively,
of coefficients over our eigenfunction sets $\psi_{pq}^{\sigma_o}$,
$\psi_{pq}^{\sigma_i}$, and $\psi_{pq}^{\sigma_\star}$.  The
convolution can then be expressed as a matrix relation
\begin{eqnarray}
{\bf b}^o & = &  {\bf C}({\bf b}^\star) \cdot {\bf b}^i \\
b^o_{p_o q_o} & = & \sum C^{p_i q_i p_\star q_\star}_{p_o q_o}
	b^i_{p_i q_i} b^\star_{p_\star q_\star}.
\end{eqnarray}
We will effect the deconvolution by inverting a truncated version of
the matrix ${\bf C}$.  Its coefficients are determined by the relation
\begin{equation}
\label{convcoeffs}
\psi_{p_i q_i}^{\sigma_i} \circ \psi_{p_\star q_\star}^{\sigma_\star}
= \sum C^{p_i q_i p_\star q_\star}_{p_o q_o} \psi_{p_o
q_o}^{\sigma_o}.
\end{equation}
The convolution is more easily expressed in $k$-space.  With the
usual definitions (denoted as ``System 3'' by \citet{Br78})
\begin{eqnarray}
\label{fourierdef}
\tilde I({\bf k}) & = & (2\pi)^{-1}\int\! d^2\!x\, I({\bf x}) e^{-i{\bf
k\cdot x}}, \\
I({\bf x}) & = & (2\pi)^{-1}\int\! d^2\!k\, \tilde I({\bf k}) e^{i{\bf
k\cdot x}}, \nonumber
\end{eqnarray}
the convolution becomes a multiplication, and the matrix coefficients
in Equation~(\ref{convcoeffs}) can also be expressed as
\begin{equation}
\label{kconv}
 2\pi \tilde \psi_{p_i q_i}^{\sigma_i} \tilde \psi_{p_\star q_\star}^{\sigma_\star}
= \sum C^{p_i q_i p_\star q_\star}_{p_o q_o} \tilde \psi_{p_o
q_o}^{\sigma_o}.
\end{equation}
We now make use of another remarkable property of the
Laguerre-exponential eigenfunctions, which is that {\em they are their
own Fourier transforms}.  First we note that $\psi_{00}^\sigma$ is a
two-dimensional Gaussian, and the transform (\ref{fourierdef}) is
easily executed to yield a new Gaussian,
\begin{equation}
\label{kpsi00}
\tilde \psi_{00}^\sigma = {1 \over {\sqrt{\pi}}} e^{-k^2\sigma^2/2}.
\end{equation}
The functional form of $\tilde\psi_{00}(k)$ matches that of
$\psi_{00}(x)$, except that we must send $\sigma\rightarrow 1/\sigma$.
Next we can use the definition of the raising operators
(Equation~[\ref{raiseops}]) to note that
\begin{equation}
\label{kraise}
\widetilde{a_p^{\sigma\dagger}\psi} =
{ {-i} \over 2}\left[ \sigma({k_x+ik_y}) - 
	{1 \over \sigma}
	\left( {\partial \over {\partial k_x}} + i{\partial \over
	{\partial k_y}}	\right) \right] \tilde\psi
\equiv \tilde a_p^{\sigma\dagger}\tilde\psi
\end{equation}
Thus the $k$-space raising operator has the same form as the $x$-space
raising operator, save the $\sigma\rightarrow 1/\sigma$ transformation
and an additional factor of $-i$.  The same is true of $a_q^\dagger$.
Since $\psi_{00}$ and the raising operators are each unchanged by the
Fourier transform, it must be true that all the $\psi_{pq}$ are their
own Fourier transforms as well, with appropriate scaling factors of
$i$ and $\sigma$.  More precisely, we have ({\it cf.} [\ref{rlaguerre}])
\begin{equation}
\label{klaguerre}
\tilde\psi_{pq}^\sigma(k,\phi)  =  
	{ {(-i)^m} \over {\sqrt\pi}}
	\sqrt{ {q!} \over {p!} }
	\left( k\sigma \right)^m e^{im\phi}
	e^{-k^2\sigma^2/2}
	L_q^{(m)}(k^2\sigma^2) \qquad (p\ge q)
\end{equation}
So the problem of convolving two of our eigenfunctions is reduced to
the simpler problem of multiplying the same two eigenfunctions.  
We may reach several conclusions immediately:
\begin{itemize}
\item The convolution of $\psi^{\sigma_i}_{N_i m_i}$
with $\psi^{\sigma_\star}_{N_\star m_\star}$ will produce an observed
image with azimuthal order $m_o=m_i+m_\star$.  Furthermore the
multipole phase of the observed image must be the sum of the phases
for the original and PSF images' coefficients.
\item If we choose the scale size $\sigma_o$ for the observed image
eigenfunctions such that
\begin{equation}
\label{deconvsizes}
\sigma_o^2 = \sigma_i^2 + \sigma_\star^2,
\end{equation}
then the convolution contains components $\psi^{\sigma_o}_{N_o m_o}$
only for $N_o\le N_i+N_\star$.  Recall also that we must have
$N_o\ge m_o=m_i+m_\star$, so it must be true that
$\psi^{\sigma_i}_{p_iq}\circ\psi^{\sigma_\star}_{p_\star q}\propto
\psi^{\sigma_o}_{(p_i+p_\star)q}$ for $q=0$.
\end{itemize}

In Appendix~\ref{convolveapp} we give a recursion relation to
calculate any of the elements $C^{p_i q_i p_\star q_\star}_{p_o q_o}$
that we need to calculate the convolution, and thus the deconvolution.

\subsubsection{Deconvolution, Noise Amplification, and Truncation}
\label{laguerredeconv}
With the formulae for the convolution matrix ${\bf C}$ in hand, we can
investigate the nature of the trade-off between the fidelity of the
deconvolution---{\it i.e.} the extent to which effects of the PSF upon
the shape are removed---and the noise level of the deconvolution.
We must truncate the ${\bf b}$ vectors at some finite order $N$ to
implement the deconvolution; higher $N$ will remove more of the PSF
effects, but also increase the noise level.  We can illustrate this
phenomenon by considering the simplest case of convolution by a unit
Gaussian PSF, $b^\star_{pq}=2\sqrt\pi\delta_{p0}\delta_{q0}$.  Since the PSF has
only $m_\star=0$ terms, we will have $m_o=m_i$, so the convolution
matrix is block-diagonal and we may deconvolve each $m$ independently.
If we choose
$\sigma_o^2 = \sigma_i^2 + \sigma_\star^2$, then the convolution matrix 
coefficients are zero for $N_o>N_i$, and hence each block matrix is
also upper-triangular.  This makes the inversion easy.  Using the
results of Appendix~\ref{convolveapp}, defining a deconvolution
parameter $D = \sigma_i^2/\sigma_o^2 = 1 - \sigma_\star^2/\sigma_o^2$,
and using
Equation~(\ref{bpqcov}) for the covariance of the measured moments,
the deconvolved value of $b^i_{00}$ and its variance are
\begin{eqnarray}
\label{invert00}
b^i_{00} & = & \sum_{p=0}^\infty
\left({ {1-D} \over D}\right)^p (-1)^p b^o_{pp}  \\
\label{var00}
\Rightarrow \quad {\rm Var}(b^i_{00}) & = & n\sigma_o^2 \sum_{p=0}^\infty
\left({ {1-D} \over D}\right)^{2p} = n\sigma_i^2 { D^2 \over {2D-1}}
\qquad (D>1/2).
\end{eqnarray}
In practice we could not measure the $b^o_{pp}$ to infinite $p$ and we
must truncate the sum (\ref{invert00}) at some finite $p$.  The more terms
we include, the more accurately we describe the deconvolved
$b^i_{00}$, but each additional term in the deconvolution
adds more noise.  For $D>1/2$, the added noise in successive terms
drops as $p$ increases, and we could in principle do a complete
deconvolution of $b^i_{00}$ with finite variance.  For $D\leq 1/2$, however,
the noise in the deconvolved moment diverges for $p\rightarrow\infty$,
and we are forced to truncate the deconvolution matrix.

For ellipticity measurements we are primarily interested in the
$b^i_{20}$ moment.  For this case of a Gaussian PSF, the general
$b^i_{m0}$ moment deconvolution is
\begin{eqnarray}
\label{invertm0}
b^i_{m0} & = & \sum_{q=0}^\infty
{1 \over D^{m/2}} \left({ {1-D} \over D}\right)^q (-1)^q 
\sqrt{ {q+m}\choose m}
b^o_{(q+m)q}  \\
\Rightarrow \quad {\rm Var}(b^i_{m0}) & = & { {n\sigma_o^2}\over {D^m}}
\sum_{q=0}^\infty
\left({ {1-D} \over D}\right)^{2q} {{q+m}\choose m} \\
& = & n\sigma_i^2 D \left( {D \over {2D-1}} \right)^{m+1}
\qquad (D>1/2).
\end{eqnarray}
We note that these moments also demonstrate the properties that they
are noisier than the observed moments, and infinitely so if $D<1/2$
and the deconvolution is not truncated.  The noise increases as $m$
increases, as we would expect, since higher spatial frequencies must
be recovered.

For a purely Gaussian PSF with $D>1/2$, it is possible to complete the
deconvolution with finite noise.  A real PSF must have components
beyond $b^\star_{00}$, since the transfer function vanishes above a
critical $k$ value.  When we invert the convolution matrix for such a
PSF we will find that the variance series akin to (\ref{var00}) would
diverge for any $D<1$.  The best truncation value for the matrix will
depend upon the form of the PSF and the accuracy to which we demand
correction of PSF effects.  We will examine some specific cases in
\citet{Paper2}.

\subsubsection{Analytic or Kernel Correction for PSF Bias?}

There are two general means of eliminating the shape
{\em bias} induced by the PSF.  One alternative is to measure the
anisotropy of the PSF carefully and apply analytic corrections to the
measured objects, as occurs naturally within the deconvolution
framework described in the preceding paragraphs.  The KSB formalism
contains an approximate analytic correction.  The other method is to
convolve the image with a spatially-varying kernel which removes the
anisotropy from the PSF, as first demonstrated by \citet{FT97} and
further advocated in K00.  Removal of the PSF bias is the most
critical task for a weak lensing method: the PSF {\em dilution} is only a
calibration error or signal modulation, but the PSF {\em bias} introduces a
first-order false signal to the lensing analysis.

Ideally, analytic correction is preferred.  The convolution kernel
will always degrade the image resolution to some extent.  Many PSFs
cannot be ``rounded out'' unless the kernel is of size comparable to
the PSF itself (K00).  Furthermore the convolution will perturb the
noise spectrum of the image, complicating error estimation.  The
kernel method, however, can for simple kernels be faster than an
involved analytic correction, especially if eliminating the bias is
more important than calibrating the dilution for the task at hand.

The kernel method may offer peace of mind.  The demands for
rejection of the PSF bias are very stringent:  PSF ellipticities of 10\%
are not uncommon, yet state-of-the-art lensing surveys require
systematic errors below 0.1\%, so the method we choose must reduce the
PSF bias by over 2 orders of magnitude.
If the kernel method can make the PSF truly round, then
symmetry principles preclude any artificial coherent ellipticity.
We flesh out this statement in \S\ref{kernel}.
We may have more trust in the symmetry principle than we do in an
analytic correction, especially when finite sampling is taken into
account. Note, however, that the symmetry principle requires that the
noise power spectrum is also isotropic; \S\ref{centroidbias} below
demonstrates how anisotropic noise can bias shape measurements
through {\em centroid bias.}

We therefore will describe an alternative procedure:
eliminate the bias with a kernel correction; measure the shapes,
either isotropizing the noise or making a correction for centroid bias;
then correct them for dilution with a heuristic formula such as
Equation~(\ref{unweightedR}).  In \S\ref{kernel} we describe a
rounding-kernel methodology, and in Appendix~\ref{rfactorapp} we
derive a higher-order version of the dilution correction.  These
methods were used by \citet{F00}, \citet{Sm01}, and the forthcoming
CTIO lens survey.

\subsection{Pixellated Data}
\label{pixels}
Real astronomical data is integrated into pixels and sampled at finite
intervals, and the continuum limit that we have assumed in our
analysis is not strictly valid.  A formal description of the process
is that the PSF is altered by convolution with the
pixel response function (PRF) to produce the {\em effective PSF}
(ePSF), which is then sampled at the pixel pitch $a$---see
\citet{L99a} or \citet{Be01} for further elaboration.  If the pixel
pitch is coarser than the Nyquist interval of the PSF ($\lambda/2D$
for a diffraction-limited image), then there is aliasing and we cannot
unambiguously reconstruct the original image or the true $b_{pq}$
coefficients.

In the case where the pixel size $a$ is small compared to the PSF size,
the $b_{pq}$ may be estimated by converting the integral
\eqq{inversion} into a sum over pixels.  But the formally correct
procedure is to fit the vector of sampled pixel fluxes $I_i$ with a
model galaxy with free $b^o_{pq}$.  This is a standard minimization
problem for the $\chi^2$ parameter
\begin{equation}
\label{pixelfit}
\chi^2 = \sum_i { \left( I_i - \sum_{pq}b^o_{pq} \psi^\sigma_{pq}({\bf
x}_i) \right)^2 \over {\rm Var}(I_i) }.
\end{equation}
Since the model is linear, the solution for $b^o_{pq}$ and its
covariance matrix have a closed form, but of course the number of
$b^o_{pq}$ coefficients allowed in the model must be less than the
number of pixels being fit.  With the best-fit coefficients and their
covariance matrix in hand, we may proceed with the methods outlined
above.   It is the ePSF rather than the PSF that has convolved the
source image, but since we in fact measure the ePSF for stars, the
deconvolution procedures are unchanged.
Dithered exposures can be handled by extending the pixel sum over
multiple exposures, as long as the PSF is unchanging.  Poorly sampled
images will manifest the aliasing as strong degeneracies in the
solution for the $b_{pq}$.

A potential time-saving procedure would be to fit the pixel data
to a Laguerre expansion that is already convolved with the local PSF,
minimizing 
\begin{equation}
\label{pixelfit2}
\chi^2 = \sum_i { \left[ I_i - \sum_{p_oq_o} \psi^\sigma_{p_oq_o}({\bf
x}_i)\sum_{p_i q_i} C_{p_oq_o}^{p_i q^i} b^i_{p_iq_i} \right]^2 
	\over {\rm Var}(I_i) }.
\end{equation}
Here $C$ is the Laguerre-coefficient convolution matrix for the local
PSF. The intrinsic coefficients $b^i_{pq}$ are the unknowns in this linear
fit.  This is very reminiscent of the method of \citet{Ku99}, except
that our model for the intrinsic galaxy shape may depart from circular
symmetry, and our Laguerre formalism will allow the fit to proceed as
an matrix solution, with matrices rapidly built by recursion.
Aside from its suitably to poorly-sampled data, this direct-fitting
approach has two further advantages over the 
deconvolution-matrix method of the previous section.  First, 
when the noise is not flat and \eqq{bpqcov} does not apply, the
direct-fitting method more directly produces the full covariance
matrix for the ${\bf b}^i$ vector.  Second, the direct fit may make
use of images that are partially contaminated by invalid pixels, {\it
e.g.} cosmic rays, since they may be excluded from the $\chi^2$ sum.
It is this method which we consider most promising for real data.

\section{Rounding-Kernel Methods}
\label{kernel}
In this section we develop a method for producing a convolution kernel
that symmetrizes the PSF to some desired degree.  As discussed above,
this is a potentially efficient means of reducing or eliminating PSF
bias from the shear determination. By taking advantage
of the Laguerre decomposition, the derivation and application of the
spatially-dependent kernel can be efficiently implemented.

\subsection{The Target Transfer Function}

As discussed in \S\ref{ponderpsf}, the effect of seeing is to convolve
the initial image, $I^i$, with a PSF transfer function $T$ to produce
an observed image, $I^o$.  One could theoretically remove 
all the effects of seeing by convolving the observed image 
with a kernel $K$ whose Fourier transform $\tilde K = 1/\tilde T$ to
produce a final image, $I^f$.  

However, as discussed above, this is impossible to carry out in practice, 
because $\tilde T({\bf k}) = 0$ for all $k$ above some critical value.
We can avoid this problem by ignoring
the high order moments of $I^o$ and $I^f$.  This is straightforward when
one does an eigenfunction expansion of the images
using the eigenfunctions introduced in
\S\ref{laguerre}, since the expansion can be truncated at some value of
$N = p + q$.  This captures most of the real information about the 
PSF without introducing the noise from the higher order moments.  
After the convolution, we then have a target transfer function, $T^\prime$,
which is not exactly a delta function, but which can be made to match a 
delta function up to some order $N$.  Most generally we have
\begin{eqnarray}
I^o & = & T \circ I^i \\
T & = & \sum_{pq} b^\star_{pq} \psi^\sigma_{pq}(r,\theta) \\
I^f & = & T^\prime \circ I^i = K \circ T \circ I^o \\
\label{tprimedef}
T^\prime & = & \sum_{pq} b^{\star \prime}_{pq} 
	\psi^\sigma_{pq}(r,\theta) 
\end{eqnarray}
For the ideal case of $T^\prime = \delta({\bf x})$, 
\begin{equation}
\label{fullkernelreq}
b^{\star \prime}_{pq}  = {(-1)^p \over \sqrt \pi} \delta_{pq} 
\end{equation}
Thus, if $T^\prime$ satisfies \eqq{fullkernelreq} 
up to some cutoff order, $N$, $I^f$ 
will be approximately identical to $I^i$ up to the same order. 
If we can find a kernel $K$, which produces this target transfer
function, $T^\prime$, we will 
be able to remove all the effects of seeing as well as possible
given our ignorance of the high order terms.  

How stringently do we need to satisfy \eqq{fullkernelreq}?
Less strict requirements on $T^\prime$ make it
easier to find an appropriate, compact kernel.
Our present goal is to create a transfer function $T^\prime$ which
does not produce any shear bias.  If the original scene $I^i$ is
unlensed, and we represent the shear measurement process as some
operator $\delta(I)$, then the isotropy of the Universe guarantees that
$\delta(I^i)=0$ (we consider $\delta$ as a complex number $\delta_+ +
i\delta_\times$). Our demand on the transfer function is that
\begin{equation}
\label{nobias1}
\delta(I^f) = \delta(T^\prime \circ I^i) = 0.
\end{equation}
Most generally the results of the shear measurement process can be
expanded as a power series in the coefficients of the $T^\prime$ PSF:
\begin{equation}
\label{nobias2}
\delta(T^\prime \circ I^i) = \sum_{p_1q_1} a_{p_1q_1}
b^{\star\prime}_{p_1q_1} 
+ \sum_{p_1q_1,p_2q_2} a_{p_1q_1,p_2q_2}
b^{\star\prime}_{p_1q_1} 
b^{\star\prime}_{p_2q_2} 
+ \cdots
\end{equation}
where the $a_i$ are some coefficients that depend upon the
measurement process, the image characteristics and the size of the
PSF.  We now examine the consequence of rotating the image and the PSF
by some angle $\beta$.  The measured shear and the $T^\prime$
coefficients behave as
\begin{equation}
\label{nobias3}
\delta(I^f) \rightarrow e^{2i\beta}\delta(I^f); \qquad
b^{\star\prime}_{pq} \rightarrow e^{(p-q)i\beta}
b^{\star\prime}_{pq}.
\end{equation}
The individual galaxies in the original scene $I^i$ are not invariant
under rotation, but any statistical measure of their {\em collective}
properties must be invariant under rotation, {\it i.e.} $I^i$ is
invariant under rotation in the same sense that the Universe is
isotropic.  Therefore the $a_i$ coefficients of \eqq{nobias2}
are unaffected by the rotation, and to satisfy the conditions
(\ref{nobias3}) we must have
\begin{equation}
\label{nobias4}
\begin{array}{rcl}
a_{p_1q_1} = 0 & {\rm for} & p_1-q_1\ne 2; \\
a_{p_1q_1,p_2q_2} = 0 & {\rm for} & p_1-q_1+p_2-q_2\ne 2; \\
 & \vdots &
\end{array}
\end{equation}
Thus only PSF terms with $m=p-q=2$ can cause a shear bias, to first
order.  The primary goal of our kernel, therefore, will be to set
\begin{equation}
\label{bpqstar1}
b^{\star\prime}_{pq} = 0 \qquad (m=p-q=2)
\end{equation}
With this condition satisfied, shear bias can only be of order
$(b^\star_{pq})^2$ where $p\ne q$.
Higher-order $a$'s are non-zero only if $m_1+m_2+\cdots=2$.
Many of these elements are zero as well; for example by considering
the invariance of $\delta(I)$ under infinitesimal translation,
one can demonstrate that $a_{10,21}$ vanishes, but $a_{10,10}$ 
must exist.  Satisfying (\ref{bpqstar1}) does not,
therefore, guarantee the elimination of shear bias to all orders.

For absolute assurance that a shear bias is absent, we can set
\begin{equation}
\label{bpqstar2}
b^{\star \prime}_{pq}  =  0 \qquad (p-q \neq 0\ [{\rm mod}\ 4]),
\end{equation}
{\it i.e.} insure that the PSF has fourfold symmetry.  In this case
the condition $\sum m_i=2$ can never be satisfied for non-vanishing
coefficients of $T^\prime$.  Note that it would also suffice to
enforce {\em any} $m$-fold rotational symmetry on the PSF beyond
$m=2$, {\it e.g.} the diffraction pattern of a triangular
secondary-support structure cannot by itself cause shear bias.

In practice we produce a kernel $K$ which enforces condition
(\ref{bpqstar1}) up to some order $N$.  We can set
$b^{\star\prime}_{10}=0$ by appropriate choice of PSF center.  The 
remaining shear biases must be of order $(b_{21}/b_{00})^2$, 
$b_{41}b_{01}/b^2_{00}$, etc., which are
generally quite small.

\subsection{The Components of the Kernel}
Since we are going to convolve the kernel with a pixellated image,
we must construct the kernel as a 2d array instead of a continuous
function. The simplest kernel is the identity kernel,
composed of an array of 0's with a 
single 1 in the middle.  This kernel conserves flux; 
we want a flux-conserving kernels, so we only consider kernels which
are the identity kernel plus a kernel for which the elements sum to
zero. 

Next note that taking the derivative of an image is usually approximated by
a discrete difference.  For example: 
\begin{equation}
{\partial I \over \partial x} = {(I(x+dx) - I(x-dx)) \over 2 dx}
\end{equation}
Another way to write this equation is as a convolution:
\begin{equation}
{\partial I \over \partial x} = \left(
\begin{array}{ccc}
0 & 0 & 0 \\ -1/2 & 0 & 1/2 \\ 0 & 0 & 0
\end{array}
\right)
\circ I
\end{equation}
Similarly,
\begin{equation}
{\partial I \over \partial y} = \left(
\begin{array}{ccc}
0 & 1/2 & 0 \\ 0 & 0 & 0 \\ 0 & -1/2 & 0
\end{array}
\right)
\circ I
\end{equation}

In fact, all partial derivatives of any order in $x$ and $y$ can be 
represented as a convolution by a zero-sum kernel.  First and second order
derivatives can be effected using 3x3 kernels; third and fourth order
derivatives require 5x5 kernels, and so on.  One can therefore think of 
the kernel as being made up of a sum of these derivatives 
$(\partial/\partial x)^i (\partial/\partial y)^j$ including, of course, 
the identity kernel, $(\partial/\partial x)^0 (\partial/\partial y)^0$.

The eigenfunction expansion of our images actually suggests a slightly
different set of components for the kernel.  Namely, 
\begin{equation}
\label{Kdef}
K = \sum_{ij} k_{ij} D_{ij}
\end{equation}
\begin{eqnarray}
\label{idealdijd}
D_{ij} & = &
\left({\partial \over \partial x} + i {\partial \over \partial y}\right)^i 
\left({\partial \over \partial x} - i {\partial \over \partial y}\right)^j \\
& = & {1 \over \sigma^{i+j}} \left(a_q^\sigma - a_p^{\sigma\dagger}\right)^i
\left(a_p^\sigma - a_q^{\sigma\dagger}\right)^j 
\label{idealdija}
\end{eqnarray}

These components make it easy to use the raising and lowering operators
to determine how a given kernel will act on an image.  Note, however, that
$D_{ij}$ are complex, which means that to end up with a real image after
the convolution, we require that $k_{ji} = \bar k_{ij}$.  

The 3x3 kernel components are:
\begin{eqnarray}
\label{d10def}
D_{10} & = & \left(
\begin{array}{ccc}
0 & i/2 & 0 \\ -1/2 & 0 & 1/2 \\ 0 & -i/2 & 0
\end{array}
\right) \\
\label{d01def}
D_{01} & = & \overline{D_{10}} \\
\label{d20def}
D_{20} & = & \left(
\begin{array}{ccc}
-i/2 & -1 & i/2 \\ 1 & 0 & 1 \\ i/2 & -1 & -i/2
\end{array}
\right) \\
\label{d02def}
D_{02} & = & \overline{D_{20}} \\
\label{d11def}
D_{11} & = & \left(
\begin{array}{ccc}
0 & 1 & 0 \\ 1 & -4 & 1 \\ 0 & 1 & 0
\end{array}
\right) 
\end{eqnarray}

We note at this point that the FT97 method is equivalent to use of
just the $D_{20}$ and $D_{02}$ kernel elements.
The 5x5 and larger kernel components can be found by convolving
the 3x3 components.  For instance, $D_{30} = D_{10} \circ D_{20}$, and
$D_{41} = D_{11} \circ D_{30}$.

\subsection{Calculating the Kernel}

\subsubsection{The Kernel for Infinitesimal Pixels}
\label{kernelinfpix}

The kernels given by \eqq{d10def} - \eqq{d11def} are only equal to the
continuous derivatives (\eqq{idealdijd}) to lowest order in $1/\sigma_\star$.
Typically, the PSF is only a few pixels in size, so this is not that good
an approximation.  However, it is a good place to start, as most of the 
technique will apply to the case of finite pixels.

Combining Equation~(\ref{Kdef}) with the definition of $T^\prime$ 
(Equation~[\ref{tprimedef}]), 
\begin{equation}
\label{bpqstar3}
{\bf b}^{\star\prime} = \sum_{ij} k_{ij} {\bf D}_{ij} {\bf b}^\star
\end{equation}

The operator matrix ${\bf D}_{ij}$, in the limit of infinitesimal
pixels, is easily calculated using Equations~(\ref{idealdija}) 
and (\ref{raiseops}), and there is a fast recursion
\begin{eqnarray}
\label{recursivedij}
{\bf D}_{00} {\bf b}^\star & = & {\bf b}^\star \\
\nonumber
{\bf D}_{(i+1)j} {\bf b}^\star & = & {1 \over \sigma^\star}
(a_q^{\sigma^\star} - a_p^{\sigma^\star \dagger}) ({\bf D}_{ij} {\bf b}^\star) 
\\
\nonumber
{\bf D}_{i(j+1)} {\bf b}^\star & = & {1 \over \sigma^\star}
(a_p^{\sigma^\star} - a_q^{\sigma^\star \dagger}) ({\bf D}_{ij} {\bf b}^\star) 
\end{eqnarray}
Since ${\bf b}^\star$ is measured and the ${\bf D}_{ij}$ are fixed
matrices, 
we have a matrix equation for the unknown kernel coefficients
$k_{ij}$, given the chosen constraints on ${\bf b}^{\star\prime}$ ({\it
e.g.} Equation~[\ref{bpqstar1}]): 
\begin{equation}
\label{mk_eq_b}
{\bf M k} = {\bf b}^{\star\prime}
\end{equation}
where ${\bf k} = \{k_{ij}\}$ and the $ij$ row of ${\bf M}$ is given
by $({\bf D}_{ij} {\bf b}^\star)^T$.  Thus, given ${\bf b}^\star$ and 
${\bf b}^{\star\prime}$, one calculates ${\bf M}$ using 
Equations~(\ref{recursivedij}), and 
then simply solves Equation~(\ref{mk_eq_b}) for ${\bf k}$.

The problem with this method is that ${\bf b}^{\star\prime}$ will not
usually be completely specified.  Neither \eqq{bpqstar1} nor
(\ref{bpqstar2}) fully constrains ${\bf b}^{\star\prime}$.
The easiest way to deal with this is to set $k_{pq} = 0$ for each 
unspecified $b_{pq}^{\star\prime}$.  Then the kernel will be as 
simple as possible while still satisfying all of the requirements for
${\bf b}^{\star\prime}$.  A more sophisticated technique for dealing with
this issue is described below in \S\ref{minimizingkernel}.

\subsubsection{The Kernel with Pixelization}
\label{kernelwpix}
The same general technique applies to the case of finite pixels in 
that we want to solve \eqq{mk_eq_b} for $\bf k$.  The only 
difference is the calculation of ${\bf D}_{ij} {\bf b}^\star$.
In this case, we must use Equations~(\ref{d10def}) - (\ref{d11def}) 
rather than \eqq{idealdija}.

Consider the version of $D_{10}$ given in \eqq{d10def}:
\begin{eqnarray}
\nonumber
D_{10} {\bf b}^\star (x,y) & = & 
{1 \over 2} \left ( {\bf b}^\star (x+1,y) - {\bf b}^\star (x-1,y) \right ) \\*
&  & + {i \over 2} \left ( {\bf b}^\star (x,y+1) - {\bf b}^\star (x,y-1) \right ) \\
& = & {1 \over 2} \left ( {\bf T}_{z1} {\bf b}^\star - {\bf T}_{-z1} {\bf b}^\star \right )
+ {i \over 2} \left ( {\bf T}_{z2} {\bf b}^\star - {\bf T}_{-z2} {\bf b}^\star \right )
\end{eqnarray}
where $z1 = 1/\sigma$, $z2 = i/\sigma$ and ${\bf T}_z$ is defined in \S\ref{Translation}.

Calculations for the other ${\bf D}_{ij}$ are similar.

\subsubsection{The Kernel in a Distorted Frame of Reference}
\label{kernalwdistort}
Most telescopes, especially those with large fields of view, have fairly 
significant distortion.  To deal with this correctly, the shape measurements should
be made in the undistorted world coordinates rather than the chip coordinates.
The kernel's pixel grid, however, is still is in the distorted
coordinate system. The calculation
of ${\bf D}_{ij} {\bf b}^\star$ must therefore take into account the
two different coordinate systems.

If the world coordinates are $(u,v)$ and the pixel coordinates are
$(x,y)$ then the correct values for $z1$ and $z2$ are:

\begin{eqnarray}
z1 & = & {1 \over \sigma} \left ( {\partial u \over \partial x} 
+ i {\partial v \over \partial x} \right ) \\
z2 & = & {1 \over \sigma} \left ( {\partial u \over \partial y} 
+ i {\partial v \over \partial y} \right )
\end{eqnarray}

\subsection{Minimizing the Noise from the Kernel Convolution}
\label{minimizingkernel}

The motivation for having a compact kernel is to
minimize the noise added by the convolution.  A large kernel
will use data with significant noise but little signal, adding 
to the noise in the convolved image.  

Therefore, let's consider the noise in the convolved image.
Define $K(m,n)$ to be the total convolution mask.  

\begin{equation}
\label{kmndef}
K(m,n) = \sum_{ij} k_{ij} D_{ij}(m,n)
\end{equation}

\begin{equation}
I^f(x,y) = \sum_{mn} K(m,n) I^o(x+m,y+n)
\end{equation}

If the noise in the image is dominated by sky noise, then we can
define $n^2$ to be the variance in each pixel of the observed image.
The noise in each pixel of the convolved image is

\begin{equation}
\label{convnoise}
n_f^2 = \sum_{mn} K(m,n)^2 n^2 = n^2 \sum_{mn} \left (\sum_{ij} k_{ij} D_{ij}(m,n) \right)^2
\end{equation}

In \S\ref{kernelinfpix}, we set some kernel coefficients to 0 to account
for the unspecified components of ${\bf b}^{\star\prime}$.  
Really, each of these coefficients is arbitrary, so the solution
to \eqq{mk_eq_b} will be 

\begin{equation}
\label{generalksolution}
{\bf k} = {\bf k_0} + {\bf A Y}
\end{equation}
where $\bf k_0$ is a specific solution, $\bf A$ is a matrix, and $\bf Y$ is an arbitrary 
vector.  The dimension of $\bf Y$ is equal to the number of unspecified components of 
${\bf b}^{\star\prime}$.  

We can then find the particular $\bf Y$ which minimizes the noise added to 
the image by the convolution.  For this derivation ${\bf D}(m,n)$ is the 
vector of each ${\bf D}_{ij}(m,n)$.  So, $K(m,n) = {\bf D}(m,n) {\bf k}$.

\begin{eqnarray}
0 & = & {\partial n_f^2 \over \partial {\bf Y}} \\
\nonumber
  & = & {\partial \over \partial {\bf Y}} \left( n^2 \sum_{mn} \left( {\bf D}(m,n) {\bf k} \right)^2 \right)\\
\nonumber
  & = & 2 n^2 \sum_{mn} \left( {\bf D}(m,n) {\bf k} \right ) \left( {\bf D}(m,n) {\bf A} \right) \\
0 & = & \sum_{mn} \left( {\bf D}(m,n) ({\bf k_0} + {\bf A Y} \right) 
        \left ( {\bf D}(m,n) {\bf A} \right )
\end{eqnarray}
\begin{equation}
\label{finalyeqn}
\sum_{mn} \left( {\bf D}(m,n) {\bf A} \right) \left( {\bf D}(m,n) {\bf A Y} \right )
 = - \sum_{mn} \left( {\bf D}(m,n) {\bf A} \right) \left( {\bf D}(m,n) {\bf k_0} \right )
\end{equation}

This is now a matrix equation which can be solved for $\bf Y$, which then
gives you the solution for {\bf k} which minimizes the noise.

It turns out that if this method is implemented exactly as described, one ends
up with a fairly large kernel, which is essentially a smoothing filter.  
While this solution will work, it is not exactly what we want from the kernel.
We would rather have a smaller kernel which minimally changes the image.
So, rather than minimize the noise in $I^f$, we minimize the noise in
$I^f - I^o$.  In other words, we leave out the ${\bf D}_{00}$ term in the 
vector products of \eqq{finalyeqn}.  This results in fairly compact kernels
which vary smoothly across the image.

\subsubsection{Additional Kernel Components}

There are 9 free parameters in a general real $3\times3$ kernel.
Equations~(\ref{d10def}) 
- (\ref{d11def}) define 5 kernels.  The identity kernel is another.  
Thus, there are 3 more independent 3x3 kernels we could construct.
Two of these can be made to approximate discrete versions of the 
derivatives $D_{10}$ and $D_{01}$.  The other can be made to approximate
a discrete version of $D_{11}$.  These alternate versions are:

\begin{eqnarray}
\label{altd10def}
AltD_{10} & = & \left(
\begin{array}{ccc}
(-1+i)/4 & 0 & (1+i)/4 \\ 0 & 0 & 0 \\ (-1-i)/4 & 0 & (1-i)/4
\end{array}
\right) \\
\label{altd01def}
AltD_{01} & = & \overline{AltD_{10}} \\
\label{altd11def}
AltD_{11} & = & \left(
\begin{array}{ccc}
1/2 & 0 & 1/2 \\ 0 & -2 & 0 \\ 1/2 & 0 & 1/2
\end{array}
\right) 
\end{eqnarray}

We can construct alternate higher order kernels in the same way as we 
constructed the regular higher order kernels.  Namely, 
$AltD_{30} = AltD_{10} \circ D_{20}$, $AltD_{41} = AltD_{11} \circ D_{30}$, etc.

These extra kernel components are useful if one is minimizing the 
noise as described above, since they add extra degrees of freedom for
the minimization, and therefore can result in a smaller, less noisy
kernel.

\subsection{Interpolating Across an Image}

The above discussion explains how to find the appropriate kernel
given a particular PSF.  However, in a real image, the PSF varies
across the chip.  Thus, the kernel will also vary across the chip.

There are two potential ways to deal with this.  One can find
the appropriate kernel for each star in the image.  Then 
fit the kernel components $k_{ij}$ as functions of $(x,y)$.  Alternatively, 
one can fit the coefficients $b^\star_{pq}$ of the measured PSF
decomposition 
as functions of $(x,y)$ and then solve for the appropriate kernel 
at each point.

We choose the first method in our analysis for two reasons. 
First, the derived kernel can be directly applied to each star
to make sure that it really does make the star round.  
Occasionally, a star will have significant high order components
due to crowding or an uncorrected cosmic ray.  When this happens, the
derivation above fails,  
and we reject this kernel from the fit.  
It is a little cleaner to recognize these outliers with kernel
interpolation than with PSF interpolation.

The second reason to prefer fitting the kernel rather than the
PSF is computational efficiency.  The appropriate kernel must
be calculated at pixel.  It is 
significantly faster to evaluate a function than to solve a 
matrix equation.  On a 2K x 4K chip, there are 8 million 
kernel evaluations.  Both methods gain significantly by 
using a locally linear approximation to the spatial variation, but there
is still a significant difference in computation for 
the two methods.

The kernel scheme described here allows an substantial speed gain over
a typical convolution method.  Fourier methods are fastest for large
convolutions, but are not practical for spatially-varying kernels.  In
our scheme, the output image can be written as 
\begin{equation}
I^f = \sum_{ij} k_{ij}(x,y) * (D_{ij} I^o).
\end{equation}
Instead of calculating the kernel at each output pixel, we can instead
produce the images $D_{ij}I^o$ (which are constant-kernel
convolutions), and then accumulate sums of these images with spatially
varying weights $k_{ij}(x,y)$.  
More importantly, the $D_{ij}I^o$ can be produced by recursive
application of the $3\times3$ kernels.  Hence the convolution by a
$7\times7$ kernel can, for example, be reduced to three successive
applications of $3\times3$ kernels.  

Note that interpolation of the PSF elements across the image is
required for the
analytic deconvolution described in \S\ref{laguerredeconv}.

\subsection{Dilution Correction}
Once a kernel has been applied to the image to symmetrize the PSF, the
measured galaxy ellipticities need to be scaled by some resolution
factor $R$ to account for the PSF dilution.  Appendix~\ref{rfactorapp}
describes the scheme we use for estimating $R$, which closely
resembles the KSB ``smear polarizability'' derivation.

\section{Centroid and Selection Biases}
\label{centroidbias}
Even if the asymmetries of the PSF have been perfectly removed by a
deconvolution or other method, there are two effects which can cause
the estimate of the mean shape to be biased in the direction of the
original PSF orientation.  The first is a {\em selection bias}, first
noted by K00.  The second is a measurement bias which arises in the
presence of anisotropic 
noise or anisotropic PSF, which we term the {\em centroid bias.}  K00
discusses an effect whereby the errors in centroid appear in second
order as biases in the ellipticity, and concludes that such errors are
probably negligible.  We show below that this bias is significant,
and in fact just one of a class of noise-rectification biases that can
occur. 

\subsection{The Selection Bias}

Kaiser's selection bias operates as follows in the presence of an
anisotropic PSF:  if the PSF is elongated in the $x$ direction
($e^\star_+>0$), then
objects with intrinsic shape $e_+<0$ cover a larger area after PSF
convolution than do objects with intrinsic $e_+>0$.  On the observed
image, therefore, such objects have both lower surface brightness and
lower significance $\nu$.  As most detection algorithms involve some
cut in surface brightness or $\nu$, the detected population will be
biased toward $e_+>0$.  The mean $e_+$ of the population will be
biased, therefore, even if all the detected objects are perfectly
corrected for the PSF anisotropy.  K00 demonstrates that the bias will
scale roughly as $e^\star \sigma_\star^2 / \sigma_i^2\nu^{2}$, 
where $\nu$ is the detected significance.

The selection bias may be defeated by careful definition of the sample
of target galaxies.  The key is to produce some significance statistic
$\tilde\nu$ which is independent of the shape of the object.  For an
image of flux $f$ covering area $A$ in an image with white noise
density $n$, the significance is normally $\nu\propto f/\sqrt{nA}$.  But the
observed area $A$ is shape-dependent in an anisotropic fashion if
the PSF is elliptical.  We may instead define $\tilde\nu\propto
f/\sqrt{n\tilde A}$, where $\tilde A$ is the object's area on a version of
the image which has been deconvolved, or at least had the PSF
rounded by a convolution kernel, as described in the previous
section.  We then define the target sample by some cutoff limit
$\tilde\nu_{\rm min}$ to this ``isotropic'' significance.

Detection algorithms generally have some statistic which is used as a
cutoff.  For KSB's {\tt imcat}, this is $\nu$.  For {\tt SExtractor}
and {\tt FOCAS}, this is the number of pixels $N$ above some isophotal
threshold in a filtered version of the image.  To eliminate the
selection bias, we make a scatter plot of the detection statistic,
{\it e.g.} $N$, vs. the ``isotropic'' significance $\tilde\nu$.  The
selection bias is eliminated if we choose our cutoff $\tilde\nu_0$
sufficiently high that no members of the selected population approach
the cutoff $N_{\rm min}$ of the detection statistic.  This is illustrated in
Figure~\ref{nuvsN}. In this way we insure that our selected population
is free of any anisotropic selection criterion that may have been
created by the original detection algorithm.  

\begin{figure}
\plotone{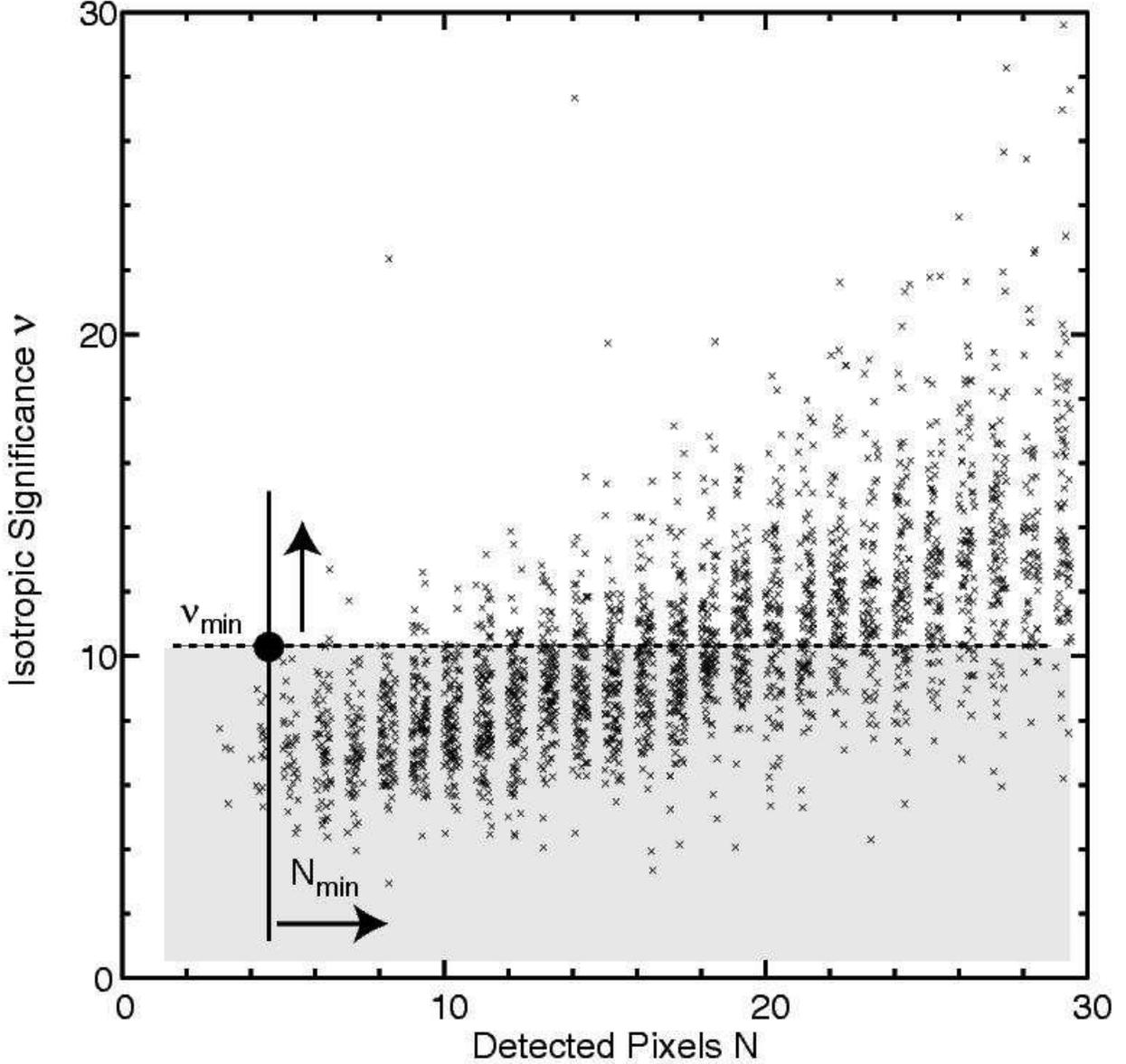}
\caption[]{
\small The scatter plot of isotropic significance $\tilde \nu$ vs
pixels above the 
SExtractor thresholds, $N_{\rm pix}$, indicates the criteria necessary
to avoid the selection bias.  To avoid selection bias, we want the
galaxy selection to be entirely on the basis of the shape-independent
statistic $\tilde\nu$, and not influenced by the potentially
shape-biased SExtractor threshold $N_{\rm min}=5$ [some galaxies with
$N<5$ are present due to object splitting].  We must therefore set our
$\nu_{\rm min}$ threshold at $\approx 10$, which is the lowest value
for which the galaxy population does not extend to $N<N_{\rm min}$.
Galaxies in the shaded region are therefore excluded.
}
\label{nuvsN}
\end{figure}

\subsection{Centroid Bias}
The centroid bias, discussed in K00, can be qualitatively explained as
follows: 
an error $\delta x$ in centroid determination along the $x$ axis inflates
the measured second moment $I_{xx}$ by some amount $\propto \delta
x^2$, whereas a $y$-axis error inflates
$I_{yy}$.  If the centroid errors are isotropic, the mean effect upon
$e_+$ is zero.  But if the $x$ centroid errors $\langle \delta x^2\rangle$
exceed those in $y$,
there is a net tendency to measure a positive $e_+$.  
If the PSF has $e^\star_+>0$, then the PSF---and consequently the mean
galaxy image---is more extended in the $x$ direction, and hence
centroids are less accurate in $x$ than in $y$.  The noise in centroid
estimation therefore pushes the mean measured ${\bf e}$ in the
direction of ${\bf e}^\star$.  The centroid errors scale as
$\nu^{-1}$, and therefore the centroid bias in ${\bf e}$ scales
roughly as $e^\star \nu^{-2}$.  Faint galaxies at low $\nu$ 
have a much larger centroid bias than the bright, high-$\nu$
stars used as PSF templates, so the effect is not properly removed by
the KSB ``smear polarizability'' corrections.

Deconvolving the image eliminates the PSF anisotropy but does not
eliminate the centroid bias, because the deconvolved image will have
anisotropic noise.  There will be more noise power along $k_x$ than
$k_y$ after deconvolution, hence there will still be anisotropic
centroid errors and a bias toward the original ${\bf e}^\star$.

The same situation arises if we apply a convolution kernel to
symmetrize the PSF.  This kernel will smooth the image slightly in the
direction perpendicular to the PSF, reducing the noise level in that
direction somewhat.  The convolved image will therefore again have a
higher centroid error along the original PSF axis, and a consequent
bias in ${\bf e}$.

A quantitative understanding of the centroid bias can be gained from
Equation~(\ref{etanonlinear}), the second-order expression for $\eta$
in the simultaneous solution for shape and centroid (and possibly size).
Let us presume that each $b_{pq}$ has an true value plus
some measurement noise $\delta b_{pq}$.  The measurement noise has
mean value of zero since $b_{pq}$ is a linear function of the intensity.
If the object is
intrinsically round, centered, and the weight is properly sized, then
we have $b_{00}=b_{10}=b_{20}=0$, and we would indeed measure
$\eta=0$.  In the presence of noise, however, the second-order terms
in Equation~(\ref{etanonlinear})---for instance the term $\propto
b_{10}^2/b_{00}^2$ in the second line---may have non-zero mean values.  If
$\langle b^2_{10}\rangle =  {\rm Var}(b_{10}) \ne 0$, there will be a
non-zero $\langle \eta \rangle$.  

Note that there are a number of second-order terms in the $\eta$
solution, arising not just from the solution for object center, but
also from the solutions for object size and ellipticity.  We
apply the term ``centroid bias'' to all of these effects.

According to Equation~(\ref{bpqcov}), ${\rm Var}(b_{10})=0$ in the
presence of a white noise spectrum $P(k)=n$ (recall that $b_{10}$ is a
complex number).  If we have a PSF of size $\sigma_\star$ and
ellipticity ${\bf e}^\star$ along the $x$ axis, we may attempt to
round out the  kernel by smoothing along the $y$ axis a little bit.
This will produce a noise power spectrum $P^\prime(k)\approx n(1-2
e^\star \sigma_\star^2 k_y^2)$.  The $k_y^2$ term in the noise power
produces a non-zero value for  ${\rm Var}(b_{10})$, and also non-zero
covariances between all the other second-order elements of
Equation~(\ref{etanonlinear}).  The leading term in the expression for
centroid bias takes the form
\begin{equation}
\label{centbias1}
\langle {\bf e} \rangle = K { {\bf e}^\star \over \nu^2}
	{\sigma^2_\star \over \sigma^2_o} 
	\approx {K{\bf e}^\star (1-R) \over \nu^2},
\end{equation}
where $K$ is a constant, $e_o$ is the ellipticity measured on the
convolved image, $\sigma_o$ is the size observed on this image, and
$R$ is the resolution factor of \S\ref{psf}.
Note that the bias is in the measured shape, before correction for PSF
dilution, which would add a factor $R$ in the denominator.
The value of $K$ depends upon some higher-order moments of the typical
galaxy, and upon the details of the PSF-correction and measurement
procedure---in some cases $K$ can be negative.  Note also that the
functional dependence of the centroid bias is essentially the same as
that for the selection bias, hence they are difficult to distinguish.

Figure~\ref{cbias} demonstrates the existence of the centroid bias in
a very numerical simulation, in which we convolve a circular Gaussian
source, with an elliptical Gaussian weight, then measure the ellipticity and
centroid with a fixed-size circular Gaussian weight.  It is clear that
the mean measured ellipticity depends upon the significance ({\it
i.e.} $S/N$ level) as described by \eqq{centbias1} with $K=-2$.  
If the weight size is iterated to maximize significance, the bias
increases to $K\approx-6$. For
other measurement algorithms or galaxy shapes, $K$ will differ, so
we must determine $K$ empirically.

\begin{figure}
\plotone{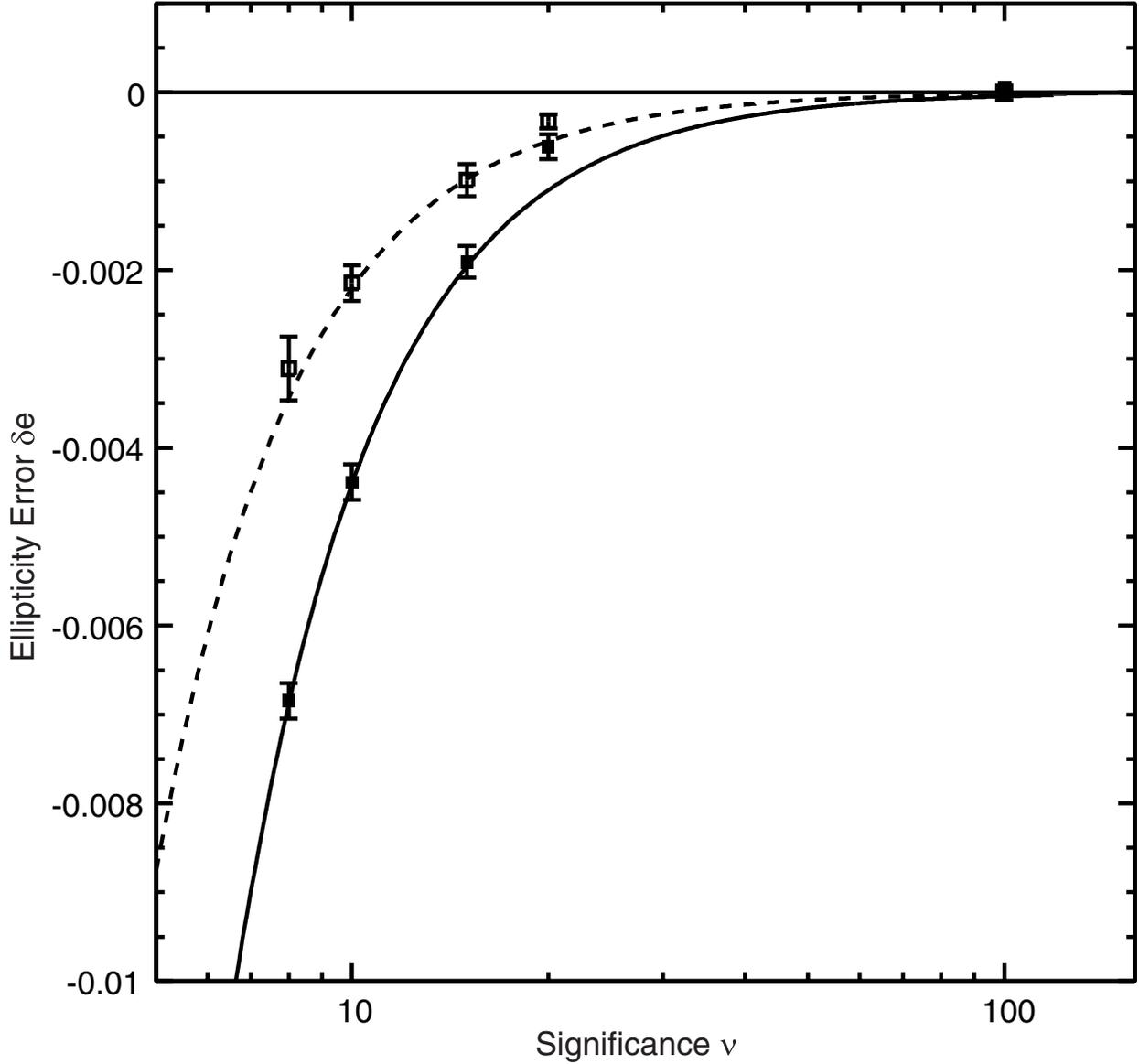}
\caption[]{
\small The effect of centroid bias is demonstrated by a simple
numerical test in which a circular Gaussian in measured in the
presence of an elliptical PSF and white noise.  High-significance
detections are observed at the correct ellipticity and would yield
$e=0$ when corrected for the PSF ellipticity.  But the measured
ellipticity drops as the significance $\nu$ decreases.  The test
results are well described by \eqq{centbias1} with $K=-2$, which is
plotted as the solid (dashed) line for the tests with
$e^\star(1-R)=0.2$ ($0.1$). 
}
\label{cbias}
\end{figure}

The bias in \eqq{centbias1} will be equal to the shape-noise
uncertainty in the mean of $N$ galaxies when
\begin{equation}
{ K e^\star(1-R) \over \nu^2} \approx {0.3 R \over \sqrt{N}}
\qquad \Rightarrow \qquad
N \approx \left({0.3 \over Ke^\star} {R \over 1-R} \right)^2 \nu^4.
\end{equation}
For typical PSF ellipticities and $K$ values, with galaxies
somewhat resolved ($R\approx0.5$) and detected at $\nu=10$,
we find the bias equal to the shape noise when $N\sim10^4$.  The
effect clearly cannot be ignored in present-day surveys.
If the PSF ellipticity varies, then the bias will induce false power
into the distortion power spectrum, with a total fluctuation power of
$\approx \langle (e^\star)^2 \rangle K^2/\nu^4$, for $R\approx0.5$.
If we take the RMS value of $e^\star$ to be about 3\%, take
$|K|\approx 5$, and $\nu=8$, we see that the bias power is about 5\%
of the typical cosmic signal, for which the RMS distortion is $\approx
1\%$.  In unfavorable parts of the power spectrum, the ratio of PSF
bias power to cosmic signal will be worse; hence the bias cannot be
ignored for cosmic shear studies which aim to move beyond mere
detection to true precision measurements.

There are several possible strategies for defeating the centroid
bias.  The simplest is to find $K$ empirically, then apply an
appropriate bias correction to each object using
\eqq{centbias1}.  Another approach, useful if one has applied
a PSF-rounding kernel to the image, is to add noise back to the image
to recreate the original isotropic noise spectrum.  If both the PSF
and the noise spectrum are isotropic, there is no way the mean shape
can be biased.  Unfortunately, creating the appropriate noise field to
add to the image is computationally expensive for all
but the simplest convolution kernels.

Symmetrizing the noise is easier if we have used the Laguerre
decomposition method to deconvolve the images.   If we have properly
propagated the original covariance matrix for the $b_{pq}$, then we
know ${\rm Var}(b_{10})$---and all the other relevant covariances---in
the deconvolved image.  We can add noise to the deconvolved $b_{pq}$
elements in order to zero out the asymmetric elements of the
covariance matrix. 

\section{Procedures for Shape Measurement with PSF Correction}
\label{procedures}
We now have all the necessary tools for a 
procedure for measuring shapes and shear in the presence of a PSF
convolution.  Figure~\ref{flowchart} is a flowchart for this
procedure, and we elaborate on each step below.  
First note that
there are two branch points.  The first is
the decision whether to measure the shapes on a summed image, or to
measure PSF-corrected moment information from each exposure and
combine the moments.  The former is easier if all exposures cover the
same sky area, but the latter is recommended whenever the coverage is
interlaced on the sky.  The second branch point is deciding whether to
use an analytic or a kernel correction for PSF bias.
The remainder of this section delineates the overall data reduction
procedure for weak lensing measurements.

\begin{figure}
\epsscale{0.6}
\plotone{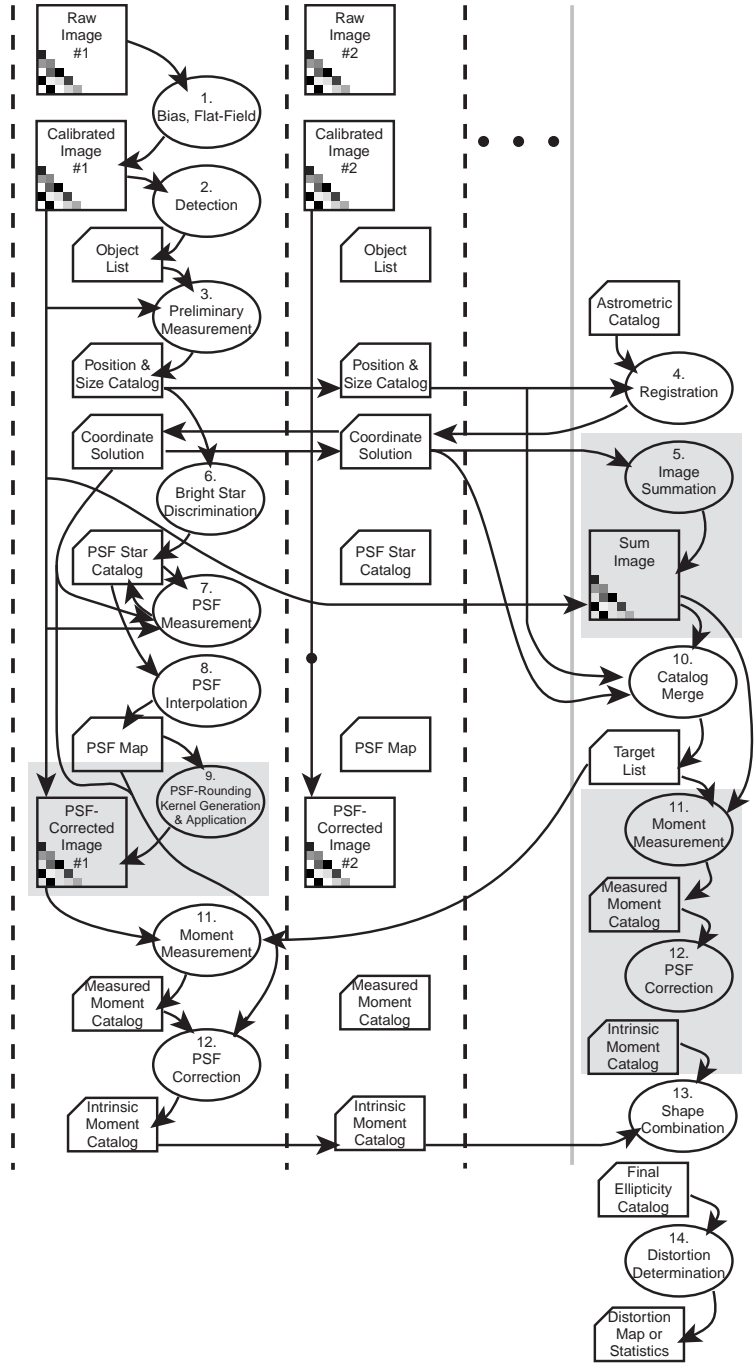}
\epsscale{1.}
\caption[]{
\small The data flow from raw images through a final lensing
distortion map or statistic is illustrated here, with each step
described in \S\ref{procedures}.  Shaded regions represent steps that
may not be used in all circumstances.  Implementation details will be
presented in \citet{Paper2}
}
\label{flowchart}
\end{figure}

\subsection{Exposure Processing}
The steps listed in this section must be performed on each 
exposure. For mosaic detectors, these steps should be done separately
for each channel of the detector if the PSF may change abruptly
at the boundaries between detectors.
\begin{enumerate}
\item {\em Bias Subtraction and Flat-Fielding:} These can be done in
the usual fashion.  Note that field distortion leads to significant
variation in pixel area in many CCD cameras.  If the flat-field image
is, as usual, obtained by exposing a source of uniform brightness,
then the flattened pixel data are a properly calibrated map of the
{\em surface brightness} of the sky rather than a map of the {\em
flux} from the sky.  This is actually what we want, though it means
that simple aperture photometry will lead to incorrect flux estimates.
\label{flatstep}
\item {\em Object Detection:} Available detection packages such as
{\sc FOCAS} \citep{FOCAS}, {\sc SExtractor} \citep{SEx}, or
{\sc ProFit} (P. Fischer, private communication) can be used to
identify all objects on 
the exposure with significance $\nu\gtrsim 3$ and produce a catalog
with  preliminary position, size, and ellipticity estimation.  Each of
these packages also produces a useful estimate of total magnitude,
which we will preserve.  The choice of object detection package is not
critical, because the objects that will have useful shape information
have sufficiently high significance that any decent detection
algorithm will work.
\label{sexstep}
\item {\em Preliminary Measurement:} Our shape-measurement algorithm
is applied to each detected object in order to obtain an accurate
Gaussian-weighted estimate of centroid, size ({\it i.e.} the $\sigma$
which maximizes significance), and shape.  Objects with saturated or
bad pixels are flagged at this point.
\label{prelimstep}
\item {\em Registration:} The map from pixel coordinates ${\bf x}_p$
to the global celestial coordinate grid ${\bf x}$ is established by
fitting the measured centroid of the bright objects to a collective
catalog ({\it e.g.} the average of all individual exposures' catalogs)
and/or an astrometric catalog.  This step is
critical as even slight misregistrations will produce a systematic
coherent ellipticity to the summed images.  The photometric offsets
between images are determined at this point as well.
\item {\em Image Summation (optional):} The exposures can now be
remapped to a common grid and photometric scale and summed to give a
deeper image.  If the flattened images are surface-brightness maps as
described in Step~\ref{flatstep}, then a simple interpolation can be
done when remapping the images---no Jacobian flux corrections are
needed. Recall the caveats about image combination in
\S\ref{combineimage}.  If shapes are to be measured from the summed
image, then we must at this point reconstruct the preliminary catalog
by executing Steps \ref{sexstep} and \ref{prelimstep} on the summed
image. 
\item {\em Identification of Stars:} As many unsaturated stellar images as
possible should be identified on the image for use in PSF
determination.  Stars are typically identified on the size-magnitude
plane.  This is a difficult task to execute without excessive human
intervention;
we want our star-finding algorithm to be flexible enough to identify
stars successfully in the presence of a spatially varying PSF, but
strict enough 
to insure that galaxies are not counted as stars.  Failure on either
count will lead to an erroneous estimate of the PSF size and shape on
some part of the exposure, which will lead to a false
feature in the shear field. Our algorithms for
star identification are described in \citet{Ja01}.  The algorithms are
particularly suited to identifying stars in the presence of a
spatially varying PSF size.
\item {\em PSF Measurement:} The shape-measurement algorithms are now
applied to the identified stars.  The integrals in
Equation~(\ref{inversion}) give ${\bf b}^\star$ at the location of each
star.  Note that the integral can readily be executed in the global coordinate
system ${\bf x}$ because we know the map from pixel coordinates, and
because pixel data are already in terms of surface brightness $I$.
Thus the effects of image distortion are removed at this step.
\label{psfmeasstep}
\item {\em PSF Interpolation:} With the PSF ${\bf b}^\star$ measured at
the locations of the stars, we need to fit a function ${\bf
b}^\star({\bf x})$ which describes the PSF at any location on the
image.  We use a polynomial to describe the variation of each
$b^\star_{pq}$ across the image.  Note that for complex or undersampled
PSFs (such as those on HST), interpolation of the 
$b^\star_{pq}$ components is much easier and more accurate than
interpolating a pixel-wise representation.  In any case this step is
again critical:  the interpolation scheme must be flexible enough to
follow PSF variation, but must remain well-behaved at the image edges and
in regions where PSF stars are sparse.  There must also be some form
of outlier rejection for PSF stars that have cosmic rays or near
neighbors contaminating the measurement, but a true excursion of the
PSF in some part of the image must not be rejected.
\label{psfinterpstep}
\item {\em PSF-Rounding Kernel (optional):} At this point we derive
and apply the kernel to remove the anisotropy from the PSF, if
desired.  Our method for doing so is described in \S\ref{kernel}.  
After the convolution it is necessary to repeat Steps
\ref{psfmeasstep} and \ref{psfinterpstep} so as to have an updated map
of the  PSF.
\label{kernelstep}
\newcounter{mysteps}
\setcounter{mysteps}{\value{enumi}}
\end{enumerate}

\subsection{Object Measurement}
We next outline the procedures which yield a shape estimate for every
object in the field. 
\begin{enumerate}
\setcounter{enumi}{\value{mysteps}}
\item {\em Creation of Target Catalog:} A list of all targets for shape
measurement must be compiled.  If there are $\lesssim5$ exposures of
each object, then a master target list can be produced by taking the
union of the individual exposure catalogs produced in
Step~\ref{sexstep}. One can use a very low detection threshold on the
individual exposures, then guard against noise detections by demanding
coincidence on 2 or more exposures.  If there are $\gtrsim 5$
exposures per object, then it will be necessary to run a detection
algorithm on a summed image in order to find all the potentially
useful target galaxies.  The target catalog should include an estimate
of the centroid (in global coordinates) and some observed size $s_o$
of each object. 
To avoid Kaiser's selection bias (\S\ref{centroidbias}), the criterion
for acceptance into the catalog must be some shape-independent
statistic.
\label{targetstep}
\setcounter{mysteps}{\value{enumi}}
\end{enumerate}

The following steps are performed for each object in the target
catalog:

\begin{enumerate}
\setcounter{enumi}{\value{mysteps}}
\item {\em Measurement of Observed Moments:} Given the global
coordinates of the object, one can determine all of the exposures on
which it should appear.  Note that for low-significance objects, this
may include exposures for which the object was missed in the
preliminary detection of Step~\ref{sexstep}.  We then use
Equation~(\ref{inversion}) to measure the Laguerre expansion ${\bf
b}^o$ of the image as observed on each exposure.  Once again the
integrals are performed in the global coordinate system to remove the
effects of optical distortions.  The integrals require choice of a
centroid and a size parameter $\sigma_o$ for the basis functions
$\psi_{pq}^\sigma$. We may use the approximate centroid from the
target catalog, and set the weight size $\sigma_o$ equal to the
typical observed object size $s_o$ from the target catalog.  The
covariance matrix for the ${\bf b}^o$ vector has the diagonal form
given in Equation~(\ref{bpqcov}) for sky-dominated galaxies.
Measurements contaminated by
saturated pixels or other defects may be rejected at this step.
\label{measurestep}
\item {\em Correction for PSF Effects:} There are two methods
available here.  The first sequence may be used if the rounding kernel
has been applied to remove the PSF bias.  In this case, if the
significance on the individual exposure is $\gtrsim 3$, we may proceed
as follows:
\begin{enumerate}
\item {\em Shape Measurement:} The image can be shifted, dilated,
and sheared until it passes the centroid, significance, and roundness
criteria embodied by Equations~(\ref{bcentroid})--(\ref{bsig}).  This
yields an optimal measure of the observed ellipticity ${\bf e}^o$.  If
the object is of significance $\nu_0\gtrsim3$ on the individual
exposure, then we can solve for centroid and size on each exposure
independently and still ignore the higher-order terms in the
uncertainty of Equation~(\ref{singexp}).
\item {\em Centroid Bias Correction:} Each measured ellipticity must
be corrected for the centroid bias (\S\ref{centroidbias}) using an
empirical value of $K$ in Equation~(\ref{centbias1}).
\item {\em Dilution Correction:} The PSF is interpolated to the
position of this object to determine the resolution parameter $R$
(Equation~[\ref{unweightedR}]) or the higher-order version derived in
Appendix~\ref{rfactorapp}. The correction for dilution to give the
image-plane (pre-seeing) ellipticity is then simply ${\bf e}^i = {\bf
e}^o/R$. The uncertainty ellipse for ${\bf e}$
(Equation~[\ref{errore}]) is therefore also scaled by $1/R$.  Note
that it is possible to obtain $e^i>1$ if the noise makes the target
appear smaller than the PSF.
\item {\em Averaging of Exposures:} The ${\bf e}^i$ from the
collection of exposures are averaged using the weighting procedures of
\S\ref{combineshapes}. Some form of outlier rejection is necessary to
remove objects contaminated by cosmic rays or other defects.
\end{enumerate}

If we are to perform the PSF bias correction analytically, or if the
significance per exposure is low, then it is best to average the
deconvolved moments rather than deriving ${\bf e}^i$ for each
exposure:
\begin{enumerate}
\item {\em Deconvolution of Moments:} The PSF decomposition ${\bf
b}^\star$ is interpolated to the position of the object, and the form
of the convolution matrix ${\bf C}({\bf b}^\star)$ is calculated using
the coefficients in Appendix~\ref{convolveapp}.  This matrix is
truncated at some order $p+q\le N$ and inverted; the deconvolved
(pre-seeing) Laguerre decomposition is ${\bf b}^i={\bf
C}^{-1}\cdot{\bf b}^o$.  Since this is a linear operation on ${\bf
b}$, the covariance matrix for ${\bf b}^i$ can be propagated from
the simple diagonal covariance matrix for ${\bf b}^o$
(Equation~[\ref{bpqcov}]). There is a subtlety involved in the choice
of weight scale $\sigma_o$:  in \S\ref{ponderpsf} we determined that
the ideal weight scale for the {\it deconvolved} image is
$\sigma_i^2=s_i^2+\sigma_\star^2=s_o^2$, where $s_i$ and $s_o$ are the
sizes of the pre- and post-seeing objects, and $\sigma_\star$ is the
size of the PSF.  The typical value of $s_o$ was placed in the target
catalog in Step~\ref{targetstep}.  According to
Equation~(\ref{deconvsizes}), our deconvolution formulae are simplest
if we choose the weight scale by
\begin{equation}
\label{convscale}
\sigma_o^2=\sigma_i^2+\sigma_\star^2=s_o^2+\sigma_\star^2. 
\end{equation}
Thus in Step~\ref{measurestep} we in fact want to use a weight scale
somewhat larger than the $s_o$ which maximizes the significance of
detection. 
\item {\em Combination of Moments:} From each exposure we have
estimated a ${\bf b}^i$ deconvolved moment vector, with known
covariance matrix.  We average these vectors to obtain a single best
estimate of ${\bf b}^i$.  We have a choice of weights to apply in
producing the average; an obvious choice is to weight each exposure
inversely with ${\rm Cov}(b_{20}^i \bar b_{20}^i)$, since $b_{20}$
carries most of the shape information.  Again some form of outlier
rejection is necessary at this step.
\item {\em Symmetrization of Noise:} Certain elements of the
propagated covariance matrix for the ${\bf b}$ vector must be zero in
order to avoid noise anisotropies that produce centroid bias
(\S\ref{centroidbias}).  This can be achieved by selectively adding
Gaussian noise to various elements of the combined moment vector.
\item {\em Determination of Shape:} Given the average of ${\bf b}^i$
from all exposures, we find the translation and shear that must be
applied to satisfy the centroid and roundness conditions
$b_{10}=b_{20}=0$.  The formulae of Appendix~\ref{finitetransforms} are
used for this.  The covariance matrix for ${\bf b}^i$ can be
propagated through the transformations as well.  The uncertainty in
the shape \bfe\ is then the square root of 
${\rm Var}(b_{20} \bar b_{20})/(b_{00}^2)$.
Note that we do not want to maximize the significance
by dilating to set $b_{11}=0$.  Our optimization criterion is that 
the shape have minimal error
after our transformations.  One could dilate the image to
satisfy this desire, but our choice of $\sigma_i$ should already have
us close to the optimum, according to the arguments of
\S\ref{ponderpsf}.
\end{enumerate}

After either of these procedures, we have a measure of the deconvolved
shape along with its uncertainty. 

\item {\em Combination of Different Wavelengths:} If we have imaged
the field in a variety of filters, then we will have obtained a shape
measurement in each filter.  The galaxy's shape and moments may depend
upon wavelength so we don't want to average together images or moments
measured in different filters.  We can, however, use the methods of
\S\ref{combineshapes} to produce a wavelength-averaged ${\bf e}^i$
which is maximally sensitive to shear.  Weighting each filter by the
error in its measured shape insures that we obtain the most
sensitivity from each galaxy regardless of its color.
\setcounter{mysteps}{\value{enumi}}
\end{enumerate}

After completion of all these steps, we have a catalog of all objects
in the field, specifying their 
location, magnitude, optimal shape measurement, and shape uncertainty.

\subsection{Determination of Shear}
With the shape catalog in hand we are close to the scientific goals.
The remaining step is
\begin{enumerate}
\setcounter{enumi}{\value{mysteps}}
\item {\em Generation of Shear Data:} The target galaxies are binned
by position, etc., into subsets for which we wish to determine a
shear.  The shapes (and their uncertainties) may have to be rotated,
{\it e.g.} into tangential coordinates about some mass center,
depending upon the shear statistic under study.
The formulae of \S\ref{shearsection} take the collection of
shapes and uncertainties and allow us to create an optimal shear
estimate, as well as to propagate the uncertainties in shape to the
shear measurement.
\end{enumerate}

\section{Conclusions}
We have attempted to produce, as rigorously as possible, an end-to-end
methodology for measuring gravitational distortion that has optimally
low noise, with calibration and noise levels derivable entirely from
the observations themselves.  We have succeeded in many, but not all,
aspects of the problem:
\begin{itemize}
\item  The measurement of individual galaxy shapes
appears to be optimal and has traceable noise characteristics; indeed
there even appears to be a straightforward way to handle undersampled
data and retain proper covariance information for the intrinsic
Laguerre coefficients $b^i_{pq}$ (\S\ref{pixels}).  The Gaussian
weights underlying the Laguerre decomposition are nearly optimal for
sky-dominated exponential-profile galaxies.  We may wish to reexamine
this scheme for the case of when the galaxy is
brighter than the sky background.

\item The correction of measured moments for the distorting and
diluting effects of the PSF can be effected to arbitrary accuracy
using the Laguerre-decomposition methods.  There will be a tradeoff
between elimination of systematic errors---which pushes toward
inclusion of higher-order terms in the deconvolution---and the
minimization of measurement noise from high-frequency terms.  It is
not clear if the Laguerre method is optimal with, for example, Airy
PSFs with sharp cutoffs in $k$ space; but the method should be better
than those yet applied.

\item We have identified methods to work around two sources of bias
that arise from PSF ellipticities even in the presence of perfect
deconvolution. 
\item Measurements of galaxy shapes from different exposures or filter
bands can be optimally combined with standard least-squares
techniques, since we know the ${\bf e}$ uncertainty from each individual
exposure. 
\item The determination
of lensing distortion from the ensemble of galaxy shapes has a
straightforward, optimal, calibratable solution in the case of no
measurement noise.  In the presence of measurement noise, however, it
is necessary to have some knowledge of the noiseless shape distribution
to get the calibration factor exactly correct.  Our approximations,
however, seem to suffice to get accuracies of 5--10\%.
\end{itemize}

A detailed performance comparison of our methods with other authors'
is beyond the scope of the paper.  We can guess, however, that the
reduction of measurement noise relative to a carefully weighted
implementation of KSB will be slight, perhaps a factor of 1.5.  But
our methods, like those of K00 and \citet{Re01}, are formally valid
for any reasonable PSF, and hence we expect to have much-reduced
systematic errors.  Indeed it is likely that the biases of
\S\ref{centroidbias} have not been extensively tested because they
were lost under the larger errors in PSF correction (as also noted by
K00).

Our methods share elements with many previous proposals.  An
aesthetic difference is that we retain the geometric meaning of
ellipticity by, in effect, using adaptively shaped weights instead of
fixed circular weights.  This makes the ``shear polarizability'' a
purely geometric effect.  As a consequence we can examine the $P(e)$
distributions and find ways
to exploit surface brightness, $b_{22}$, or color information
to separate the spheroid galaxies and weight them more heavily to
reduce the shape noise.

It will be of interest to see how our method compares to the
commutator method of K00.  One would hope that the two independent
methods could be applied to the same dataset and yield the same
results, bolstering our confidence in these very difficult measurements.

In a succeeding paper \citep{Paper2} we will present some of the
implementation details for the analytical methods here, and test the
methods on real and simulated data.  \citet{F00, Sm01, Wi00}  make use
of portions of this methodology, so the systematic-error tests and
calibration tests in those papers already serve as demonstrations.
Upcoming precision measurements of cosmic shear will make even greater
demands upon the systematic-error reduction and accurate calibration
that our methods offer.

\acknowledgements
This work was supported by grant AST-9624592 from the National Science
Foundation.  We would like to thank Phil Fischer, Deano Smith, Tony
Tyson, Jordi Miralda-Escud\'e, and Dave Wittman for many discussions
on these methodologies, 
and help in implemented various image-processing algorithms.  Thanks
also to all our collaborators who have waited several years for 
a coherent explanation of the methods we are using.

\newpage
\appendix
\section{Point Transformation for Laguerre Expansions}
\label{finitetransforms}
In \S\ref{xformsec} we derived the mapping matrices for the vector
${\bf b}$ of Laguerre-expansion coefficients when the underlying image
$I$ is transformed by an infinitesimal translation, dilation, or
shear.  In this Appendix we derive the coefficients of the mapping
matrices for finite transformations.  We have implemented these
transformations as methods for {\tt C++} classes that represent the
Laguerre expansions.

\subsection{Translation}
\label{Translation}
We defined the transformation ${\bf T}_z$ on the image by
\begin{eqnarray}
{\bf T}_z I(x,y) & = & I(x-x_0, y-y_0) \\
z & \equiv & { {x_0+iy_0}\over\sigma}.
\end{eqnarray}
We first define a translated raising operator $\hat
a_p^{\sigma\dagger}$ by
\begin{eqnarray}
\hat a_p^{\sigma\dagger} ({\bf T}_z I) & \equiv & {\bf T}_z
(a_p^{\sigma\dagger}I) \\
\label{transraise}
\Rightarrow \quad \hat a_p^{\sigma\dagger} & = & a_p^{\sigma\dagger} -
{z \over 2}.
\end{eqnarray}
The second line is apparent from examining the form of the raising
operators in \eqq{raiseops}.
We decompose the images into our eigenfunctions as $I=\sum
b_{pq}\psi_{pq}^\sigma$ and ${\bf T}_zI=\sum
b^\prime_{pq}\psi_{pq}^\sigma$; then we can express ${\bf T}_z$ as a
matrix operation on ${\bf b}$:
\begin{eqnarray}
{\bf b^\prime} & = & {\bf T}_z \cdot {\bf b} \\
b^\prime_{p^\prime q^\prime} & = & \sum T_{p^\prime q^\prime}^{pq}
b_{pq} \\
\label{transpq}
{\bf T}_z \psi_{pq}^\sigma & = & \sum T_{p^\prime q^\prime}^{pq} 
\psi^\sigma_{p^\prime q^\prime}. \\
\label{transint}
\Rightarrow \quad T_{p^\prime q^\prime}^{pq} & = & \sigma^2
\int\!d^2\!x\, ({\bf T}_z \psi_{pq}^\sigma) \bar\psi^\sigma_{p^\prime
q^\prime}. 
\end{eqnarray}
A direct integration of (\ref{transint}) yields the first
coefficient
\begin{equation}
T_{00}^{00} = e^{-|z|^2/4}.
\end{equation}
Applying the translated raising operator (\ref{transraise}) to the
definition (\ref{transpq}) of $T_{p^\prime q^\prime}^{pq}$ yields the
recursion relation
\begin{equation}
\label{precurtrans}
\sqrt{p^\prime} \,T_{(p^\prime-1)q^\prime}^{pq} 
- {z \over 2} \,T_{p^\prime q^\prime}^{pq} =
\sqrt{p+1} \,T_{p^\prime q^\prime}^{(p+1)q}.
\end{equation}
The same procedure using the $q$ raising operator and the lowering
operators gives the recursions
\begin{eqnarray}
\label{qrecurtrans}
\sqrt{q^\prime} \,T_{p^\prime(q^\prime-1)}^{pq} 
- {\bar z \over 2} \,T_{p^\prime q^\prime}^{pq} & = &
\sqrt{q+1} \,T_{p^\prime q^\prime}^{p(q+1)} \\
\sqrt{p^\prime+1} \,T_{(p^\prime+1)q^\prime}^{pq} 
- {\bar z \over 2} \,T_{p^\prime q^\prime}^{pq} & = &
\sqrt{p} \,T_{p^\prime q^\prime}^{(p-1)q} \\
\sqrt{q^\prime+1} \,T_{p^\prime(q^\prime+1)}^{pq} 
- {z \over 2} \,T_{p^\prime q^\prime}^{pq} & = &
\sqrt{q}\, T_{p^\prime q^\prime}^{p(q-1)} 
\end{eqnarray}
These two relations allow us to generate any $T_{p^\prime q^\prime}^{pq}$
recursively from $T_{00}^{00}$.  In particular, using the first 2 we
have
\begin{equation}
T_{00}^{pq} = { {(-1/2)^{p+q}} \over \sqrt{p!q!}} z^p \bar z^q
e^{-|z|^2/4}. 
\end{equation}
One can derive a closed-form expression for the general $T_{p^\prime
q^\prime}^{pq}$ using further recursion, but the expression involves a
double sum and is not particularly illuminating.  The most efficient
algorithm for computing all the coefficients is to note that the
generator for translation (Equation~[\ref{transgenerator}]) separates
into $p$- and $q$-dependent components.  The general matrix element
can therefore be expressed as
\begin{equation}
T_{p^\prime q^\prime}^{pq} = f(p,p^\prime) \bar f(q,q^\prime),
\end{equation}
with the above recursion relations for $T_{p^\prime q^\prime}^{pq}$
leading to
\begin{eqnarray}
f(p,0) & = &  { {(-z/2)^p} \over \sqrt{p!}} e^{-|z|^2/8}, \\
f(p,p^\prime+1) & = & \left[ \sqrt{p}\,f(p-1,p^\prime)
	+ { \bar z \over 2} f(p,p^\prime) \right ] / \sqrt{p^\prime+1}
\end{eqnarray}

\subsection{Dilation}
The dilation operation is defined by
\begin{equation}
{\bf D}_\mu I(x,y) = I(e^{-\mu}x, e^{-\mu}y) 
\end{equation}
As for the translation, we can define a dilated raising operator and
use it to derive a recursion relation for the coefficients of ${\bf
D}$:
\begin{eqnarray}
\hat a_p^{\sigma\dagger} ({\bf D}_\mu I) & = & {\bf D}_\mu
(a_p^{\sigma\dagger}I) \\
\label{dilateraise}
\Rightarrow \quad \hat a_p^{\sigma\dagger} & = &
\cosh\mu\,a_p^{\sigma\dagger} - \sinh\mu\,a_q^\sigma\\
\label{dilaterecur}
\Rightarrow \quad 
\sqrt{p+1}\,D_{p^\prime q^\prime}^{(p+1)q} & = & 
\cosh\mu\,\sqrt{p^\prime}\,D_{(p^\prime-1) q^\prime}^{pq}
- \sinh\mu\,\sqrt{q^\prime+1}\,D_{p^\prime(q^\prime+1)}^{pq}
\end{eqnarray}
Using the recursion operator and its $q$ equivalent, we can generate
any desired coefficient from $D_{pq}^{00}$.  With the direct
integration analogous to Equation~(\ref{transint}), we derive
\begin{equation}
D_{pq}^{00} = e^\mu {\rm sech}\mu (\tanh\mu)^p \, \delta_{pq}  .
\end{equation}
In deriving this we make use of the identity \citep{Abram}
\begin{equation}
\label{laguerrealpha}
L_q^{(m)}(\alpha x) = \sum_{k=0}^q { q+m \choose q-k } \alpha^k
(1-\alpha)^{q-k} L_k^{(m)}(x).
\end{equation}
In fact with this identity one can derive any $D_{p^\prime
q^\prime}^{pq}$ by direct integration, but the closed form is a sum
over $k$ that is again not particularly useful, as the recursion
relation (\ref{dilaterecur}) is a faster way to generate the
coefficients.

\subsection{Shear}
A shear $\eta$ oriented on the $x$ axis gives the transformation
\begin{equation}
{\bf S}_\eta I(x,y) = I(e^{-\eta/2}x, e^{\eta/2}y).
\end{equation}
When we define the transformed raising operator $\hat
a_p^{\sigma\dagger}$ as for the translation and dilation, we find the
operator and consequent recursion relation to be
\begin{eqnarray}
\label{shearraise}
\Rightarrow \quad \hat a_p^{\sigma\dagger} & = &
\cosh(\eta/2)\,a_p^{\sigma\dagger} - \sinh(\eta/2)\,a_p^\sigma\\
\label{shearrecur}
\Rightarrow \quad 
\sqrt{p+1}\,S_{p^\prime q^\prime}^{(p+1)q} & = & 
\cosh(\eta/2)\,\sqrt{p^\prime}\,S_{(p^\prime-1) q^\prime}^{pq}
- \sinh(\eta/2)\,\sqrt{p^\prime+1}\,S_{(p^\prime+1)q^\prime}^{pq}.
\end{eqnarray}
The recursion again simplifies by noting
that the shear generator in
Equation~(\ref{sheargen}) separates into $p$ operands and $q$
operands.  Hence the matrix elements must be expressible as
$S^{pq}_{p^\prime q^\prime} = f(p,p^\prime)f(q,q^\prime)$.  We can
determine the function $f(p,0)$ by direct integration of the
$S_{00}^{p0}$ matrix element, which yields (for $p$ even)
\begin{eqnarray}
S_{00}^{p0} & = & { \sqrt{p!} \over {(p/2)!}} {\rm sech} (\eta/2)
\left( { {-\tanh(\eta/2)} \over 2} \right)^{p/2} \\
\Rightarrow \quad
S_{00}^{pq} & = & { \sqrt{p!q!} \over {(p/2)!(q/2)!}} {\rm sech} (\eta/2)
\left( { {-\tanh(\eta/2)} \over 2} \right)^{(p+q)/2}.
\end{eqnarray}
The coefficients vanish if $p$ or $q$ are odd.
The recursion relation then generates any desired coefficient.
For a shear oriented at position angle $\beta$, the coefficients
acquire an additional phase factor
$\exp[i(p^\prime-q^\prime-p+q)\beta]$.

\section{Convolution of Laguerre Expansions}
\label{convolveapp}
We wish to derive the coefficients which express the convolution of
two eigenfunctions as a new sum over eigenfunctions, as defined in
Equation~(\ref{convcoeffs}). This will be easier if we work in $k$
space, with the convolution turned into a multiplication as in
Equation~(\ref{kconv}), with the $k$-space eigenfunctions given in
Equation~(\ref{klaguerre}).  

A rapidly computable recursive formulation of the coefficients is
again derivable from the raising operator.  From the form of the
$k$-space raising operator (\ref{kraise}) we see that if
$\sigma_o^2 = \sigma_i^2 + \sigma_\star^2 $, then
\begin{eqnarray}
\sigma_o \tilde a_p^{\sigma_o\dagger} \tilde I_o & = & 
(\sigma_i \tilde a_p^{\sigma_i\dagger} \tilde I_i) \tilde I_\star
+ \tilde I_i (\sigma_\star \tilde a_p^{\sigma_\star\dagger} \tilde
I_\star) \\
\label{convolverecur}
\Rightarrow\quad
\sigma_\star \sqrt{p_\star+1}\, C_{p_o q_o}^{p_i q_i (p_\star+1) q_\star}
& = & \sigma_o \sqrt{p_o}\, C_{(p_o-1) q_o}^{p_i q_i p_\star q_\star}
 - \sigma_i \sqrt{p_i+1}\, C_{p_o q_o}^{(p_i+1) q_i p_\star q_\star}.
\end{eqnarray}
An equivalent manipulation with the lowering operator yields the
recursion relation
\begin{equation}
\label{convolvelower}
\sigma_o \sqrt{p_o}\, C_{(p_o+1) q_o}^{p_i q_i p_\star q_\star}
 =  \sigma_i \sqrt{p_i}\, C_{p_o q_o}^{(p_i-1) q_i p_\star q_\star}
 + \sigma_\star \sqrt{p_\star}\, C_{p_o q_o}^{p_i q_i (p_\star-1) q_\star}.
\end{equation}
These recursion relations, and their $q$ equivalent, will allow us to
derive any coefficient if we know $C_{p_o q_o}^{p_i q_i 0 0}$,
{\it i.e.} if we can calculate the effect of multiplication by the
Gaussian $\tilde\psi_{00}^{\sigma_\star}$.  This is straightforward if
we recall that $m_o=m_i$ for $m_\star=0$, and use
Equation~(\ref{laguerrealpha}): 
\begin{eqnarray}
 2\pi \tilde \psi_{p_i q_i}^{\sigma_i} \tilde \psi_{00}^{\sigma_\star}
& = & 2\sqrt\pi \sum_{p_o=0}^{p_i} \sum_{q_o=0}^{q_i} 
\sqrt{ {p_i! q_i!} \over {p_o! q_o!}} 
{ {D^{(p_o+q_o)/2} (1-D)^{q_i-q_o}} \over (q_i-q_o)! }
\tilde \psi_{p_o q_o}^{\sigma_o}\\
\label{cpq00}
\Rightarrow \quad C_{p_o q_o}^{p_i q_i 0 0} & = & 
2 \sqrt{\pi} \sqrt{{p_i \choose p_o} { q_i \choose q_o}}
D^{(p_o+q_o)/2} (1-D)^{(p_i-p_o+q_i-q_o)/2}, \\
D & \equiv & { \sigma_i^2 \over \sigma_o^2} = 
{ \sigma_i^2 \over {\sigma_i^2 + \sigma_\star^2}} 
= 1 - {\sigma_\star^2 \over \sigma_o^2}.
\end{eqnarray}
The parameter $D$ is the ``deconvolution ratio,'' with $D=1$ being the
limit of perfect resolution and $D=0$ meaning a PSF much broader than
the image.  This expression and the recursion relations are both
separable into $p$- and $q$-dependent terms, so we can simplify the
computation of the matrix elements by using the expression
\begin{eqnarray}
\label{csep}
C_{p_o q_o}^{p_i q_i p_\star q_\star} & = & 2\sqrt{\pi}
\left[ \sqrt{ p_i! p_\star! \over p_o! \Delta!} G(p_o,p_i,p_\star)
\right]
\left[ \sqrt{ q_i! q_\star! \over q_o! \Delta!} G(q_o,q_i,q_\star)
\right] \\
\label{defDelta}
\Delta & \equiv & p_i + p_\star - p_o = q_i + q_\star - q_o \ge 0.
\end{eqnarray}
Terms for which the conditions in the second line are not met are
zero.  The recursion relations and specific values given above for $C$
can be recast for the function $G$ as follows:
\begin{eqnarray}
\label{grecur}
G(p_o+1, p_i, p_\star) & = & 
{ \sigma_i \over \sigma_o } G(p_o, p_i-1, p_\star) +
{ \sigma_\star \over \sigma_o } G(p_o, p_i, p_\star-1) \\
\label{ginit}
G(0,p_i,p_\star) & = & (-1)^{p_\star} 
{ p_i + p_\star \choose p_i }
\left( {\sigma_i \over \sigma_o}\right)^{p_\star}
\left( {\sigma_\star \over \sigma_o}\right)^{p_i}
\end{eqnarray}
The symmetry between initial and PSF images is clear in these equations.
There is a consequent closed form for $G(p_o,p_i,p_\star)$, but it is
again not particularly illuminating, and the recursive form is stable
and faster for computation.

In the case where the PSF is a unit-flux Gaussian,
Equation~(\ref{cpq00}) can be used to give the observed ${\bf b}^o$ in
terms of the intrinsic decomposition:
\begin{equation}
b^o_{p_o q_o}  =  \sum_{j=0}^\infty
D^{(p_o+q_o)/2} \sqrt{ {p_o+j \choose p_o} { q_o+j \choose q_o}}
 (1-D)^j  b^i_{(p_o+j)(q_o+j)}  \qquad (p_o\ge q_o).
\end{equation}
Note that the convolution matrix ${\bf C}$ is in this case block-diagonal, as
the $m$ states do not mix, and also upper triangular, as $p_o,q_o\le
p_i, q_i$ for non-zero elements.  The inverse (deconvolution) matrix
can in this case be expressed in closed form:
\begin{equation}
b^i_{p_i q_i}  =  \sum_{j=0}^\infty
D^{-(p_i+q_i)/2} \sqrt{ {p_i+j \choose p_i} { q_i+j \choose q_i}}
\left({ {1-D} \over D}\right)^j (-1)^j b^o_{(p_i+j)(q_i+j)} .
\end{equation}

\section{Second-order Formulae for PSF Dilution}
\label{rfactorapp}
Here we describe a refinement to the resolution parameter $R$ defined
in \eqq{unweightedR}.  The observed ellipticity is $R$ times the true
ellipticity in the special cases of unweighted moments or when 
the PSF is a circular Gaussian and the galaxy an elliptical Gaussian.
In these cases the factor $R$ can be expressed as
\begin{eqnarray}
\label{gaussR}
R & = & 1 - {s^2_\star \over s^2_o}, \\
s^2 & \equiv & \langle (x^2 + y^2)/2 \rangle = \sigma^2 \cosh\eta.
\end{eqnarray}
In the second line we have assumed that the methods of
\S\ref{noseeing} have been used to shear the object by 
$\eta$ to produce something that is ``round'' under a
Gaussian weight of optimal size $\sigma$.  

Here we derive a form for $R$ which is applicable to the case in which
the galaxy has homologous elliptical isophotes with $\eta\ll 1$, but
with an arbitrary radial profile.  In the language of our Laguerre
coefficients, we have $b_{11}=0$ by proper choice of
$\sigma$, and $b_{pp}$ are arbitrary for $p\ge 2$.  By \eqq{shearbpq}
this slightly out-of-round object has
$b_{20}=\sqrt{2}\eta(b_{00}-b_{22})/4$. All odd-indexed coefficients are
zero. 

We also take the PSF size
$\sigma_\star$ to be small compared to the intrinsic object size
$\sigma_i$.  The action of this small isotropic convolution ${\bf
C}_{\sigma_\star}$ on the moments 
is the same as a transformation in which the image $I$ is replaced by
the average of two versions displaced by $\pm \sigma_\star$ in the $x$
direction, followed by a similar infinitesimal spread in the $y$
direction.  Defining $z=\sigma_\star / \sigma$ and using a
second-order version of the generator \eqq{transgenerator}, we can
show that
\begin{eqnarray}
{\bf C}_{\sigma_\star} & = & {1 \over 4} \left({\bf T}_z + {\bf
T}_{-z}\right)
 \left({\bf T}_{iz} + {\bf T}_{-iz}\right) \\
 & \approx & 1 + {z^2 \over 2}\left[ a_p^\dagger a_q^\dagger
	+ a_p a_q - p - q -1 \right]
\end{eqnarray}
When this convolution acts upon our original slightly-elliptical
object, the resulting object has Laguerre coefficients ${\bf
b}^\prime$ with
\begin{eqnarray}
b^\prime_{00} & = & b_{00}(1 - z^2/2) \\
b^\prime_{20} & = & b_{20}(1-3z^2/2) \\
b^\prime_{22} & = & b_{22}(1-5z^2/2)
\end{eqnarray}
We then need the value $\eta^\prime$ of the shear that will make this
new object appear round.  According to \eqq{linearsoln}, this will
give
\begin{eqnarray} 
\eta^\prime & = & { \eta (b_{00}-b_{22})(1-3z^2/2) \over
	1 - z^2/2 -b_{22}(1-5z^2/2) } \\
\Rightarrow\qquad R \equiv {\eta^\prime \over \eta} & = &
1 - z^2\left( 1 + {2b_{22} \over b_{00}-b_{22}} \right) + O(z^4) \\
 & \approx & 1 - {\sigma_\star^2 \over s^2} , \\
s^2 & \equiv & \sigma^2 { b_{00} - b_{22} \over b_{00} + b_{22} }.
\end{eqnarray}
Note that the kurtosis measure $a_4$ defined in \eqq{sigeta} is the
same as $b_{22}/b_{00}$ which appears here.  We make the ansatz that
the correct form for $R$ in the case of finite dilution or finite $e$
is the Gaussian form \eqq{gaussR} with the kurtosis term added:
\begin{eqnarray}
\label{Ransatz}
{\bf e}_i & = & {\bf e}_o / R, \\
R & = & 1 - {s^2_\star \over s^2_o}, \\
s^2 & \equiv & { 1 - a_4 \over 1 + a_4 } \sigma^2 \cosh \eta .
\end{eqnarray}
Note that we apply the kurtosis correction to the PSF size measure
$s^2_\star$ in the same way as for the object, to give a well-behaved
correction for poorly-resolved objects.

\end{document}